\documentclass[twocolumn]{aastex62}

\newcommand\elec{e$^{-}$}

\graphicspath{{./}{figures/}}

\submitjournal{PASP}

\shorttitle{Palomar Gattini-IR}
\shortauthors{K. De et al.}

\begin{document}

\title{Palomar Gattini-IR: Survey overview, data processing system, on-sky performance and first results}

\correspondingauthor{Kishalay De}
\email{kde@astro.caltech.edu}

\author[0000-0002-8989-0542]{Kishalay De}
\affil{Cahill Center for Astrophysics, California Institute of Technology, 1200 E. California Blvd. Pasadena, CA 91125, USA.}

\author[0000-0001-9315-8437]{Matthew J. Hankins}
\affil{Cahill Center for Astrophysics, California Institute of Technology, 1200 E. California Blvd. Pasadena, CA 91125, USA.}

\author{Mansi M. Kasliwal}
\affil{Cahill Center for Astrophysics, California Institute of Technology, 1200 E. California Blvd. Pasadena, CA 91125, USA.}

\author{Anna M. Moore}
\affil{Research School of Astronomy and Astrophysics, Australian National University, Canberra, ACT 2611, Australia}

\author{Eran O. Ofek}
\affil{Department of Particle Physics \& Astrophysics, Weizmann Institute of Science, Rehovot 76100, Israel}

\author{Scott M. Adams}
\affil{Cahill Center for Astrophysics, California Institute of Technology, 1200 E. California Blvd. Pasadena, CA 91125, USA.}

\author{Michael C. B. Ashley}
\affil{School of Physics, University of New South Wales, Sydney NSW 2052, Australia}

\author{Aliya-Nur Babul}
\affil{Department of Astronomy, Columbia University, 550 West 120th Street, New York, NY 10027, U.S.A.}

\author{Ashot Bagdasaryan}
\affil{Cahill Center for Astrophysics, California Institute of Technology, 1200 E. California Blvd. Pasadena, CA 91125, USA.}

\author{Kevin B. Burdge}
\affil{Cahill Center for Astrophysics, California Institute of Technology, 1200 E. California Blvd. Pasadena, CA 91125, USA.}

\author{Jill Burnham}
\affil{Caltech Optical Observatories, California Institute of Technology, Pasadena, CA 91125, USA}

\author{Richard G. Dekany}
\affil{Caltech Optical Observatories, California Institute of Technology, Pasadena, CA 91125, USA}

\author{Alexander Declacroix}
\affil{Caltech Optical Observatories, California Institute of Technology, Pasadena, CA 91125, USA}

\author{Antony Galla}
\affil{Research School of Astronomy and Astrophysics, Australian National University, Canberra, ACT 2611, Australia}

\author{Tim Greffe}
\affil{Caltech Optical Observatories, California Institute of Technology, Pasadena, CA 91125, USA}

\author{David Hale}
\affil{Caltech Optical Observatories, California Institute of Technology, Pasadena, CA 91125, USA}

\author{Jacob E. Jencson}
\affil{Cahill Center for Astrophysics, California Institute of Technology, 1200 E. California Blvd. Pasadena, CA 91125, USA.}
\affil{Steward Observatory, University of Arizona, 933 North Cherry Avenue, Tucson, AZ85721-0065, USA}

\author{Ryan M. Lau}
\affil{Institute of Space \& Astronautical Science, Japan Aerospace Exploration  Agency,  3-1-1 Yoshinodai,  Chuo-ku,  Sagamihara, Kanagawa 252-5210, Japan}

\author{Ashish Mahabal}
\affil{Cahill Center for Astrophysics, California Institute of Technology, 1200 E. California Blvd. Pasadena, CA 91125, USA.}

\author{Daniel McKenna}
\affil{Caltech Optical Observatories, California Institute of Technology, Pasadena, CA 91125, USA}

\author{Manasi Sharma}
\affil{Department of Physics, Pupin Hall, Columbia University, New York, NY 10027, USA.}

\author{Patrick L. Shopbell}
\affil{Cahill Center for Astrophysics, California Institute of Technology, 1200 E. California Blvd. Pasadena, CA 91125, USA.}

\author{Roger M. Smith}
\affil{Caltech Optical Observatories, California Institute of Technology, Pasadena, CA 91125, USA}

\author{Jamie Soon}
\affil{Research School of Astronomy and Astrophysics, Australian National University, Canberra, ACT 2611, Australia}

\author{Jennifer Sokoloski}
\affil{Columbia Astrophysics Laboratory, Columbia University, 550 West 120th Street, New York, NY 10027, USA}

\author{Roberto Soria}
\affil{National Astronomical Observatories, Chinese Academy of Sciences, Beijing 100012, China}

\author{Tony Travouillon}
\affil{Research School of Astronomy and Astrophysics, Australian National University, Canberra, ACT 2611, Australia}

\begin{abstract}
Palomar Gattini-IR is a new wide-field, near-infrared robotic time domain survey operating at Palomar Observatory. Using a 30 cm telescope mounted with a H2RG detector, Gattini-IR achieves a field of view of 25 sq. deg. with a pixel scale of 8.7\,\arcsec in J-band. Here, we describe the system design, survey operations, data processing system and on-sky performance of Palomar Gattini-IR. As a part of the nominal survey, Gattini-IR scans $\approx 7500$ square degrees of the sky every night to a median 5$\sigma$ depth of $15.7$ AB mag outside the Galactic plane. The survey covers $\approx 15000$ square degrees of the sky visible from Palomar with a median cadence of 2 days. A real-time data processing system produces stacked science images from dithered raw images taken on sky, together with PSF-fit source catalogs and transient candidates identified from subtractions within a median delay of $\approx 4$ hours from the time of observation. The calibrated data products achieve an astrometric accuracy (RMS) of $\approx 0.7$\,\arcsec\, with respect to Gaia DR2 for sources with S/N $> 10$, and better than $\approx 0.35$\,\arcsec\, for sources brighter than $\approx 12$\,Vega mag. The photometric accuracy (RMS) achieved in the PSF-fit source catalogs is better than $\approx 3$\% for sources brighter than $\approx 12$ Vega mag and fainter than the saturation magnitude of $\approx 8.5$\,Vega mag, as calibrated against the 2MASS catalog. The detection efficiency of transient candidates injected into the images is better than $90$\% for sources brighter than the 5$\sigma$ limiting magnitude. The photometric recovery precision of injected sources is 3\% for sources brighter than 13 mag, and the astrometric recovery RMS is $\approx 0.9$\,\arcsec. Reference images generated by stacking several field visits achieve depths of $\gtrsim 16.5$ AB mag over 60\% of the sky, while it is limited by confusion in the Galactic plane. With a field of view $\approx 40\times$ larger than any other existing near infrared imaging instrument, Gattini-IR is probing the reddest and dustiest transients in the local universe such as dust obscured supernovae in nearby galaxies, novae behind large columns of extinction within the galaxy, reddened micro-lensing events in the Galactic plane and variability from cool and dust obscured stars. We present results from transients and variables identified since the start of the commissioning period.
\end{abstract}


\section{Introduction} 
\label{sec:intro}

Optical time domain astronomy has undergone a revolution in the last two decades due to the advent of wide field of view (FOV) telescopes equipped with large format CCD detectors. Combined with improvements in detector technology (faster readout and higher quantum efficiency), computing capabilities and the lower cost of detectors per pixel, several surveys have tiled large portions of the sky to provide exquisite time domain coverage of the optical variable sky over a large parameter space of areal coverage, depth, cadence and color. Examples include the Sloan Digital Sky Survey (SDSS; \citealt{York2000}), Skymapper \citep{Keller2007}, the Catalina Real-time Transient Survey (CRTS; \citealt{Drake2009}), the Palomar Transient Factory (PTF; \citealt{Law2009}), PanSTARRS \citep{Kaiser2010}, the All Sky Automated Survey for Supernovae (ASASSN; \citealt{Shappee2014}), Evryscope \citep{Law2015}, the Dark Energy Survey (DES; \citealt{DES2016}) the Asteroid Terrestrial-impact Last Alert System  (ATLAS; \citealt{Tonry2018}) and the Zwicky Transient Facility (ZTF; \citealt{Bellm2019}).

The first near infrared (NIR) sky survey was carried out as a part of the Two Micron Sky Survey (TMSS; \citealt{Neugebauer1969}) that covered 70\% of the sky and produced a catalog of $\sim 5700$ sources. It was followed by its deeper successor three decades later with the Two Micron All Sky Survey (2MASS; \citealt{Skrutsie2006}), which surveyed the entire sky in J, H and Ks bands down to a depth of $\approx 16, 15$ and $14$\,Vega mag respectively (for sources with a Signal to Noise ratio (SNR) of $\approx 10$). The Deep Near-Infrared Survey of the Southern sky (DENIS; \citealt{Epchtein1999}) also surveyed the southern sky to depths of 16.5 and 14 Vega mag in J abnd Ks bands respectively. The UKIRT Infrared Deep Sky Survey performed a deeper survey of a smaller fraction of the sky ($\sim 7500$ square degrees) to a depth of $K \approx 18.5$\,Vega mag \citep{Warren2007}. The VISTA hemisphere survey (VHS; \citealt{McMahon2013}), when combined data from the public VISTA surveys, will produce a deep NIR map of the entire southern sky down to $J\approx 20.2$\,Vega mag and $Ks\approx18.1$\,Vega mag. In the mid-infrared, the Wide-field Infrared Survey Explorer (WISE) all-sky survey \citep{Wright2010} created maps of the entire sky from 3.4 $\mu$m to 22 $\mu$m with 5$\sigma$ point source sensitivities ranging from $\approx 14$ AB mag (at 22 $\mu$m) to $\approx 19$ AB mag (at 3.6 $\mu$m). 

However, the time domain sky in the NIR remains largely unexplored due to limitations posed by the high sky background and detector technology. The brightness of the sky in the NIR wavebands arises from OH emission lines from the atmosphere, making ground based imaging limited by the high sky background noise. At the same time, the high cost of suitable detectors for the NIR (relative to optical CCD sensors) hinders the development of large format detectors that can perform fast imaging of large areas of the sky. Limited by the small FOV of most infrared imaging instruments, transient searches at these wavelengths have been largely limited to pencil beam surveys targeting small regions of the sky to hunt for variable and explosive events. For instance, a number of surveys have targeted luminous infrared galaxies (LIRGs) in the local universe at NIR wavelengths to search for supernovae (SNe) obscured by dust and hidden from optical surveys (e.g., \citealt{Mannucci2003, Cresci2007, Mannucci2007, Mattila2007, Kankare2008, Kankare2012, Miluzio2013, Kool2018}) owing to the high amount of star formation and dust in these galaxies. In the mid-infrared, the Spitzer Space Telescope \citep{Werner2004, Gehrz2007} has been used to conduct targeted surveys of variables (e.g., \citealt{Kozlowski2010, Freedman2011, Rebull2014, Boyer2015}) and transient phenomena (e.g., \citealt{Fox2011, Fox2012, Kasliwal2017, Jencson2019}). The Near-Earth Object WISE (NEOWISE; \citealt{Mainzer2011}) mission initially used enhanced data processing from the primary WISE all-sky survey to find solar system objects. It was subsequently repurposed as an all-sky time domain survey to study solar system objects through their mid-infrared emission \citep{Mainzer2014}. 

Given the relatively high cost of infrared detectors per pixel, two approaches have been proposed to probe the time domain sky at these wavelengths -- i) the use of fast optics to achieve a large pixel scale and field of view and ii) the use of alternative and cheaper semi-conductor detector technology (relative to HgCdTe). In the former case, the large pixel scale produces under-sampled point spread functions (PSFs) that can be reconstructed in data processing while suffering a degradation in sensitivity due to the high sky background \citep{Moore2016}. The latter case takes advantage of the high sky background in the NIR to be able to use lower cost InGaAs detectors that have higher read noise and dark current, and can be operated at higher temperatures \citep{Simcoe2019}. 

Here, we present Palomar Gattini-IR, a recently commissioned NIR time domain survey at Palomar Observatory in southern California, which serves as a working demonstration of the former approach to NIR time domain astronomy. Using a small 30-cm telescope housed in a clam-shell dome, Palomar Gattini-IR achieves a field of view of 25 square degrees and surveys $\approx 7500$ square degrees every night to a median depth of $\approx 16$\,AB mag in J band outside the Galactic plane. This produces an unprecedented 2 night cadence NIR coverage of the entire visible sky from Palomar (see \citealt{Moore2019} for an overview). 

This paper describes the instrument, survey modes and data processing system for the survey, along with on-sky performance and results from the survey commissioning phase on transients and variable science. Section \ref{sec:instrument_spec} summarizes the telescope optics, detector and its housing and readout electronics. Table \ref{tab:instrument} provides a summary of the instrument specifications, and the nominal sky survey. Section \ref{sec:ros} describes the robotic observing system and its performance. Section \ref{sec:drp} describes the data processing system designed to deliver science quality data products and transient candidates in real-time. Section \ref{sec:performace} describes the on-sky performance of the instrument derived from commissioning data. Section \ref{sec:results} provides and overview of the first results on infrared transients and variables identified since the start of commissioning operations. Section \ref{sec:summary} summarizes the survey status and planned developments.

\section{Hardware}
\label{sec:instrument_spec}
\subsection{Telescope and mount}
The requirement of a large areal survey speed required the use of a fast focal beam to achieve a large field view when feeding the single, moderate-sized, detector that we had available for the project (Section \ref{sec:detector}). The aperture of the telescope was set to 30 cm by a requirement of a single epoch 5  point source sensitivity of $\approx 16.0 - 16.5$\,AB mag in J band. Gattini-IR uses a 30 cm aperture, f/1.44 catadioptric optical telescope assembly (OTA) with 6 all-spherical elements, commercially available as the Terebizh TEC300VT by Telescope Engineering Inc. The optical design was evaluated for performance at J band combined with the 18 micron pixel size of the available detector. Fortuitously, with only a slight detector focus adjustment, sub-pixel performance was simulated to be possible across the entire field over a temperature range from 0 to 30 degree Celsius over a maximum field of 7 degrees, corresponding to a detector size of 52 mm (Moore et al. 2016 ). The OTA was assembled and tested to be diffraction limited on-axis at a wavelength of 633nm prior to delivery to Caltech. The telescope, mount and cryogenic system are located inside a clam-shell dome at Palomar Observatory, which also houses a compute server controlling the robotic observing system (`scheduler node' hereafter).  

However, design specifications of sub-pixel imaging and athermal focus quality over the entire detector plane have not been achieved during on-sky tests (Section \ref{sec:performace}). The original focusing mechanism relies on three variable screws that to be manually adjusted to best focus. The manual operation makes it difficult to settle such a fast telescope on a stable position. In addition, temporal variations in the image quality as a function of ambient temperature are not corrected dynamically during nightly operation or even revisited at regular intervals.  This issue will be corrected in the final quarter of 2019 with the addition of a robotic focusing mechanism. The system is based around a circular flexure element that is pre-stressed against three linear actuators. This design provides a level of anti-backlash to the system as well as helping with variable gravity vector effects during operation at various zenith angles. The three linear actuators will be remotely controlled and adjusted on regular basis to optimise the image PSF. The focus has been fully tested in the laboratory and gives a focus range of motion of 3\,mm and a resolution of 10\,$\mu$m.

The telescope and the cryogenic system housing the detector are attached to a to a GM3000 HPS robotic equatorial mount made by 10Micron Technology. The mount is designed to support up to 100 kg of instrument weight with a slew speed of up to 12$^\circ$/s. A pointing model was constructed for the mount after telescope installation and mount balancing, using a sample of 100 bright stars placed randomly across the visible sky. The resulting pointing model produces RMS residuals of $\approx 12$\,\arcsec\, ($\approx 1.5$ native detector pixels). 

\subsection{Detector}
\label{sec:detector}

Gattini-IR uses an engineering grade 2K $\times$ 2K Hawaii-2RG (H2RG) detector from Teledyne with a cut-off at 1.7 $\mu$m to be capable of J-band imaging and avoid the thermal infrared background beyond 2 $\mu$m. The pixel size is 18 $\mu$m, which provides a pixel scale on sky of 8.73\,\arcsec\, / pixel and a field of view of  $4.96^\circ \times 4.96^\circ$ when attached to the F/1.44 focal beam of the telescope. The mean quantum efficiency in J band is $\approx 70$\% \citep{Blank2011}.  The gain and read noise of the detector were measured in the laboratory and found to be $\approx 4.5$\,\elec/ADU and 25\,\elec respectively. The read noise together with the measured J band sky background with the instrument (at least $\sim 1000$\,\elec/s/pixel) allows for background limited imaging in exposures as short as $\approx 1$ second (the minimum allowed by the detector electronics). The detector does not have a shutter mechanism and is thus parked horizontally (facing the dome walls) during day time and poor weather conditions.

The detector non-linearity was measured using a constant light source in the laboratory and found to be $<3$\% up to 30,000 counts (136,000 \elec), which we nominally adopt as the linearity limit for the purposes of the data processing system. The detector saturates at $\approx 36,000$ counts ($\approx160000$ \elec). The number of hot pixels in the detector are measured periodically using darks inside the telescope dome.  Due to the absence of a shutter mechanism, darks are periodically acquired in the presence of observatory staff by covering the telescope tube with an aluminium coated cap and recording images. Due to the high background in our imaging application, the requirement for low dark current is not substantial. Nevertheless, the detector has been found to exhibit dark current levels of $\approx 0.9$ \elec/s under nominal operating conditions while hot pixels amount to 0.1\% of the detector pixels. 

Dead (non-linear and unresponsive) pixels are identified in the array using sky flats with different exposure times taken during twilight at the start and end of every observing night. Dead pixels amount to $\approx 2.7$\% of the total number of pixels in the detector, including the intentionally non-responsive reference pixels that are 4 deep on each edge of the detector. In additon, the detector has a triangular corner region of lower QE (measuring $700 \times 740$ pixels on the perpendicular sides, amounting to $\approx 12.4$\% of the detector area) due to the absence of an anti-reflection (AR) coating that was layered on the rest of the detector during a previous experimental phase. The region missing the AR coating was left to experimentally measure the change of the QE due to the presence of the AR coating, and has been verified to have a different zero-point (i.e. lower sensitivity) from the rest of the detector in commissioning data.

\begin{figure}
\centering
\includegraphics[width = 0.49\textwidth]{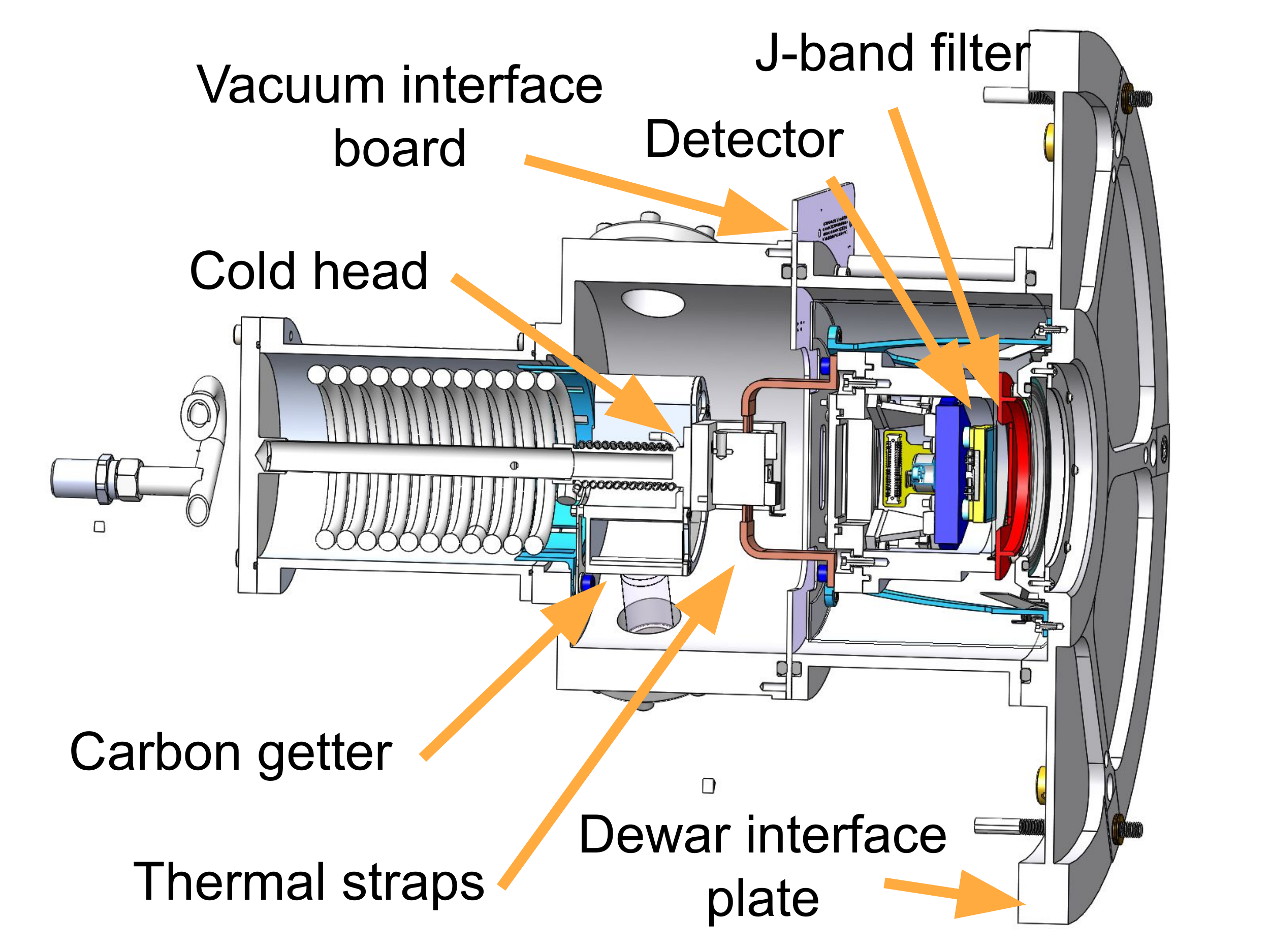}
\caption{Layout of the Gattini cryostat system}
\label{fig:cryostat}
\end{figure}

\begin{table*}
\centering
\caption{Summary of the Gattini observing system}
\textbf{Survey characteristics}\\
\begin{tabular}{ll}
\hline
\hline
Telescope & TEC300VT 30 cm\\
Location & 33$^\circ$ 21\,\arcmin\, 21\,\arcsec\, N, 116$^\circ$ 51\,\arcmin\, 54\,\arcsec\, W\\
 & Altitude 1712 m (Palomar Observatory)\\
Camera field dimensions & $4.96^\circ \times 4.96^\circ$\\
Camera field of view & 24.7 square degrees\\
Light sensitive area & 24.0 square degrees (97.2\% fill factor)\\
Filters & Gattini-J\\
Median image quality & FWHM 1.2 to 2.1 pixels\\
Median sensitivity & 15.7 AB mag outside Galactic plane ($|b| > 20^\circ$)\\
(64.8 s, 5$\sigma$)  & 15.3 AB mag in the Galactic plane ($|b| < 20^\circ$)\\
\hline
\end{tabular}\\
\vspace{0.5cm}
\textbf{Detector array}\\
\begin{tabular}{ll}
\hline
\hline
Make & Teledyne HAWAII2RG with 1.7$\mu$m cutoff\\
Size & 2048 $\times$ 2048 pixels\\
Pixel size & 18 $\mu$m/pixel\\
Plate scale & 8.73\,\arcsec/pixel\\
Gain & 4.54 \elec/ADU \\
Readout noise & 25 \elec\\
Dark current & 0.9 \elec s$^{-1}$\\
Typical sky background & 4500 \elec s$^{-1}$\\
Readout channels & 32\\
Linearity & $<3$\% up to 30,000 ADUs (136,000 \elec)\\
Saturation & $\approx 160000$ \elec\\
\hline
\end{tabular}\\
\label{tab:instrument}
\end{table*}

\subsection{Readout electronics}

The detector is read out using a detector controller system supplied by Astronomical Research Cameras, Inc. (ARC). The ARC controller chassis houses four 8-channel infrared video processor boards, a clock driver board, and a 250 MHz fiber optic timing board.  Each video board contains eight identical video processors, with each processor consisting of multiple stages having adjustable gains and offsets and an 18-bit analog to digital (A/D) converter. Together, these four 8-channel video boards read all 32 outputs of the H2RG simultaneously. The video boards also contain programmable DC supplies to supply the bias voltages to the detector. The clock driver board translates digital input signals into analog output signals for driving the clocks required to control the H2RG detector. The fiber optic timing board contains a digital signal processor (DSP) which generates the timing waveforms and communicates between the ARC controller and the host computer using a duplex fiber optic link to a PCIe interface board in the host.

The H2RG detector is read out non-destructively using conventional readout, though the pixel time has been reduced from what is typically used because previous studies \citep{Wizinowich2014} have shown improved noise performance. The pixel readout time is 6 $\mu$s, which represents 2\,$\mu$s settling and 4\,$\mu$s integration per pixel. The line overhead of 21 $\mu$s contains “pre-charge” pulses required by the H2RG as well as pulses required for initializing the ARC controller’s video board electronics. This amounts to a minimum frame time of about 834 ms.

\subsection{Detector housing}
The Gattini-IR cryostat is a custom built in-house fabricated assembly that was designed to minimize weight and volume by using 6061-T6 aluminum (Figure \ref{fig:cryostat}). The H2RG detector is mounted to a molybdenum block to match the CTE of the detector package which is cooled to about 100K by using a Brooks Poly Cold Compact Compressor charged with P-14 refrigerant. The cold head cools the activated carbon “getter” and heat is conducted away from the detector by using three braided copper thermal straps. In addition, the vacuum volume uses two room-temperature zeolite desiccants, freshly baked before pumping down the vacuum volume. Electrical connections to the vacuum volume is accomplished with the Vacuum Interface board (VIB) that carries both the detector and thermal management wiring. 

Two heaters are used -- a low power heater to stabilize the detector temperature, and a higher power (up to 50W, currently set to 8W) heater near the entrance window to guard against dew. The latter was installed towards the end of the commissioning period to remedy the accumulation of condensation on the window plate during periods of high humidity. The H2RG detector is optically filtered by a cold, J band interference filter. The dewar is connected to the compressor by 50 feet of armored flex hose with particular attention paid to maximizing the bend radius in an effort to protect the system from fatigue failure due to the observing cadence and all-sky motion envelope needed for Gattini-IR. To prevent the compressor from becoming too cold in the winter, a wooden enclosure with a thermostat controlled fan surrounds the compressor and uses the compressor waste heat to maintain the box temperature at $21^\circ$\,C.

\section{Robotic observing system}
\label{sec:ros}
\begin{figure*}
    \centering
    \includegraphics[width=0.7\textwidth]{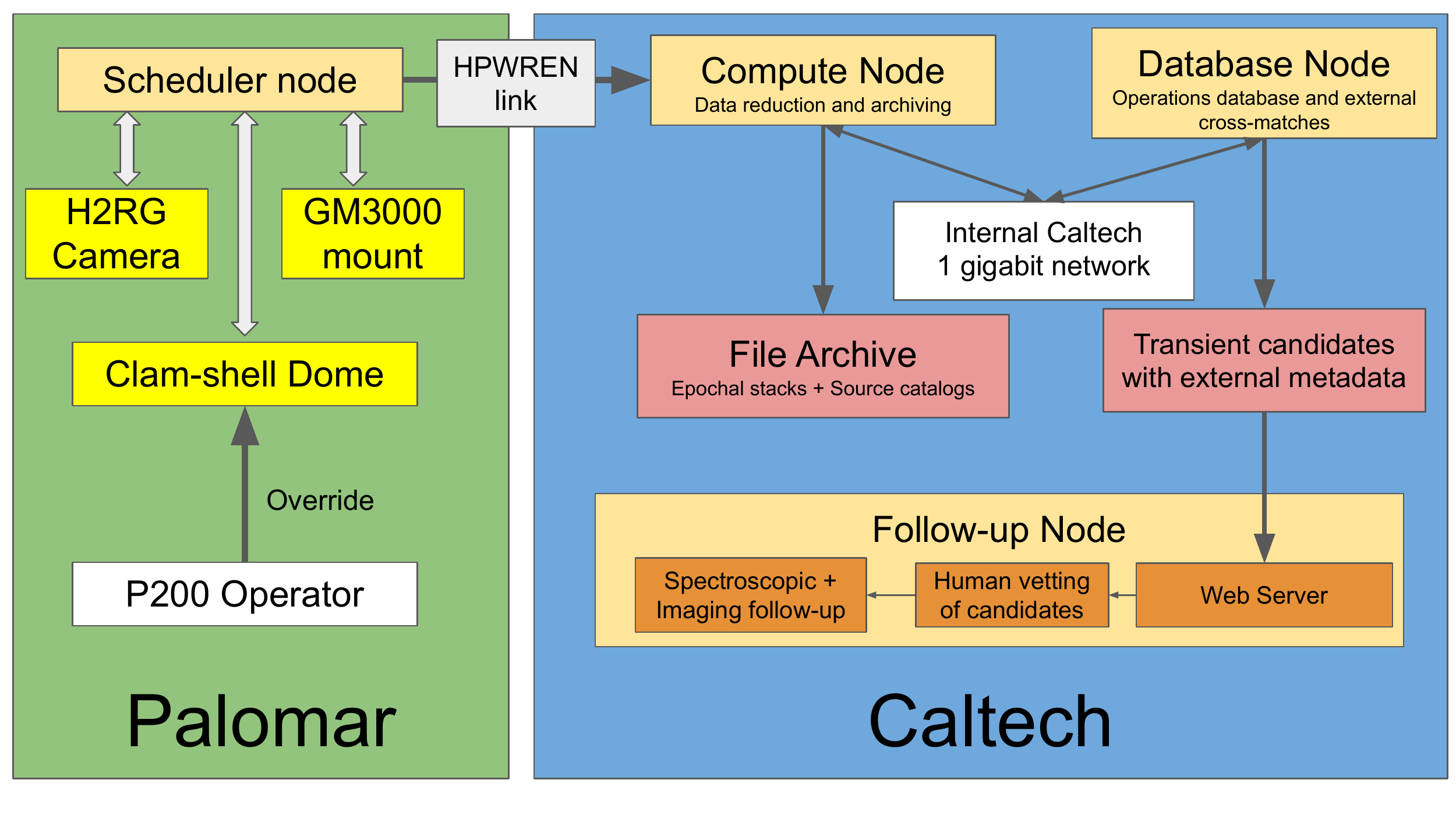}
    \caption{Overview of the Gattini observing, data reduction and follow-up system. The green box on the left show the observing components located at Palomar Observatory, while the blue box on the right shows the data reduction and follow-up components housed at Caltech. The light yellow boxes show the various computing servers involved in the observation scheduling, data reduction and archiving. The observing hardware components (in light yellow) -- telescope, camera and dome are controlled with the scheduler node at Palomar. The operator at the Palomar 200-inch telescope is capable of overriding the dome status in case of poor observing conditions, which halts the scheduler operations. The compute node at Caltech is reponsible for real-time data processing, while the database node hosts the \texttt{PSQL} server for metadata archiving and transient candidates. The compute node and data base node are linked with the internal Caltech 1-gigabit network. The red boxes show the long term science products produced in the  data processing system -- calibrated science images, source catalogs and transient candidates with external cross-match metadata. The follow-up node represents an independent web server at Caltech which allows human vetting of candidates for spectroscopic and imaging follow-up downstream. }
    \label{fig:overview}
\end{figure*}

The observing system (OS) for Gattini-IR serves as a primary control interface for the telescope, dome, mount and detector. The OS runs on a single compute server inside the telescope dome at Palomar, hosting an Intel Xeon E5-2620V3 2.4 GHz processor with 6 cores (12 threads) and 32 GB of RAM. Figure \ref{fig:overview} gives an overview of the interlinking between the OS and the data reduction server at Caltech, and the subsequent flow to science quality data products, human vetting and follow-up.

{}

\subsection{Scheduler and nominal survey operations}
\label{sec:scheduler}
Nightly operations are controlled by an automated telescope scheduler adapted from the publicly-available software\footnote{\url{https://github.com/ZwickyTransientFacility/ztf\_sim}} for the Zwicky Transient Facility \citep{Bellm2019b}. Unless the Palomar 200-inch telescope operator sets a weather override, the scheduler opens the dome and observes nightly between the times of nautical twilight. The celestial sphere north of $-28$ degrees is divided into 1329 fields with overlaps of 6\,\arcmin\, between fields. The fields are separated by an average of $\approx 4.86^\circ$ in the N-S direction and up to $\approx 4.9^\circ$ in the E-W direction, depending on declination. Under nominal survey operations, Gattini-IR observes fields over the entire visible celestial sphere from Palomar Observatory.  

\begin{figure}
    \centering
    \includegraphics[width=\columnwidth]{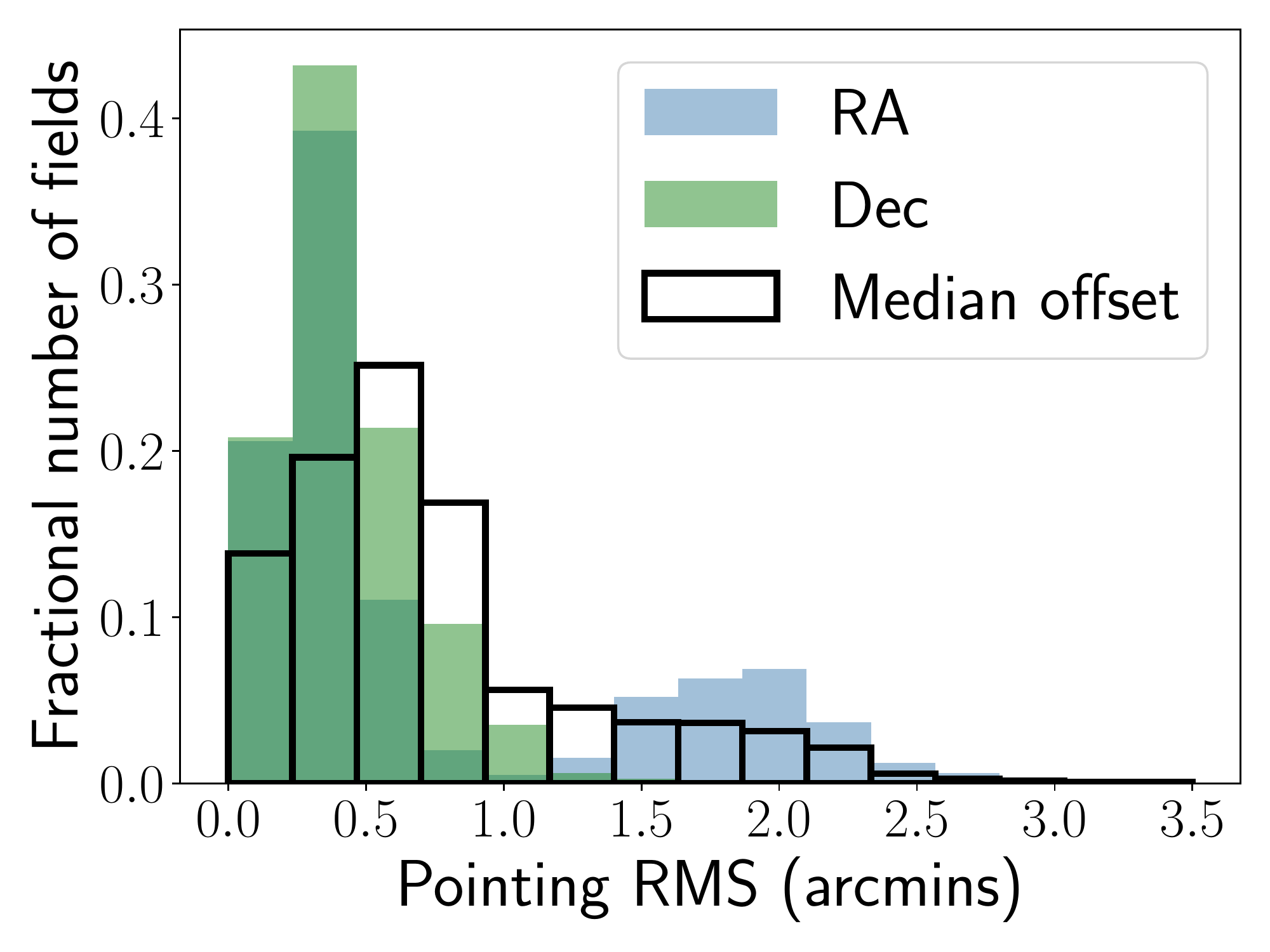}
    \caption{Histogram of the RMS residuals of the pointing accuracy of the mount across fields over the entire sky -- both in the RA axis and in the Dec axis. The white blocks show the distribution of the median total offset across all fields in the sky.}
    \label{fig:pointing_accuracy}
\end{figure}

Each field visit consists of a set of 8 dithered exposures with an exposure time of 8.1\,s each (total exposure time of 64.8 seconds per field visit). Multiple dithers were required to facilitate longer exposure times on the bright sky background, and to allow PSF reconstruction in data processing (Section \ref{sec:drp}) using the \texttt{Drizzle} algorithm \citep{Fruchter2002}. The amplitude of the dither is set to $\approx 3$\,\arcmin, which is randomized by a uniform distribution of 1\,\arcmin\, amplitude to sample random sub-pixel phases for individual point sources. Figure \ref{fig:pointing_accuracy} shows a distribution of the pointing RMS in the RA and Dec direction for fields distributed over the sky. As the dither amplitude is larger than the typical pointing RMS and median offset, the entire field region is covered during the dither sequence. A minimum number of 8 dithers was selected to obtain uniform coverage of the sub-sampled pixels across the drizzled images such that $\sigma_w / m_w < 0.15$ in the output drizzled images \citep{Gonzaga2012}, where $\sigma_w$ is the standard deviation of the output weight image from image reconstruction using \texttt{Drizzle} and $m_w$ is the median weight. The exposure time per dither and number of dithers were balanced as a trade-off between maximizing the volumetric survey speed and cadence over the sky (Section \ref{sec:volumespeed}).

\begin{figure*}
\includegraphics[width=0.49\textwidth]{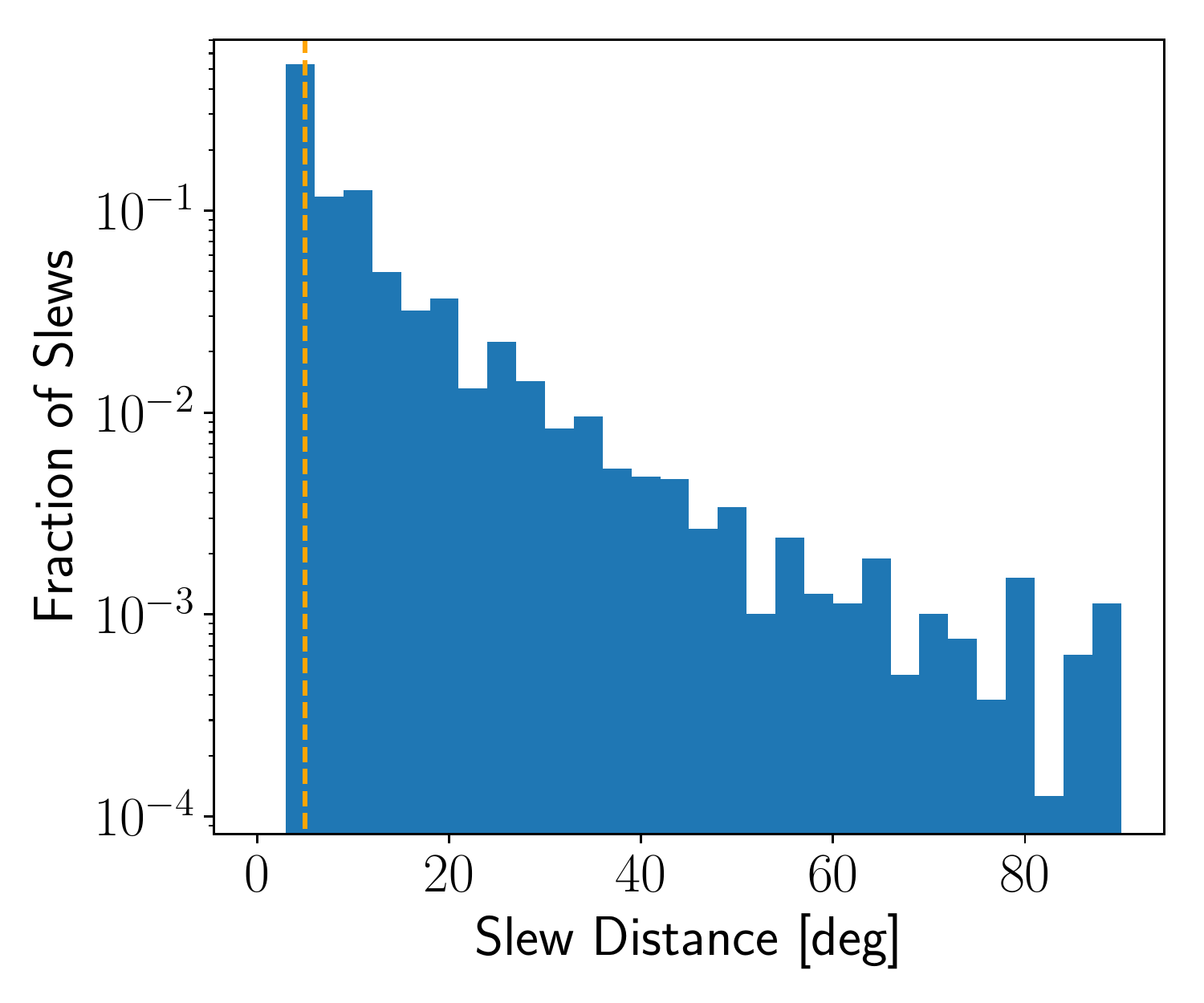}
\includegraphics[width=0.49\textwidth]{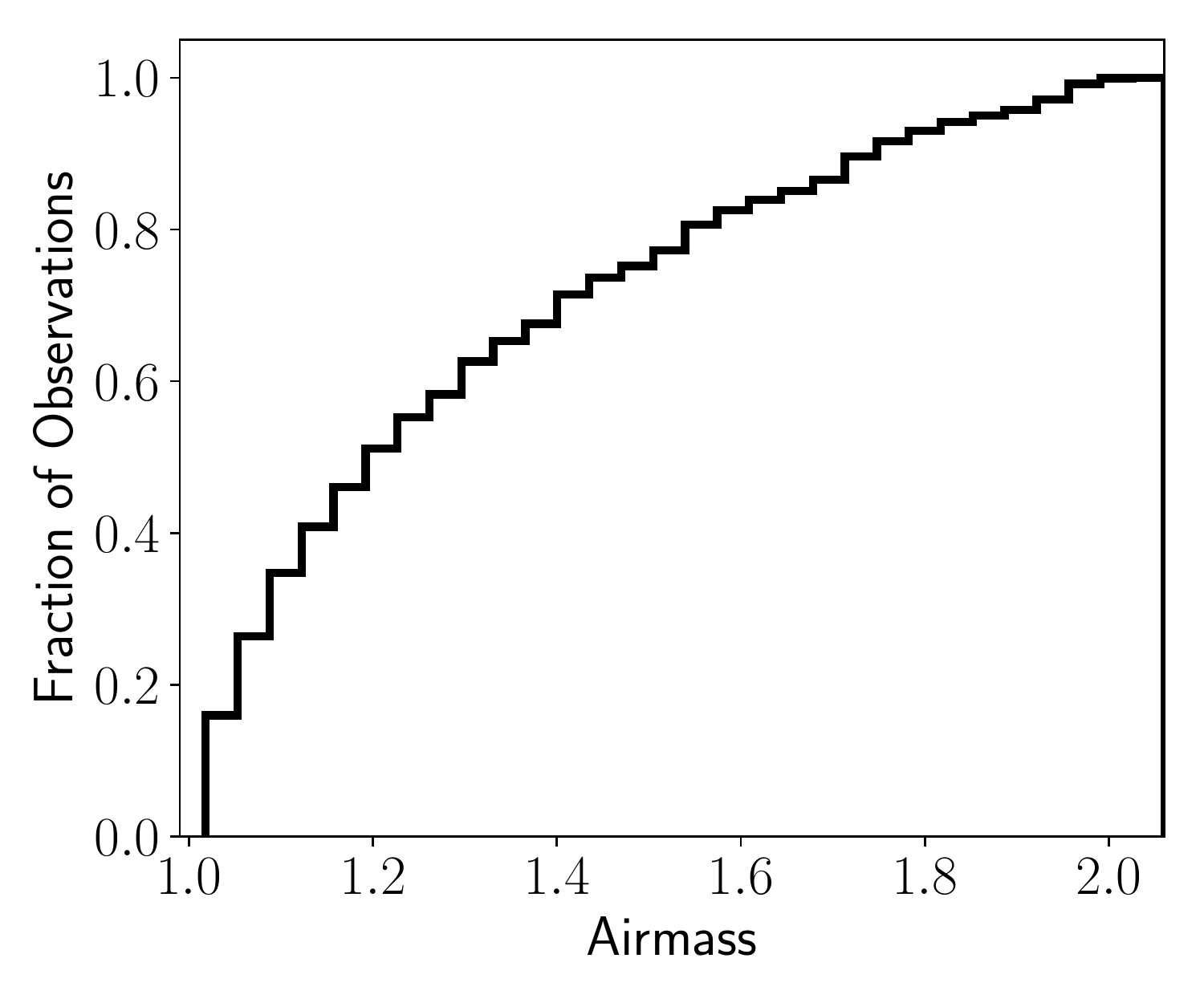}
\caption{(Left) Distribution of slew distances for the nominal survey over the course of six months of the commissioning period. The scheduler prioritizes field slews that involve smaller slews. The dashed orange line shows the average field spacing of $4.9^\circ$. (Right) Cumulative distribution of airmass for observations taken during the commissioning period. 80\% of observations are performed at airmass $< 1.6$ owing to the prioritization of fields at low airmass.}
\label{fig:scheduler_slews}
\end{figure*}

The scheduler currently employs a greedy algorithm that minimizes slew time and airmass while prioritizing fields that have not been observed as recently. This is done by selecting for the next field for observation that has the highest value of the following metric:
\begin{equation}
 \frac{t_{\mathrm{exp}}+t_{\mathrm{OH_{\mathrm{min}}}}}{t_{\mathrm{exp}}+t_{\mathrm{OH}}} \times \Delta t^{0.1} \times \frac{z_{\mathrm{min}}}{z} \mathrm{HA}_{\mathrm{weight}}
\end{equation}
where $t_{\mathrm{exp}}$ is the exposure time (in seconds), $t_{\mathrm{OH}}$ is the overhead (in seconds, including slew, settle, and camera initialization/readout) to slew from the current field to the next, $t_{\mathrm{OH_{\mathrm{min}}}}$ is the minimum field slew overhead measured in operations, $\Delta t$ is the time since the field was last observed (in days), $z$ is the airmass, $z_{\mathrm{min}}$ is the minimum observable airmass allowed by the mount, and $\mathrm{HA}_{\mathrm{weight}}$ is a weight factor based on the hour angle of the field. The exposure time and overhead factors serve to maximize the areal survey rate and the $\Delta t$ prioritizes fields in proportion to the time elapsed since the prior observation. The airmass factors are used to encourage observations to occur at the local meridian. The hour angle factor prioritizes setting fields because this would maximize the single-pass sky coverage with the original survey rate which was intended to be faster than sidereal. Figure \ref{fig:scheduler_slews} shows a distribution of the slew distances and observation airmasses from data taken during six months of the commissioning period using this scheduling algorithm. As shown, the scheduling algorithm prioritizes observations that minimize the slew distance and airmass of the observation.

\begin{figure*}
\includegraphics[width=0.49\textwidth]{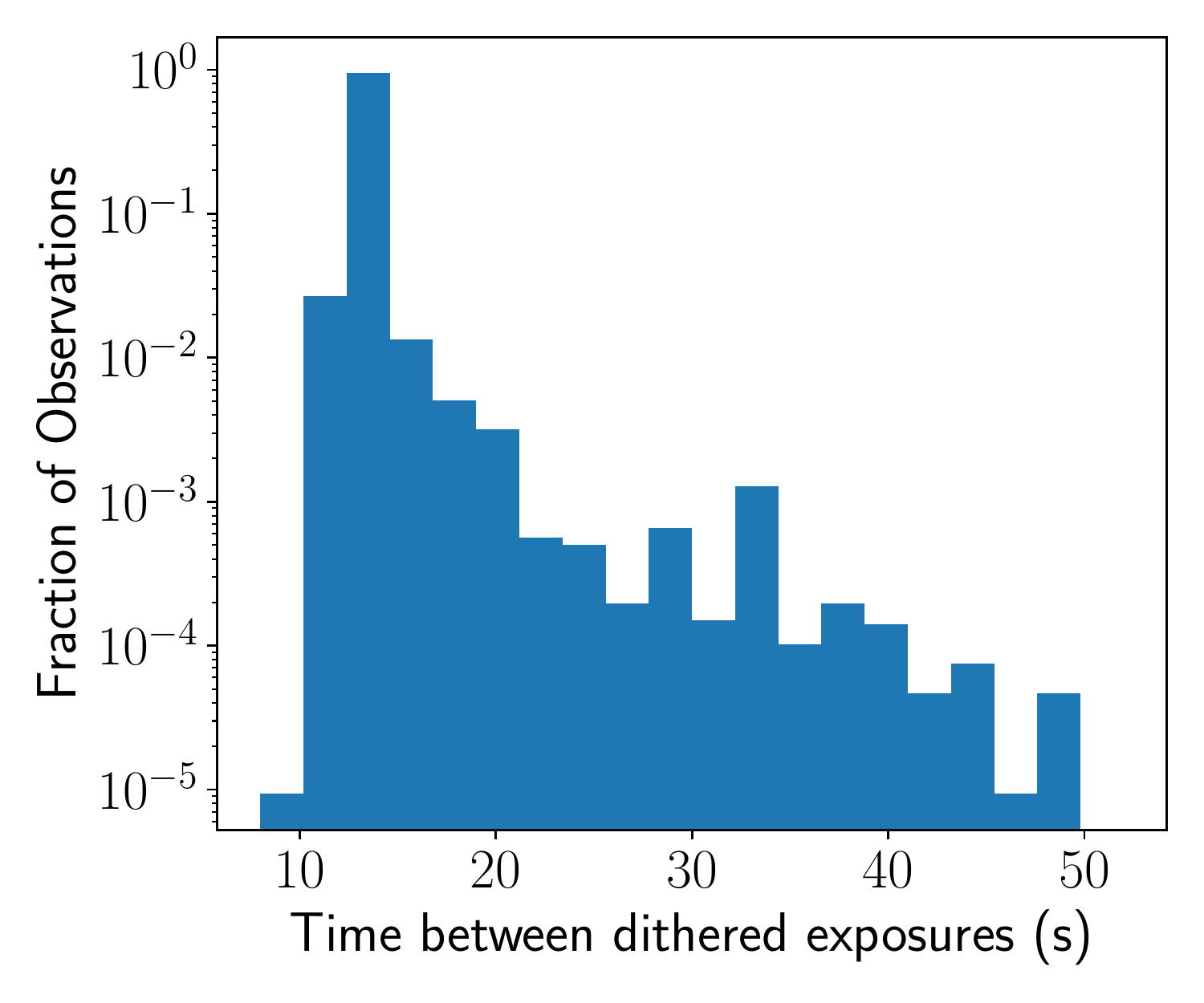}
\includegraphics[width=0.49\textwidth]{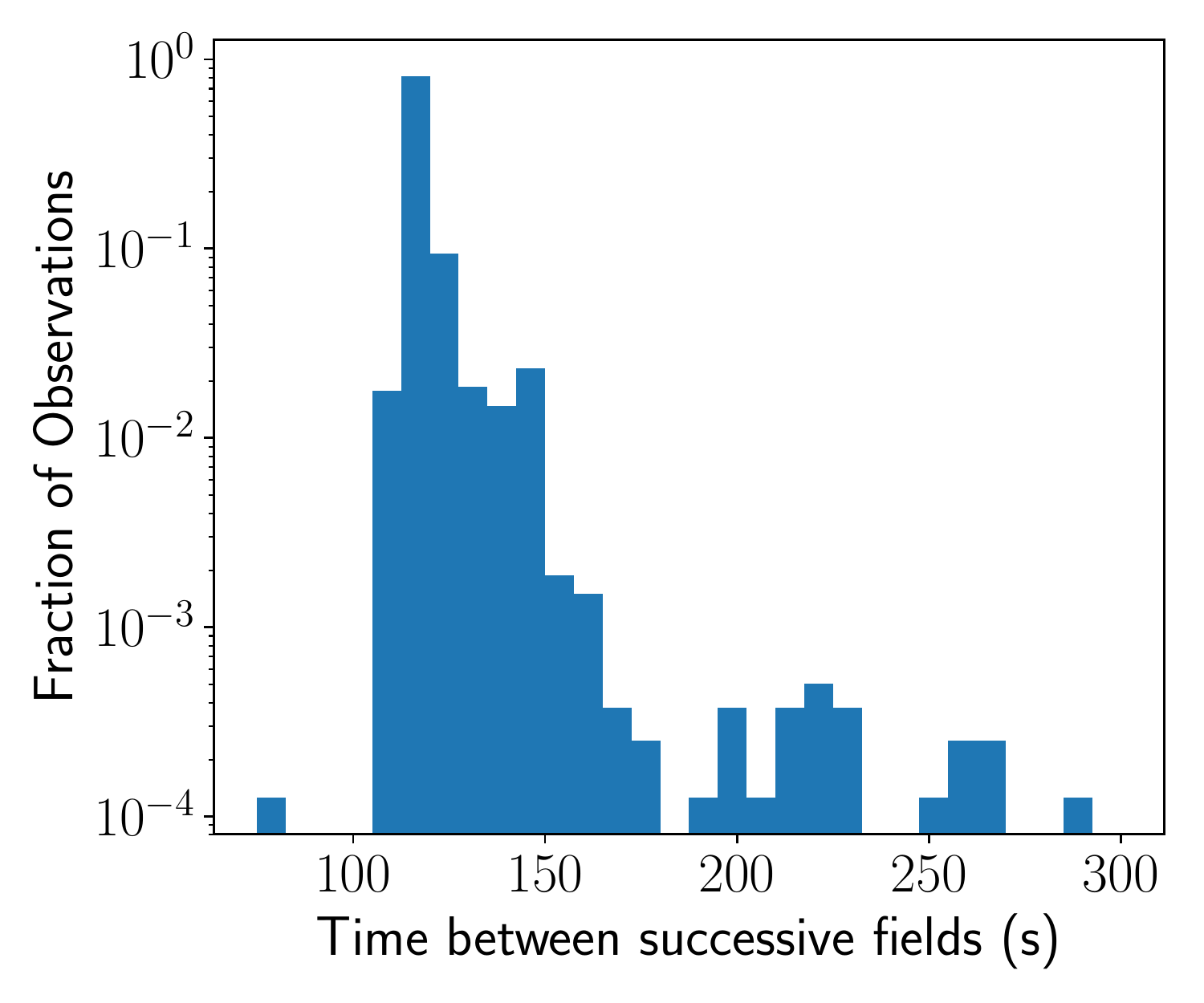}
\caption{(Left) Distribution of times between successive dithers in a field visit, including the nominal exposure time of 8.1 s. The median time between start of exposures is 13.0 s, accounting for the overhead due to dithering the telescope. (Right) Distribution of the time between the start of observations of successive field visits in nightly operations. The elapsed time includes the total exposure time (64.8 s), dithers between exposures and the slew across the successive fields. The median time between the start of successive field observations is 113.0 s, accounting for the dithers inside each field and the time to slew between successive fields. }
\label{fig:scheduler_dithers}
\end{figure*}

\begin{figure}
    \centering
    \includegraphics[width=\columnwidth]{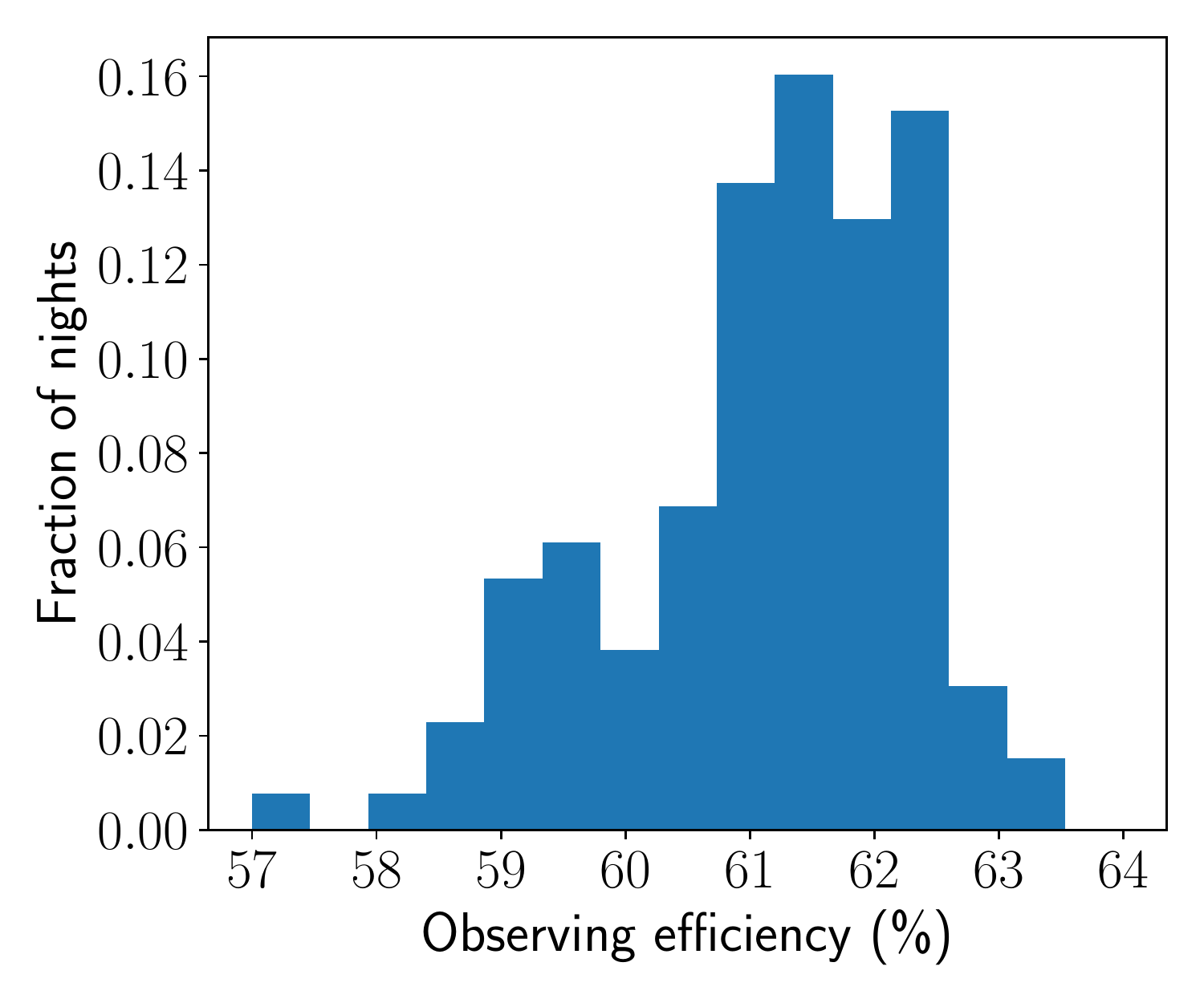}
    \caption{Distribution of the observing efficiency of the Robotic Observing System on nights during the commissioning phase. The median observing efficiency is $\approx 61$\%, accounting for the over heads due to the readout time, dithers inside each field and slewing between fields (Section \ref{sec:scheduler}).}
    \label{fig:obseff}
\end{figure}

The primary overhead during observing sequences is the time taken to move between individual dither positions in a dither sequence given the settling time of the mount, since the detector readout time is small ($\approx 0.9$\,s). Figure \ref{fig:scheduler_dithers} shows a distribution of the time between the start of subsequent dithered exposures in dither sequences taken over several nights. The distribution has a narrow peak around the median time of $\approx 13.0$\,s, including the exposure time of $8.1$\,s amounting to a dither overhead of $\approx 60$\%. Figure \ref{fig:scheduler_dithers} also shows a distribution of the time between the start of exposures of successive fields during a night, which includes the exposure time inside the field (64.8 s), the time taken to dither between dither positions inside the field and the slew time to the next field. The distribution also has a sharp peak near the median time of $\approx 115$ s. Figure \ref{fig:obseff} shows a distribution of the observing efficiency of the system, defined as the total exposure time in a night as a fraction of the time the telescope was observing. The overall observing efficiency distribution has a median of $\approx 61$\%, accounting for all the overheads for dithers and slewing between fields.

\subsection{Volumetric survey speed and cadence}
\label{sec:volumespeed}

\begin{figure}
    \centering
    \includegraphics[width=\columnwidth]{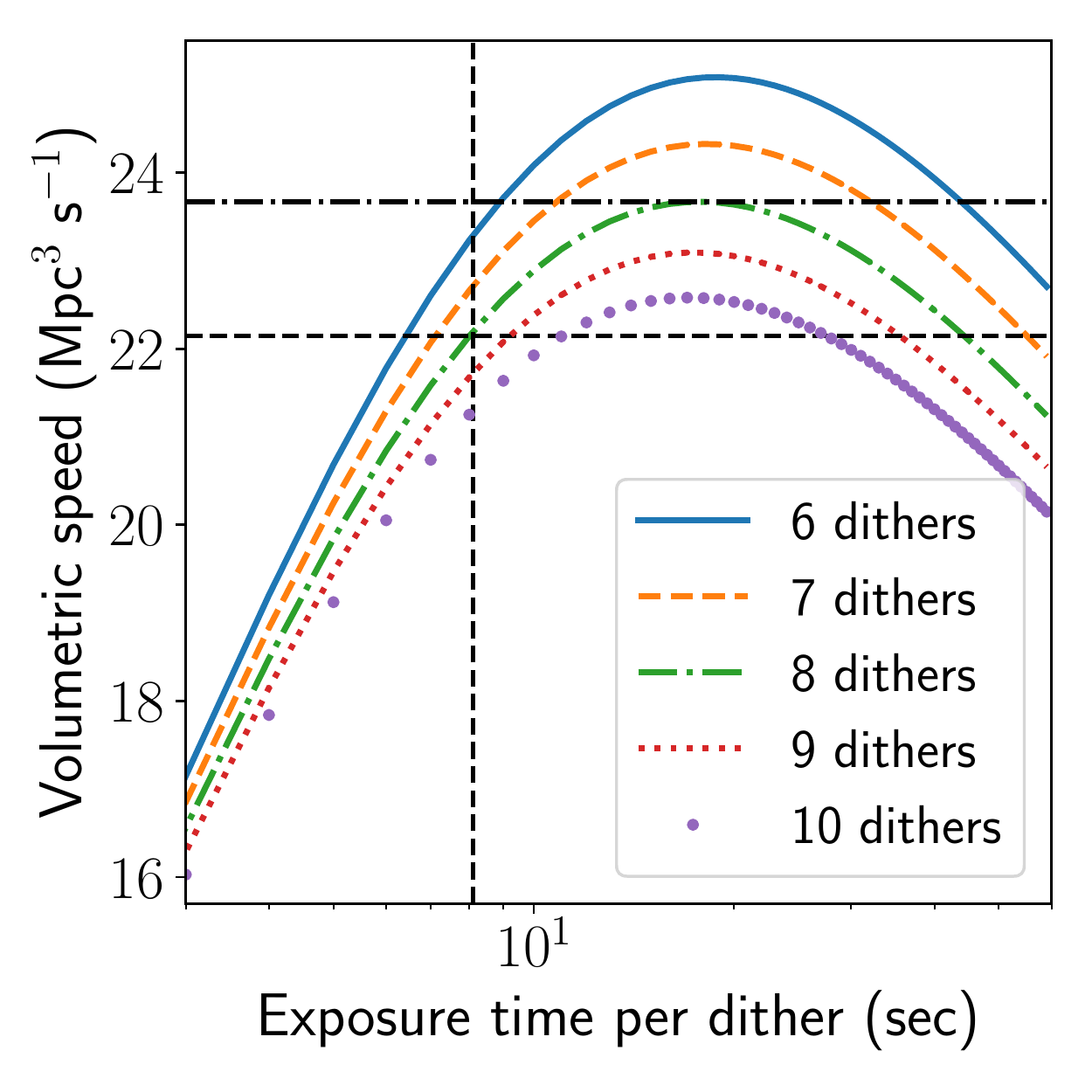}
    \caption{Volumetric survey speed of the Gattini observing system for a fiducial Type Ia SN peaking at $M = -19$, assuming a limiting magnitude of 16 AB mag. The survey speed is plotted as a function of the exposure time per dither folding in the overheads for dithering and slewing across fields. The different line styles correspond to different number of dithers as indicated in the legend. The horizontal and vertical dashed black lines correspond to the survey strategy adopted for the survey, while the black dot-dashed line shows the maximum volumetric survey speed possible for the adopted number of 8 dithers.}
    \label{fig:volspeed}
\end{figure}

We show the volumetric survey speed \citep{Bellm2016} for a Type Ia SN (peaking at $M = -19$) as a function of the exposure time per dither in Figure \ref{fig:volspeed}, for different choices of the number of dithers folding in the measured dither and field slew overheads. The volumetric survey speed increases with a smaller number of dithers and hence we select 8 dithers as a minimum to support the PSF re-construction downstream. For the choice of 8 dithers per field, the exposure time that maximizes the volumetric survey speed is $\approx 17$s per dithered exposure, and a corresponding areal survey rate of $\approx 470$ sq. deg. hr$^{-1}$ with an all sky cadence of $\approx 4$ nights. Adopting an exposure time of $\approx 8.1$s instead, we found the volumetric survey speed to be only $\approx 5$\% smaller, while providing an areal survey speed of $\approx 800$ sq. deg. hr$^{-1}$, allowing coverage of the entire visible sky over two nights (assuming an average of $\approx 9$ hours per night). On the other hand, increasing the cadence to cover the entire sky over a single night would require reducing the exposure time per dither to be $<2$ seconds, where the volumetric survey speed would be 50\% smaller than the maximum volumetric speed. 

\begin{figure}
    \centering
    \includegraphics[width=\columnwidth]{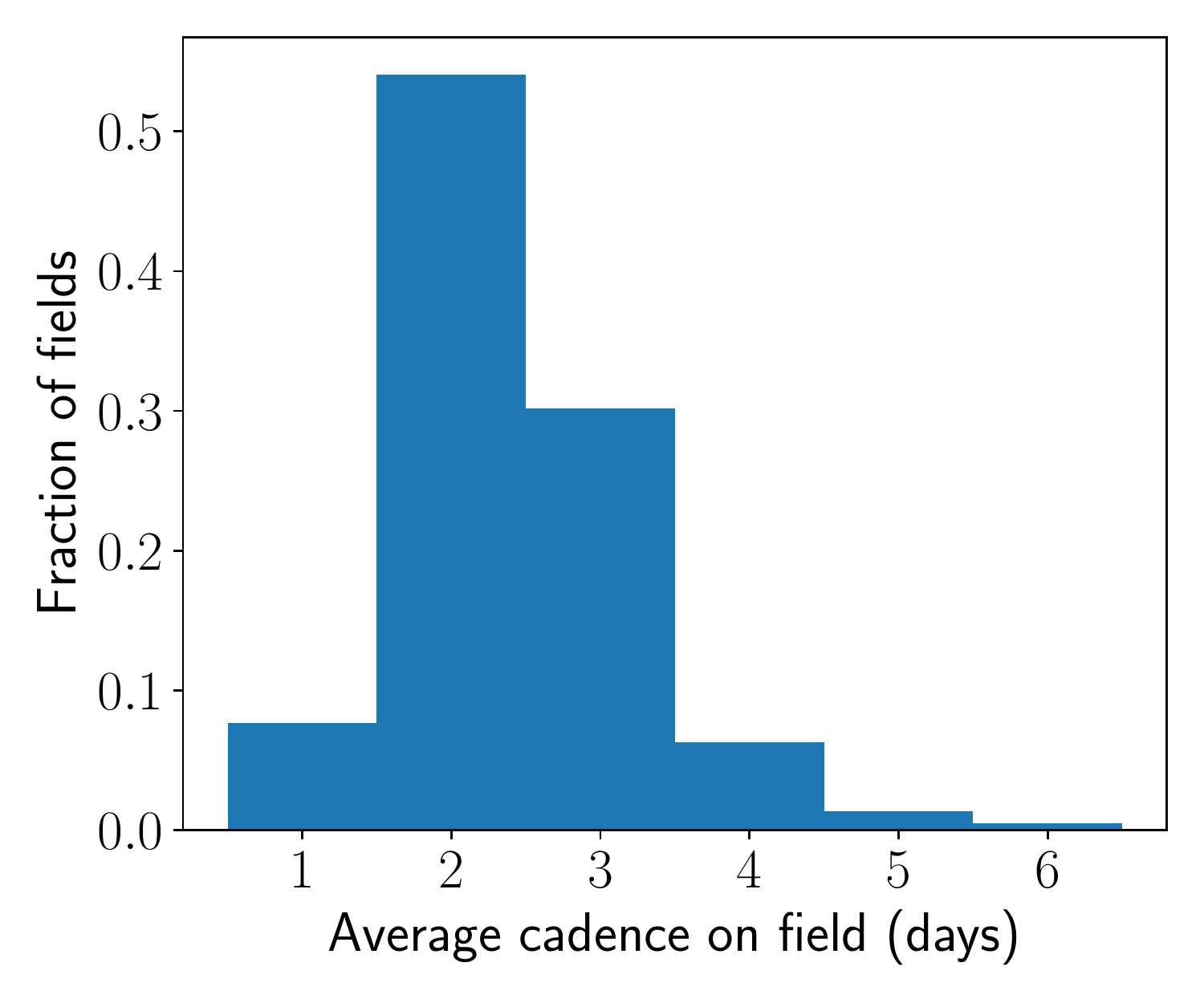}
    \caption{Distribution of the average cadence per field (in days) for the visible sky in the month of September 2019. The cadence is $\approx 1 - 3$ days over $\approx 90$\% of visible fields while it is shorter ($\approx 1$ day) for fields near the north celestial pole due to their longer visibility.}
    \label{fig:cadence}
\end{figure}

\begin{figure*}
    \centering
    \includegraphics[width=0.9\textwidth]{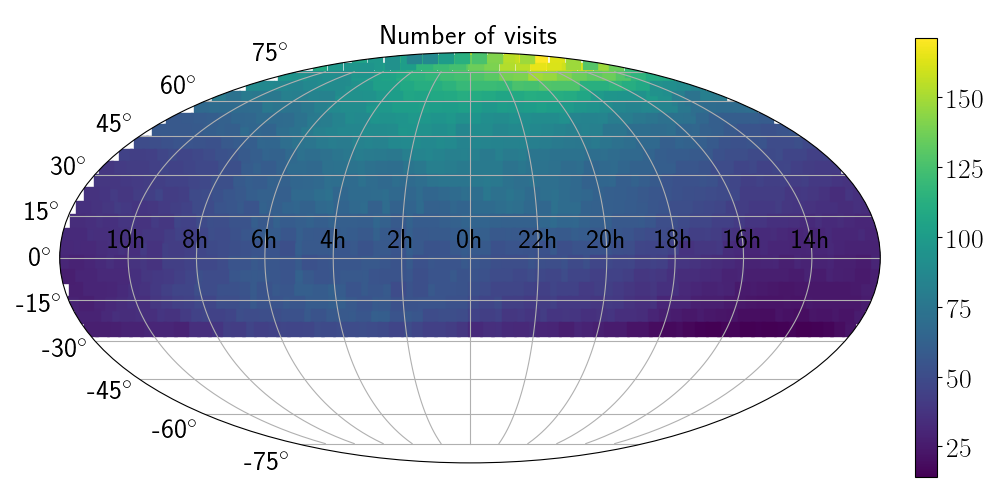}
    \caption{Sky distribution of the cumulative number of field visits since the start of the commissioning period, as of 2 October 2019. Fields near the north pole were visited the largest number of times due to their long visibility window from Palomar Observatory.}
    \label{fig:visits}
\end{figure*}

We thus adopt eight dithers of 8.1\,s each as the nominal observing strategy for the survey. Figure \ref{fig:cadence} shows a distribution of the average cadence over the visible sky for a typical month of observing. The cadence is $< 3$ nights for 90\% of visible fields in the sky, while the median cadence of two days (for $\approx 60$\% of fields). A small fraction of fields (8\%) near the north celestial pole have typically shorter cadence of $\approx 1$ day owing to their longer visibility during the night. Figure \ref{fig:visits} shows the sky distribution of the cumulative number of field visits since the start of the survey commissioning period. Fields near the north pole have the largest number of visits due to their long visibility window from Palomar Observatory.

{}

{}

\subsection{Target of Opportunity observations}

The scheduler is designed to respond to target of opportunity (ToO) triggers to respond to time critical events such as gravitational wave triggers and neutrino alerts. ToO requests are submitted as a list of fields to be observed for a specified integration time. In the case of gravitational wave and neutrino triggers, the field tiling is optimally determined using the algorithm presented in \citet{Coughlin2019a} and forwarded to the scheduling system via the GROWTH Marshal \citep{Kasliwal2019}, including a start and expiry date for the request. A \texttt{cron} job on the scheduler server periodically checks for new ToO requests every 5 minutes. In case a new ToO request is found, the scheduler interrupts the nightly survey after finishing the set of dithers on the field being observed. 

Once triggered, the ToO scheduler checks if the list of requested fields in the ToO are observable (above airmass 2.5). If this condition is satisfied, the telescope slews to the field and begins dithered exposures of the nominal 8.1\,s exposure time such that the total integration time equals the requested exposure time on the field. Each field in the submitted ToO is observed until no more observable fields are left, after which the scheduler resumes the nightly survey. The list of fields in each submitted ToO is checked every 5 minutes until all the fields in the ToO are observed or the ToO has reached its expiry date.

\section{Data processing system}
\label{sec:drp}
In this section, we describe the Gattini Data Processing System (GDPS), a highly parallelized real-time data reduction system running at Caltech to support fast delivery of science quality data products from the survey. Figure \ref{fig:overview} provides an overview of the flow of data from the scheduler node at Palomar to the GDPS at Caltech. Figure \ref{fig:drp_flow} provides an overview of the data reduction within the GDPS. The GDPS was developed completely in \texttt{python} and uses open source tools available from several \texttt{python} packages and the Astromatic suite of software \citep{Bertin1996, Bertin2006, Bertin2011}. The GDPS is supported by a \texttt{Postgre-SQL} (PSQL) based database (DB) system\footnote{\url{https://www.postgresql.org/}} storing metadata for every step in the image reduction process. The code was was derived from a multi-purpose near-infrared image reduction pipeline developed originally for the Wide field infrared camera (WIRC) on the Palomar 200-inch telescope (and later adapted to several other small field-of-view optical and near-infrared imagers), which we briefly describe in Appendix \ref{sec:appendix_drp}. 

The basic requirements for the pipeline were to deliver science quality data products, including calibrated science images from raw images taken on sky, and transient candidates from difference imaging in real-time to support timely follow-up observations of transients. In addition, the GDPS monitors quality metrics for data taken during nightly operations and maintains DBs to allow for long-term storage and efficient access of data products generated from the system. The metrics are used to support light curve generation on epochal stacks and difference images. Although the epochal stacked images, calibrated source catalogs and difference photometry are the primary data products for users, we also store a number of intermediate files to support auxiliary day-time tasks such as reference building, sky-flat generation and dead pixel masking.

The pipeline was developed to run on a dual unit Intel Xeon E5-2620V4 2.1 GHz computing node (`compute node' hereafter) with a total of 16 cores (32 threads) and 64 GB of RAM, heavily utilizing parallelized operations to speed up processing during night-time operations. The \texttt{PSQL} DB is hosted on another server (`database node' hereafter) that was initially designed for testing purposes, and hosts an Intel Xeon E5-2620V3 2.4 GHz processor with 6 cores (12 threads) and 32 GB of RAM. Tables on the DB node are updated in real-time across the internal Caltech 1-gigabit network during nightly operations (Figure \ref{fig:overview}), and are backed up to a remote server on a daily basis. These two servers support the data reduction system in addition to the scheduler node, which runs the night time scheduler controlling the dome, telescope and camera operations. 

\subsection{Processing architecture}

\begin{figure*}[!ht]
\centering
    \includegraphics[width=0.9\textwidth]{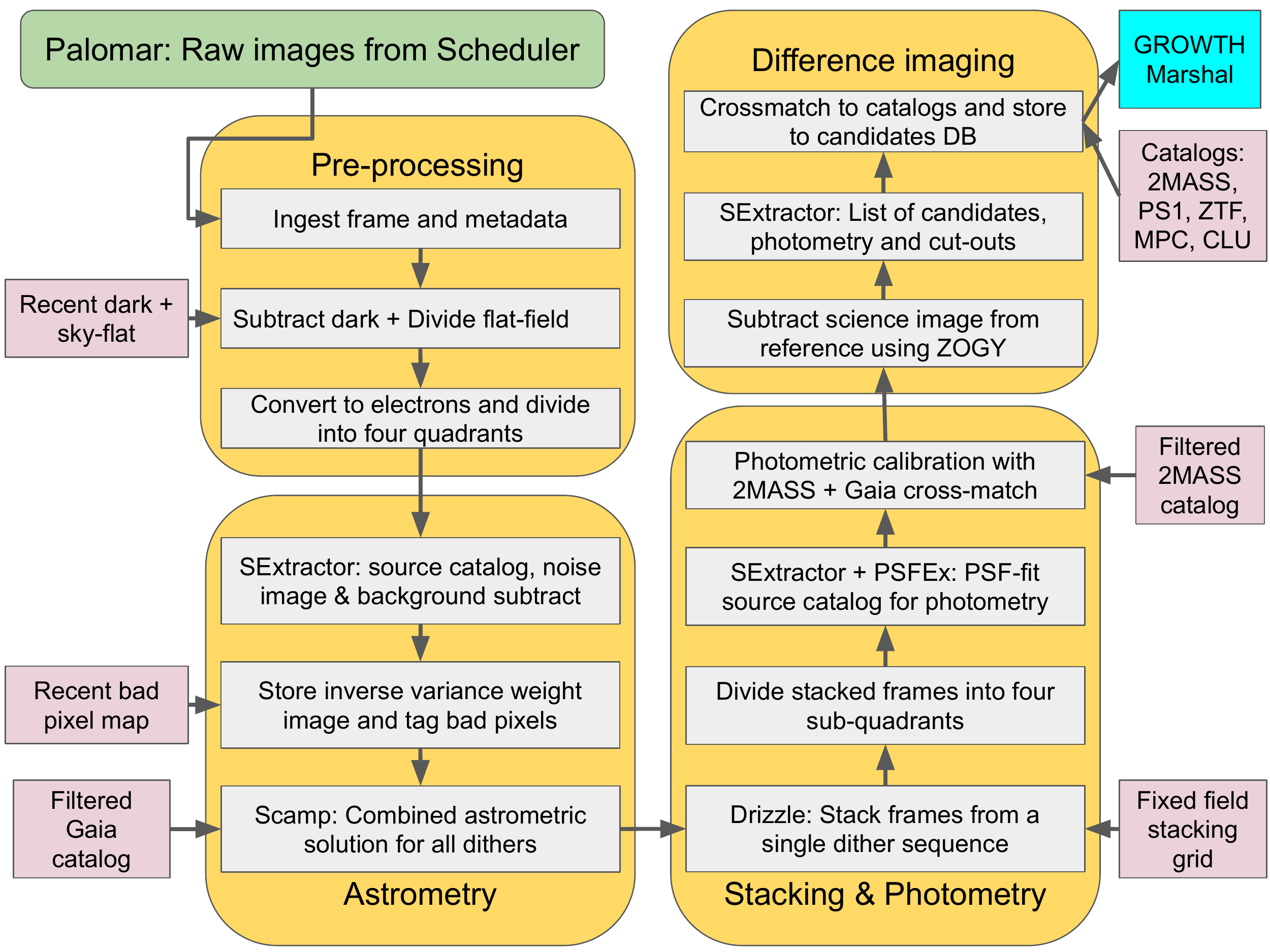}
    \caption{General reduction flow for the Gattini data reduction system. Boxes in yellow show the two automatic codes that perform the observing and reductions -- the data reduction robot and the scheduler. Boxes in grey show the major steps in the data reduction flow, while boxes in cyan show external input catalogs to facilitate various processing steps. Green boxes show the major outputs from the pipeline -- PSF-fit light curve catalog and subtraction candidates.}
    \label{fig:drp_flow}
\end{figure*}

Figure \ref{fig:drp_flow} provides an overview of the data processing flow within the GDPS. The data reduction flow for the system proceeds in five steps -- i) image pre-processing, ii) astrometric solutions, iii) stacking of dithered exposures, iv) photometric solutions and v) difference imaging, which are performed sequentially when a new set of raw images are received. Although each step is performed sequentially on the raw incoming data, the processing of multiple observed fields within each step is parallellized with 30 threads to support the requirement of real-time processing. 

Overall, the GDPS is controlled by a watchdog program that looks for new incoming images from the telescope and performs the data reduction through each of the steps mentioned above, while recording metadata for raw and intermediate data products at each step of the pipeline. At the end of the night, the watchdog accumulates the metadata and quality metrics for the data acquired during the night, including nightly sky coverage, image depths and PSF quality, while performing accountability checks on the number of files received and ingested through each step in the pipeline. These metrics are sent to the members of the project in a summary email. We summarize the processing architecture below and provide detailed descriptions in the following sections.

\begin{figure}
    \centering
    \includegraphics[width=0.49\textwidth]{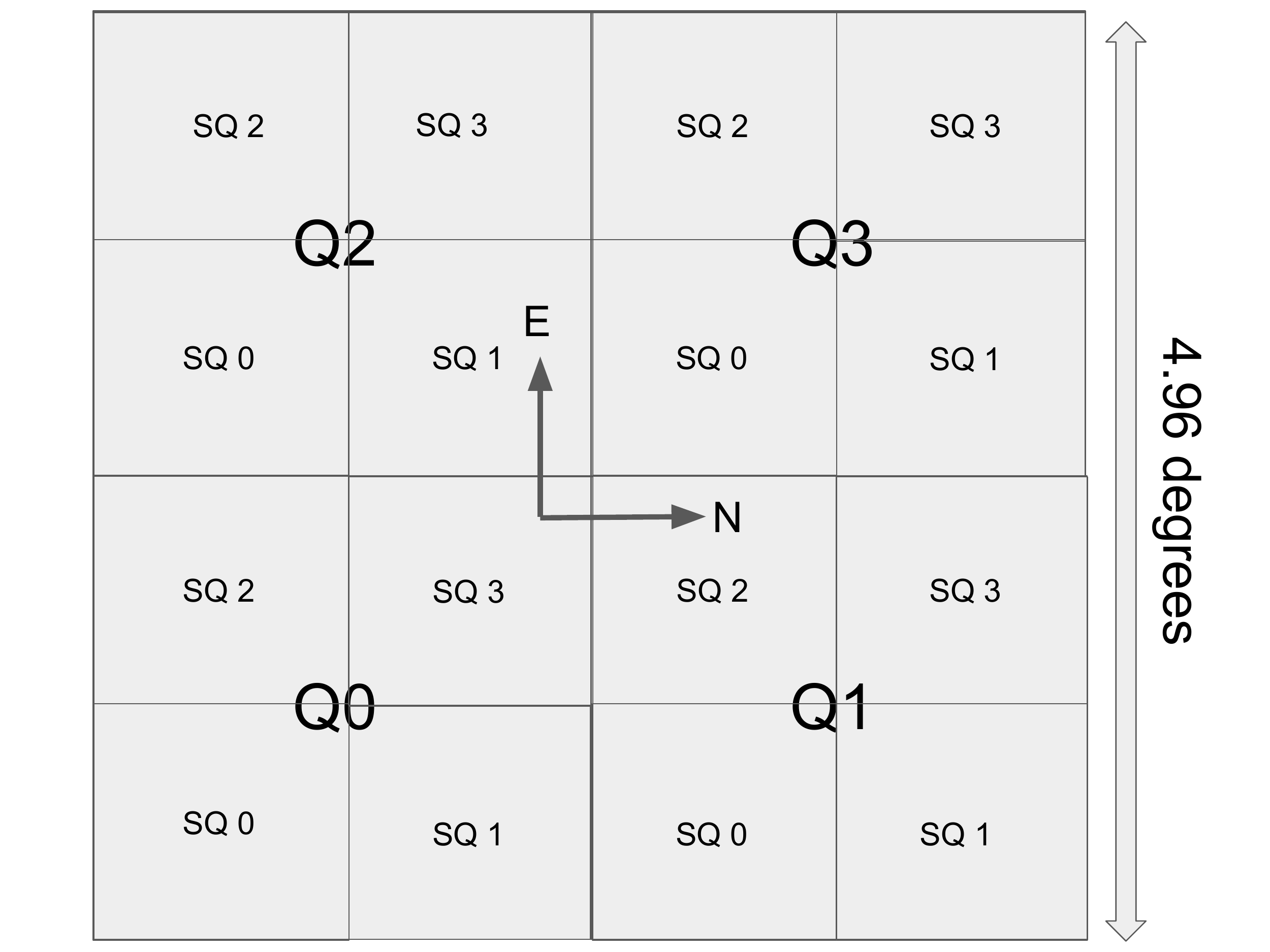}
    \caption{Detector layout for the Gattini quadrants and sub-quadrants.}
    \label{fig:det_layout}
\end{figure}

\begin{itemize}
    \item Raw data received from the scheduler node are initially de-trended and then digitally split into four quadrants of $1044 \times 1044$ pixels each (corresponding to a size of $2.5^\circ \times 2.5^\circ$ on sky), including an overlap (of 10 pixels $= 87$\,\arcsec) between the common edges to avoid missing sources that are split between the quadrant edges (Section \ref{sec:preproc}).
    \item An astrometric solution is derived for all the image quadrants produced from a single field visit i.e., 32 images from 8 dithers in four quadrants in the nominal observing strategy\footnote{Since the observing strategy involves acquiring multiple dithered exposures (8 nominally) over a single field, the subsequent processing (astromery, stacking, photometry and difference imaging) proceeds only after all of the dithers for a given field have been received from the scheduler node.} (Section \ref{sec:astrometry}).
    \item The astrometric solutions are used to stack the processed images (Section \ref{sec:stacking}) on a per-quadrant basis resampled to a pixel scale half of the raw pixel scale of the detector (thus producing stacks which are $2088 \times 2088$ pixels in size) using the Drizzle algorithm \citep{Fruchter2002}.
    \item The stacked quadrant images are again digitally split up into four sub-quadrants containing $1044 \times 1044$ resampled pixels each ($\approx 1.25 ^\circ \times 1.25^\circ$ on sky), including the same overlap between the sub-quadrants (of 87\,\arcsec\, or 20 pixels in drizzled images). Figure \ref{fig:det_layout} shows the layout of the detector plane with respect to the sky. Photometric solutions are derived on the split sub-quadrants (Section \ref{sec:photometry}).
    \item Transient candidates are identified from difference imaging  using the ZOGY algorithm \citep{Zackay2016}. These sources are passed through a machine learning (ML) classifier and cross-matched to several all-sky catalogues and known solar system objects from the Minor Planet Center. Candidates are subsequently uploaded to a web-portal for human vetting (Section \ref{sec:differencing}).
\end{itemize}

The step-wise design for splitting up the raw images into 16 sub-images as the final products was motivated by several reasons that were tested during the commissioning phase:
\begin{enumerate}
    \item Typical images exhibit a large variation of the PSF and sky-background over the large $5^\circ \times 5^\circ$ field of view, and hence better photometric solutions and difference imaging is obtained by splitting up the image into smaller sections covering 1/16 of the total detector area, where the PSF and background are locally uniform.
    \item Splitting up the raw images into four quadrants before the astrometric solution derivation produces images that are small enough such that the distortion in the field is small but also large enough so that sufficient number of sources remain to derive a robust solution (this is particularly important under non-ideal observing conditions with low sky transparency). Dividing the images into 16 pieces at the start of the processing (before astrometry) increased both the failure rate and processing time for the reductions.
\end{enumerate}{}

\subsection{Data transfer}

Raw images from Gattini are recorded as FITS images, along with comprehensive header information including meteorological conditions (temperature, pressure, humidity, wind speed/direction) and ephemeris information (airmass, sun and moon positions, estimated sky brightness). A quick astrometric and photometric solution is performed, to allow real-time correction of the telescope position and an estimate of the sky transparency. The raw FITS images contain 2048 $\times$ 2048 pixels stored as 2 byte short integers, resulting in an image size of $\approx 8.1$ MB each. The total amount of raw data acquired in a full night of observing is $\approx 20$ GB. Raw data are robotically acquired by the observing system under safe observing conditions and saved to disk on the scheduler node inside the telescope dome. Subsequently, the observing system transfers the images to the data processing server housed at Caltech (using an \texttt{rsync}-based synchronization running every minute) via the NSF-funded High Performance Wireless Research and Education Network (HPWREN) administered by the University of California San Diego. The volume of raw data is small enough to be transferred in real-time to Caltech (transfer time of $\lesssim 1$ second per image compared to the image acquisition time of $\approx 10$ seconds).

\subsection{Image pre-processing}
\label{sec:preproc}
The metadata for the raw images are ingested into the \texttt{PSQL} DB followed by de-trending of the raw images. De-trending involves subtraction of the most recently acquired dark frame from the raw image followed by flat-fielding with the most recently made sky flat (Section \ref{sec:endofnight}) by querying metadata for calibration frames stored in the DB. Images are then digitally split into four overlapping quadrants, with an overlap size of 10 pixels (corresponding to $\approx 1.5$\,\arcmin\, on sky) between quadrants sharing common edges. The image quadrants are stored to disk and metadata are recorded to the DB. Since the data processing was designed such that each quadrant corresponds to a fixed position in sky coordinates, the position of each quadrant has to be rotated by 180$^\circ$ depending on the side of the meridian of the equatorial mount i.e., the raw image is rotated by 180$^\circ$ for all images taken on the rising side of the meridian, while the orientation is left as is for observations taken on the setting side of the mount (Figure \ref{fig:det_layout}).

The pre-processed quadrants as stored as 32-bit floating point numbers, and are fed to the Astromatic package  SExtractor \citep{Bertin1996} to create a source catalog for the image and produce a corresponding background subtracted image. The detection threshold for point sources is set to 5$\sigma$ in the de-trended image using the prescription in \citealt{Zackay2017}. Background subtraction is performed using a spatial median filter using a box size of $32 \times 32$ pixels ($\approx 5 \times 5$\,\arcmin\, on sky) to remove the bright spatially varying background in J band. The source catalog is saved to support astrometric calibration in the subsequent steps. A background RMS map is also generated during the same SExtractor run and saved to disk with locations of known bad pixels (see Section \ref{sec:endofnight}) masked. The background RMS maps are used to generate inverse variance weight maps for stacking individual dithers into a single image. 

\subsection{Astrometric calibration}
\label{sec:astrometry}
\begin{figure*}
    \centering
    \includegraphics[width=0.45\textwidth]{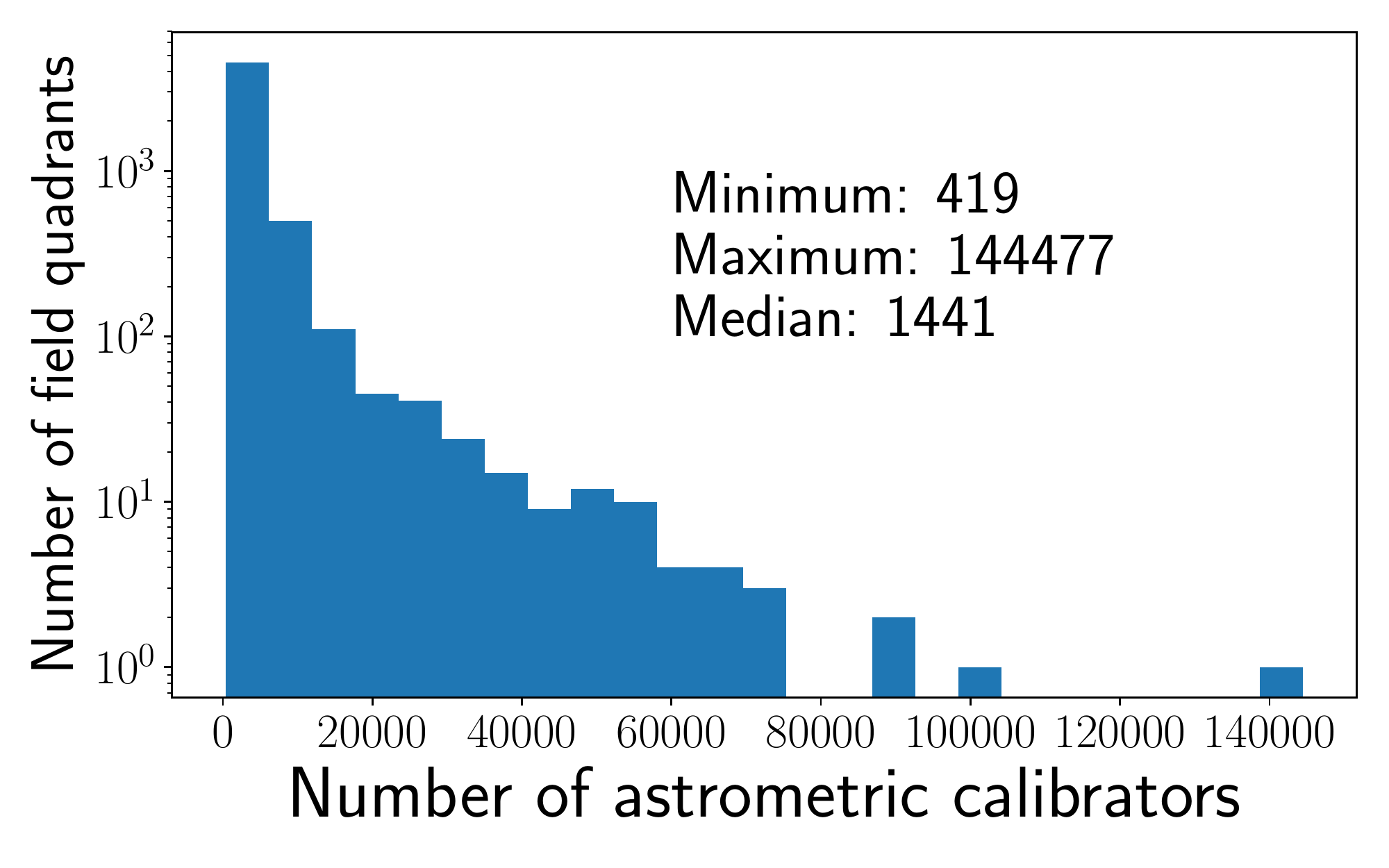}
    \includegraphics[width=0.45\textwidth]{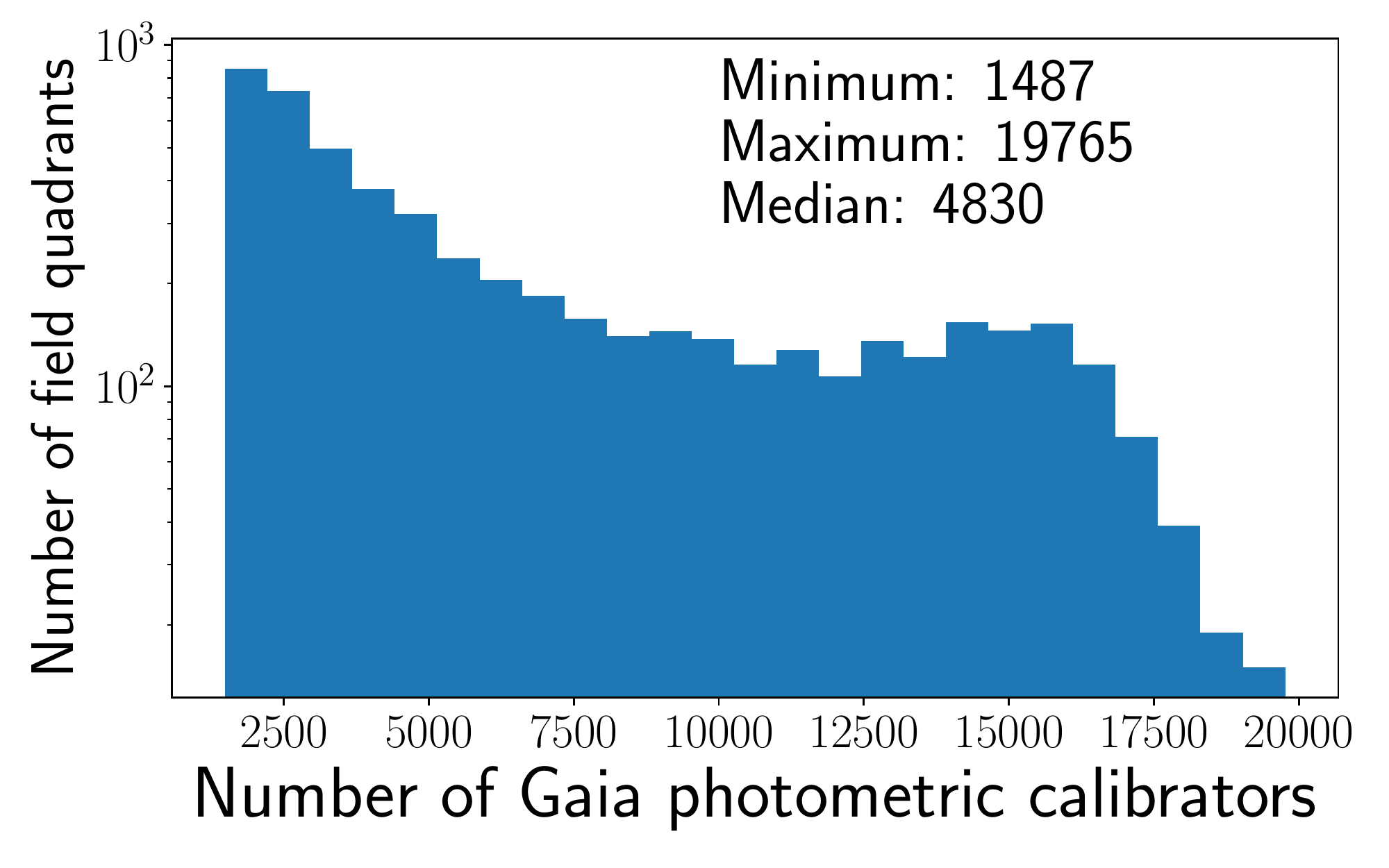}
    \caption{(Left) Histogram of number of astrometric calibrators per field quadrant, based on the Gattini pointing grid. (Right) Histogram of number of photometric calibrators per field quadrant, selecting sources that are isolated (no sources in 2MASS within 12\,\arcsec). The selection criteria for the calibrators are given in the text.}
    \label{fig:calibrator_dist}
\end{figure*}

Astrometric calibration is performed with respect to the Gaia DR2 catalog \citep{Gaia2018}, using the pre-exisiting cross-match table between 2MASS and Gaia DR2 available as a part of the Gaia DR2 release \citep{Marrese2019}. The list of astrometric calibrators are stored as pre-partitioned static binary FITS tables ordered by field numbers since the observing is performed on a fixed sky grid. Only sources with 2MASS J magnitudes between 7 and 13 are used for astrometric calibration. In order to select a pure sample of stars (i.e. point sources), we require that all astrometric calibrators have a non-zero proper motion ($>3\sigma$) in the Gaia DR2, do not have a corresponding counterpart in the 2MASS Extended source catalog \citep{Jarrett2000}, and are not confused in the Gaia - 2MASS cross-match solutions (i.e., \texttt{number\_of\_mates = 0} and \texttt{number\_of\_neighbours = 1}). Figure \ref{fig:calibrator_dist} shows a distribution of the number of astrometric calibrators per field quadrant, showing a minimum of a few hundred astrometric calibrators.

All image quadrants acquired as part of a single dither sequence on a field are astrometrically solved in a single run of \texttt{Scamp} \citep{Bertin2006}. For the nominal observing strategy of 8 dithers per field\footnote{Larger number of dithers may be used for targeted observations of ToO fields.}, this involves the solution of 32 image quadrants per execution of \texttt{Scamp}. The astrometry is solved to derive a common astrometric solution for all the dithers in a given quadrant using a third order distortion solution. The distortion coefficients are stored using the \textit{TPV} convention and written to the headers of the image quadrants, and recorded in the DB. Images for which the astrometric solutions fail are not processed further downstream. Astrometric failures are nearly zero for 70\% of nights, while the highest observed  failure rate can be $\approx 25$\% for images acquired through clouds and non-photometric conditions. We discuss the accuracy of the astrometric solutions in Section \ref{sec:astrometry_perf}.

\subsection{Stacking}
\label{sec:stacking}
Image quadrants acquired in each dither sequence are stacked using the \texttt{Drizzle} algorithm \citep{Fruchter2002}. Briefly, {Drizzle} resamples input images on to a user specified output grid with the option of shrinking each pixel by a user-defined parameter called \texttt{pixfrac}. \texttt{pixfrac} specifies the linear fraction by which each side of the input pixels are shrunk before co-addition. While smaller values of \texttt{pixfrac} produce sharper PSFs, smaller values of \texttt{pixfrac} result in uneven coverage in the output pixels if the dithers are not placed ideally. The pixel scale of the output grid is determined by the \texttt{scale} parameter, which determines the linear size of the output pixels with respect to the input pixels. \texttt{Drizzle} also produces an output weight image, which reflects the number of images that were sampled into each pixel during the resampling. In order to reduce artifacts associated with pixel shrinking and uneven coverage, \citealt{Gonzaga2012} recommend $\sigma_w / m_w < 0.15$, where $\sigma_w$ is the standard deviation in the resampled weight image and $m_w$ is the median of the resampled weight image. 

In the GDPS, the input images are resampled on to a fixed output grid with a pixel size half of the native pixels ($\approx 4.3$\,\arcsec), corresponding to a \texttt{scale} of 0.5. The output grid is determined from the fixed sky grid of fields using astrometric solutions of on-sky images from the telescope. We use a \texttt{pixfrac} parameter of 0.9, that controls the shrinkage of the pixels before resampling on to the output grid (i.e., the raw pixels are shrunk to 90\% of the size on each side before adding on the output grid). Smaller values of \texttt{pixfrac} produce uneven coverage over the smaller pixels in the output grid, resulting in a larger dispersion in the weight over the output pixels (see \citealt{Gonzaga2012} for a discussion). As the mount pointing is not accurate enough to provide sub-pixel pointing adjustments during the dither sequence, we resort to using a random sampling of sub-pixel phases during the dither sequence. A \texttt{pixfrac} of 0.9 was found to be adequate to produce Nyquist sampled images that are limited by the focusing of the optics.

The astrometry-solved and background subtracted input quadrant images are resampled to the fixed output WCS grid using a \texttt{python} implementation of the \texttt{Drizzle} algorithm\footnote{\url{https://github.com/spacetelescope/drizzle}}. In order to remove effects from cosmic rays, hot pixels, moving planes and satellites, the \texttt{Drizzle} code was modified to perform sigma-clipping on the resampled images to reject outliers on a per-pixel basis. The stacking of the images is performed using a sigma-clipped (at $2.5\sigma$), inverse-variance weighted mean of the resampled input images. The stacked image quadrants are then split further into four sub-quadrants for photometry and image-subtraction downstream. The splitting is done as shown in Figure \ref{fig:det_layout}, including a 20 pixel (10 raw pixel) overlap between the sub-quadrants, while transforming the WCS between the parent quadrant and child sub-quadrants. Both the resampled image sub-quadrants and their corresponding weight images are stored on disk as floating point 32-bit images for long term archiving in a single Multi-Extension FITS image, and quality metrics are recorded in the DB.

\subsection{Photometric calibration}
\label{sec:photometry}

Photometric calibration is performed against the 2MASS point source catalog \citep{Cutri2003} cross-matched to Gaia DR2, using a subset of the sources that were used for astrometric calibration. The photometric catalog includes additional filters to select a list of isolated stars with J magnitudes between 9 and 16 that are not saturated in images. `Isolated' stars were defined to be sources that did not have any neighbors in 2MASS (regardless of any cuts in the astrometric catalog) within a radius of 12\,\arcsec\, (3 Gattini drizzled pixels) to avoid confused sources. The stacked image sub-quadrants are fed as inputs to the photmoteric calibration pipeline. Figure \ref{fig:calibrator_dist} shows a distribution of the number of photometric calibrators per field quadrant. As in the case of the astrometric calibrators, the photometric calibrators are also stored as static pre-partitioned FITS binary tables per field on sky. 

The stacked image sub-quadrants are fed as inputs to the photomteric calibration pipeline, which first creates a SExtractor catalog of detected sources along with a set of $15 \times 15$ pixel cutouts for each detected source. The SExtractor catalog is fed to PSFEx \citep{Bertin2011} to generate a PSF model of $15 \times 15$ drizzled pixels using the sources detected in the field. The PSF model is saved to disk for supporting difference imaging further downstream. The PSFEx model is then fed to a second run of SExtractor to generate a PSF-fit photometry catalog for the drizzled stack. Aperture corrections are computed between PSF-fit magnitudes and apertures of different sizes and recorded in the FITS headers. The PSF-fit catalog is used to select a list of unsaturated sources for photometric calibration, that are at least 40 pixels away from the edges of the image, and with SExtractor parameter \texttt{FLAGS} = 0 and \texttt{FLAGS\_MODEL} = 0 and \texttt{SNR} $> 10$. The crossmatch proceeds only if there are at least 5 good cross-matched sources in the image.

The instrumental PSF and catalog magnitudes are fit with a linear solution of the form:
\begin{equation}
    m_{\text{TM, J}} - m_{\text{ins}} = ZP + c \, (m_{\text{TM, J}} - m_{\text{TM, H}})
\end{equation}
where $ m_{\text{TM, J}}$ and $ m_{\text{TM, H}}$ are the 2MASS magitudes in J and H filters respectively, $m_{\text{ins}}$ is the instrumental magnitude, $ZP$ is the zero-point of the image and $c$ is the color coefficient to convert from the Gattini system to the 2MASS (TM) system. The solution is derived by fitting a linear polynomial to the magnitude differences as a function of the source color (so the intercept is the zero-point and the slope is color coefficient). The extreme outliers (1 percentile) in the fit are eliminated first and a solution is derived. Subsequently, outliers that are more than 4 sigma away (typically $< 2$\% of the total number of stars) from the best-fit solution are clipped again and the final solution is re-derived. The photometric solution is recorded in the header of the image sub-quadrant and the DB, including quality metrics for data quality filtering. The PSF-fit source catalog is saved to disk and used to support light curve generation for sources detected across multiple-epochs. We discuss the accuracy of the photometric solutions in Section \ref{sec:photometry_perf}.

\begin{figure}
    \centering
    \includegraphics[width=0.4\textwidth]{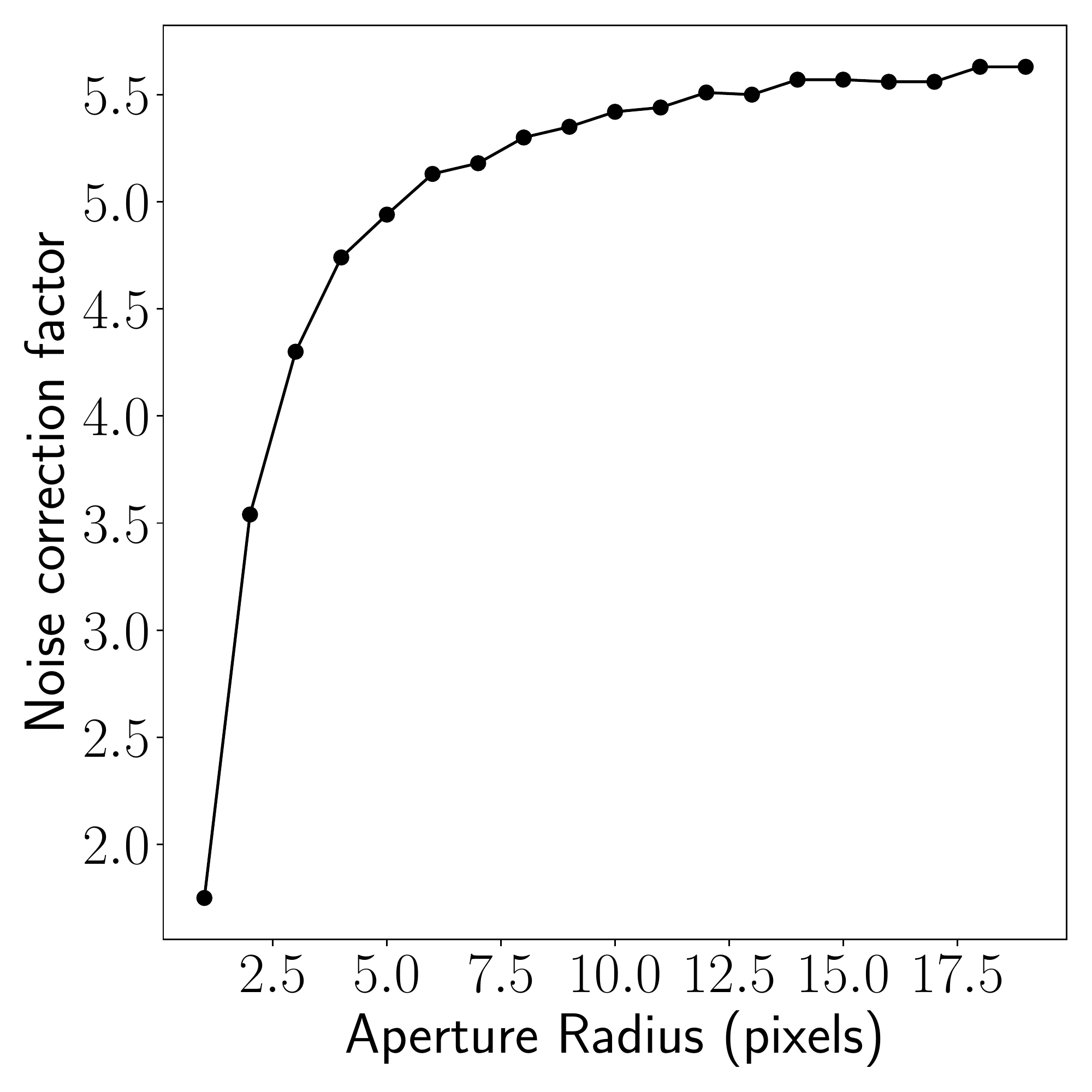}
    \caption{(Left) Noise correction (multiplication) factor as function of aperture size for typical Gattini drizzled images, as derived from Monte Carlo simulations. }
    \label{fig:corr_noise}
\end{figure}{} 

\subsection{Image depths and correlated noise}
\label{sec:correlatednoise}
The pixel reconstruction and resampling procedure used in \texttt{Drizzle} leads to correlated noise between adjacent pixels in the output image. The noise correlation leads to underestimation of the photometric uncertainties in the PSF-fit source catalogs and correspondingly, an overestimation of the depths of the images. In order to estimate the correction to the noise RMS in the images due to correlated noise, we use the prescription used for the WISE survey\footnote{\url{http://wise2.ipac.caltech.edu/staff/fmasci/ApPhotUncert_corr.pdf}}. The GDPS uses simulated white noise images that are drizzled to an output grid using the same dither pattern, \texttt{pixfrac} and \texttt{scale} parameters as in the GDPS operations to correct for the correlated pixel noise in the output photometric catalogs, estimates of image depths and the noise in the difference images produced downstream. 

Briefly, a Monte-Carlo simulation is performed for estimating the PSF-flux uncertainty correction by calculating PSF-fit fluxes at random locations in the simulated drizzled images using the PSF model for the input image. The variance in these measurements are then compared to the flux uncertainties expected from uncorrelated pixel noise, and a noise RMS rescaling factor is computed to inflate the uncertainties for the PSF-fit photometry fluxes. The same simulation is used to estimate the correction for the aperture photometry fluxes by measuring the aperture summed fluxes at random locations in the simulated drizzled images and computing the rescaling factor expected from uncorrelated noise. Figure \ref{fig:corr_noise} shows the estimated scaling of the noise RMS of the image as a function of the aperture size used for photometry. The correction factor increases for large aperture sizes but flattens beyond a radius of $\approx 5$ pixels due to the correlation length of the resampling process. The correction factors are used to correct the estimated limiting magnitude of the epochal stacked images and stored in the FITS headers of the calibrated image sub-quadrants and the DB.

\subsection{Difference imaging and transient extraction}
\label{sec:differencing}

Photometrically calibrated science images are fed to the image differencing pipeline if a good quality reference image exists for the field and if the photometric solution quality flags suggest that the image was acquired under good observing conditions (see Section \ref{sec:performace}). The difference imaging pipeline starts by preparing the input science and reference images by cross-matching the PSF-fit source catalogs to compute the relative flux-scaling and astrometric uncertainty between the two images. Since both the science and reference images are resampled on to the same fixed sky grid, only a spatial differential background is subtracted from the input images before subtraction with the ZOGY algorithm \citep{Zackay2016}. The science and reference image uncertainty maps, the corresponding PSF models and the astrometric registration uncertainty are fed as additional inputs to the image subtraction. 

The uncertainty maps are prepared by computing robust standard deviations (background noise) of the science and reference images and scaling them in regions of the image with low weight (smaller number of dithers) to inflate the noise maps accordingly. Source noise from the sources in the science and reference image are added to the uncertainty maps. Since the images fed into the ZOGY \textit{do not} have uncorrelated pixel noise (due to the resampling performed by Drizzle), the uncertainty (RMS) maps are multiplied with the RMS scaling factor derived for scaling the PSF-fit photometric uncertainties. This factor reflects the noise correlation over the size scale of a PSF and hence is appropriate for the match filtering performed to produce the Scorr image, which denotes the statistical significance for point source detection \citep{Zackay2016}. A saturated source mask is generated using a 25 pixel box at the location of any sources in the science image with pixel values above 0.95 times the saturation limit of the image. We adopt a lower value than true saturation to account for non-linearity effects near the saturation limit. An additional mask for bright sources in the field is prepared by querying the 2MASS point source catalog for sources brighter than 5th magnitude in the field. The size of the mask is adjusted such that brighter stars have larger masks around them, ranging from 81 pixel box masks for 5th magnitude sources to 181 pixel box masks for sources brighter than magnitude 0. A final bright source mask is produced by combining the saturation mask and the bright source mask from the 2MASS catalog. 

The image subtraction produces a difference image, a difference image PSF and a corresponding Scorr image (Equation 25 in \citealt{Zackay2016}), which is a match-filtered S/N image optimized for point source detection. An initial quality check of the Scorr image is performed to ensure the ZOGY run did not fail before proceeding (i.e., there are not a significant number of NANs in the image\footnote{Occasionally `good' quality Scorr images can have NANs values due to edge effects. For these data, we adopt a more aggressive edge masking at the location of the bad values before proceeding.}). Further quality checks of the subtraction are determined later in the pipeline using candidate counts. The final bright source mask is applied to the difference and Scorr image to remove saturation artifacts. The Scorr image is then fed to SExtractor to generate a list of sources with S/N $> 5$ in the match-filtered image, corresponding to sources that have peak Scorr values of $>=5$ in at least one pixel. The candidates are filtered to exclude image pixels with low weights (e.g., if the dither pattern did not sample the edge of a field uniformly, or if the region of the image is populated with many bad pixels). Additionally, candidates within 20 pixels of the edges of the image are excluded since they are already included in the adjoining sub-quadrant (or field) due to the overlap between the images earlier in the processing.

Quality metrics for the candidates are computed to detect bad subtractions from PSF-variation and astrometric residuals, including\footnote{These criteria were tested and modifies starting from the prescription used for the ZTF pipeline in \citealt{Masci2019}, and presented in \url{https://irsa.ipac.caltech.edu/data/ZTF/docs/ztf_pipelines_deliverables.pdf}}:
\begin{enumerate}
    \item The ratio of magnitudes measured in a 4-pixel aperture and 8-pixel aperture is required to be between [0.4,1.5] for a candidate to be saved. This criterion rejects yin-yang residuals resulting from large PSF variation across the images, as well as hot pixels.
    \item The ratio of the sum of pixels over the sum of their absolute values in a $3\times3$ median filtered image using a $7\times7$ pixel cutout centered at the source location is required to be $\geq0.4$ for a candidate to be saved. This criterion is effective at rejecting yin-yang residuals from large PSF variation in the images. An additional ratio of pixel sums using a variable box size which depends on the image FWHM, PSF asymmetry, and relative astrometric uncertainty at the source location is also calculated and stored for later filtering. 
    \item The number of pixels 5-$\sigma$ below the median value of the difference image contained in a $9\times9$ pixel box centered at the source location is determined. Candidates are rejected if the count is $>1$ and the ratio of pixel sums with variable box size in \#2 is $<0.8$. This cut is only applied for sources with measured magnitudes $>10$ to avoid rejecting bright sources which can produce significant `ringing' in the difference image as a result of the noise decorrelation process in ZOGY.  
    \item The ratio of the flux of the candidate in the difference image and the reference image. Cross matching between sources in the difference and reference image is performed using a variable radius that is a function of the image FWHM, PSF asymmetry, and relative asymmetric uncertainty at the source location. No cuts are performed solely using this metric.
\end{enumerate}
Values for these metrics were determined by performing tests with injected fake sources, which is described in section \ref{sec:recovery}. This process is also performed for the corresponding negative difference and Scorr image to find sources that have faded from the reference image. An additional filtering of positive-negative source pairs is performed to remove extended residuals caused by astrometric residuals or extreme cases of PSF-variation between the new and reference image. Cross matching between the positive and negative source catalog is performed using a variable radius that is a function of the image FWHM, PSF asymmetry, and relative asymmetric uncertainty at the source location. Candidates are discarded if a positive-negative cross match is found and the ratio of the source flux in the difference and reference image (item 4) is less than 1.

PSF-fit fluxes are measured for each source in the Scorr image by fitting the difference image PSF model on the location of the source detected in \texttt{SExtractor}. The position of the source is refined in the fit by $\chi^2$ minimization of the residuals from the PSF model. Although the ZOGY algorithm is designed to produce difference images with uncorrelated pixel noise, this does not hold in cases where the input images have correlated noise in them. Hence, in order to correctly estimate the difference photometric uncertainties, we perform a Monte Carlo simulation by estimating the variance of PSF-fit fluxes over a simulated difference image containing only noise, and scale the PSF-fit photometric uncertainties from the difference image accordingly. As an additional filter of bad quality candidates, we require the absolute value of the difference between the measured PSF-fit magnitude and 8-pixel aperture magnitude be $\leq0.4$. Last, a final quality check is performed on the difference image using the total number of `good' candidates, both positive and negative, found as compared to the total number of objects in the source catalog of the science image. If this value is $>0.2$, the image is flagged as poor quality and no candidates are saved. Otherwise, the values for aperture photometry, PSF fitting, and other quantities described in this section are recorded in the DB for downstream filtering. Image cutouts ($61 \times 61$ pixels; $4.4$\,\arcmin\, on each side) are recorded in the DB around the location of the source in the science, reference and difference image for machine learning based classification and human vetting externally. 

\subsection{Machine Learning classification}

To automatically distinguish between an astrophysical source and image subtraction artifacts, we use a Machine Learning (ML) based real-bogus (RB) classification scheme. The GDPS uses a real-bogus classifier scheme implemented through supervised Deep Learning where features are extracted from an input set of candidate sources using many-layered perceptrons (artificial neural networks). The classifier was trained by assembling a training set of separately labelled real and bogus data by human classifiers. Bogus candidates were compiled using a labeling scheme on Zooniverse, a citizen science web portal which allows set up of individual projects usually pertaining to classification and data visualization\footnote{\url{https://www.zooniverse.org/}}. Real sources were selected based on a sample of known variable stars, supernovae and asteroids found with human vetting during the commissioning period.

We used a two-layer Convolutional Neural Network (CNN) for our Deep Learning model, as CNNs are
commonly used in analyzing visual imagery and have numerous benefits over standard multi-layer
perceptrons\citep{Krizhevsky2012}. This model is implemented using the TensorFlow package \citep{Abadi2016} and the high-level Keras API\footnote{\url{https://keras.io/}}. The model has two convolutional layers, one flatten layer and two fully-connected layers. The first convolutional layer uses 32 $3\times3$ filters with a Refined Linear Unit (ReLU) activation function, and is followed by a maxpooling layer of size $2\times2$. The second convolutional layer uses 64 $5\times5$ filters with a ReLU activation function, and is followed by a maxpooling layer of size 4x4. Dropout layers with rates of 0.25 are included after each convolutional layer to minimize over-fitting. After a flatten layer, there is one fully-connected layer of size 32 with a ReLU activation function, followed by a dropout layer of rate 0.40. The final output layer is a fully-connected layer with a sigmoid activation function for binary classification (real or bogus), amounting to a total of 5 layers.

\begin{figure}
    \centering
    \includegraphics{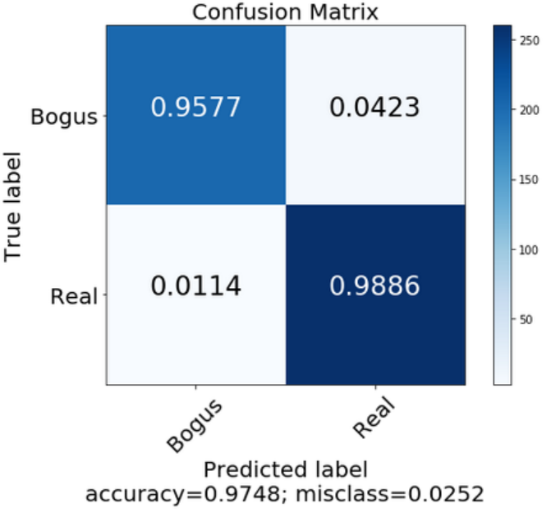}
    \caption{Confusion matrix for the deep learning based real-bogus classification system used in the subtraction pipeline of the GDPS.}
    \label{fig:rbconf}
\end{figure}

The model is trained and run on stacks of images of shape $61\times61\times3$, consisting of science, reference, difference cutouts (of size $61\times61$ each). We utilized a training/validation/test split of a 72\%/8\%/20\%. While training the model, we used a batch size of 30 and utilized the early stop method at 20 epochs as there was no improvement in validation accuracy and an increase in validation loss past that point. We used K-fold cross validation technique with k=10 to reduce bias and prevent over-fitting. The performance of the model was evaluated using the following metrics: accuracy on the test set of 0.975, a Matthews correlation coefficient of 0.949 and an F1 score of 0.977. Figure \ref{fig:rbconf} presents the normalized confusion matrices for the model, showing a false positive rate (FPR) under 5\% and false negative rate (FNR) under 1.5\%. The ML model is currently being tested on unseen data and will be refined before final deployment.

{}
 
\subsection{External cross-matches and human vetting}

Transient sources recorded in the DB are cross-matched to external catalogs for supporting follow-up prioritization using the a dedicated database server at Caltech \citep{Duev2019}. In order to filter on variable stars, extracted sources are cross-matched to the 2MASS point source catalog \citep{Cutri2003} and the distances and J magnitudes of the three nearest sources are recorded. The same is recorded for the three nearest sources in the reference image for the respective field sub-quadrant. Sources are also cross-matched to the PS1 DR2 catalog \citep{Chambers2016} including the star-galaxy classification scheme described in \citet{Tachibana2018}, to store the distances, magnitudes and star-galaxy classification score of the three nearest objects. Additional recorded metadata include cross-matches to known solar system objects using the \texttt{astcheck}\footnote{\url{https://www.projectpluto.com/astcheck.htm}} utility, the ZTF public alert archive \citep{Masci2019}, the Census of the Local Universe (CLU) catalog of nearby galaxies \citep{Cook2019}, Gaia DR2 and previously known objects in SIMBAD. 

The cross-matches are performed on the database node and recorded in the DB, which is accessible across the Caltech internal network for subsequent human filtering and vetting. Human vetting is performed via a dedicated scanning page on the GROWTH marshal \citep{Kasliwal2019} showing the science, reference and difference image cutouts of detected sources together with their metadata. The scanning page is created via read-access to the DB on the database node, and allows the user to enable several metadata-based filters on the detected transients, including the distance to the nearest 2MASS source (to reject variable stars), ML real-bogus score, the presence of a ZTF counterpart and proximity to a nearby galaxy in the CLU catalog. Candidates are assigned a survey name once they are saved by a human scanner, and are followed up with imaging and spectroscopy via requests on the GROWTH Marshal.

\subsection{Offline tasks}
\label{sec:endofnight}
\subsubsection{Reference image generation}
\label{sec:referencegen}

Reference images are generated per field sub-quadrant and serve as a historical average snapshot of the sky over the duration when the images were acquired for creating the reference. Difference imaging in the nightly operations are performed against the reference image for each field sub-quadrant. Reference images in the GDPS are created using a minimum of 40 dithered images (5$\times$ the number of images for the nominal survey) on the field. The minimum number of images was set by the requirement to quickly build up references after a large period of poor weather during the commissioning period of the GDPS and trigger validation of the image subtraction and transient extraction pipelines. As more images are acquired over the first year of GDPS operations, higher quality references are expected to be built by the end of the first year.

The input images that are stacked into the reference co-add are filtered for several quality cuts by querying the historical record of images in the DB. These include quality cuts on the astrometric and photometric solutions (e.g., to reject images affected by clouds or high humidity), the limiting magnitudes of the images and a moon-phase dependent cut on the moon distance of the observation (i.e., distance cuts are relaxed near new moon, but strict near full moon). Since the FWHM of the images are limited by the quality of the instrumental focus and not by the local seeing at the time of the observations, we do not put additional cuts on the FWHM of the images. Since the quality of the PSF focus changes from one side of the mount to the other (as the detector rotates with respect to the plane of the sky), both in terms of PSF FWHM (from 1.4 to 2.1 pixels) and ellipticity / orientation (from 0.02 to 0.25), the possibility of generating a separate set of reference images for each side of the mount is being investigated. 

Reference images creation is triggered at the end of the night for fields that \textit{do not} already have reference images. If the number of image quadrants that pass the quality cuts for reference generation is larger than the minimum number, the pipeline proceeds with reference image creation. In this process, the input images and their corresponding noise variance maps are flux-scaled to a common zero-point, and then individually drizzled to the pre-fixed drizzle field grid on the sky using the same \texttt{scale} and \texttt{pixfrac} parameters as the nightly survey stacks. The resampled images are combined as a sigma-clipped weighted mean of the input images, where the weighting is done with the scaled inverse variance maps. The sigma-clipping removes cosmic rays and artificial tracks from satellites and planes. The stacked image is stored to disk and metadata recorded to the DB. 

The reference image is fed to the same photometric calibration pipeline described in Section \ref{sec:photometry}, and a PSF model and photometric solution is derived for the stacked image, recorded to the DB and stored in the FITS header. As in the case of the epochal source catalogs, the photometric calibration produces a PSF-fit source catalog and aperture corrections for the reference image, which are used as seeds for generating light curves of all sources detected in the epochal stacks.

\subsubsection{Flat image and dead pixel mask generation}

The pixel to pixel responsivity of the detector (with respect to unity median over the image) is corrected using a flat image for the detector. Flat calibrations are generated using science exposures taken on sky during the night -- after applying quality cuts for the image counts, moon distance and local humidity. The flat generation pipeline is executed at the end of every night to query a list of raw images that qualify these cuts and are ordered by increasing humidity to select images taken during lowest humidity. The flat generation proceeds only if there are at least 200 images that satisfy the quality selection criteria. 

The most recent dark image is subtracted from each of the raw sky images, the input images are normalized by their own median and the flat image is created as a $\sigma$-clipped median (at $2.5 \sigma$) of the normalized images to reject outliers from stars, satellites and other sources on sky. The new flat images are saved to disk and their metadata are recorded to the DB along with a quality flag. A pixel-wise $\sigma$ and `count' image storing the number of pixels that contributed to the median are also recorded to disk and metadata recorded to DB. Nightly data reduction proceeds by querying the most recent good quality flat image constructed out of sky images, which usually corresponds to the sky flat constructed from images taken the previous night in the absence of bad weather.

The robotic scheduler was designed to acquire sky images with different exposure times (4\,s and 8\,s nominally) at the start and end of each night to support the creation of dead pixel masks. At the end of the night, the robotic watchdog queries for dead pixel calibration images taken within the last 2 weeks, and mask creation proceeds only if at least 100 good quality images are available in this time range. If available, the calibration images are normalized and median combined separately for the two exposure times and divided. The ratio image of the two exposure times are used to flag unresponsive and non-linear pixels by measuring the distribution of pixels in the ratio image. A dead pixel mask is created from this distribution flagging pixels that are more than n$\sigma$ away from unity. The dead pixel mask and its corresponding ratio image is stored to disk, and the metadata are recorded to the DB and the FITS headers. The nightly operations use the most recently constructed dead pixel mask for flagging bad pixels in the reduced images.

\subsubsection{Dark image and hot pixel map generation}

The thermal background of the detector is corrected using a dark image frame. Since the telescope and detector do not have a robotic shutter system, darks are acquired by manually covering the telescope tube with a cap back-coated with aluminium foil in the presence of observatory staff. Images of the dark optical beam are recorded with the same exposure time as used in the survey operations and then fed to the dark calibration pipeline. The dark calibration is created as a $\sigma$-clipped median of the input dark frames, with 20 input frames at a time. The median dark image is recorded to disk along with a `count' image denoting the number of images that survived the $\sigma$-clipping in each pixel and a $\sigma$-image (corresponding to the standard deviation of the pixel values across multiple dark frames) after the pixel clipping.

The median dark frame generated is used to flag hot pixels in the detector that have dark current levels more than $20\sigma$ larger from the median dark current in the detector. The $20\sigma$ cut is used since the high background imaging application of the detector does not require exceptionally low dark current, whereas the cut removes the worst outlier hot pixels. Pixels where the measured $\sigma$ (i.e., the standard deviation across multiple dark frames) in the dark frames is more than 5$\times$ the median $\sigma$ in the constructed dark are also flagged as these are noisy pixels. The final hot pixel mask is created as a logical OR between the high dark current and noisy pixels, and recorded to disk. Quality metrics and metadata for the median dark frame and hot pixel masks are stored to the FITS headers and recorded to the DB. The nightly operations use the most recently acquired dark frame and hot pixel mask for data calibration.
                                                                                    
\subsubsection{Match file generation}                                                

Light curves for every source detected in the single epoch field visits are stored in HDF5 format using match files, one for every field sub-quadrant. These are created using the reference image source catalog as seeds for  cross-matching detected sources across multiple visits of the same field. The files are generated manually (usually once a month) during day time when the processors are not occupied with night time processing. The pipeline proceeds by querying the complete list of stacked images for a given field that were acquired under photometric conditions, and uses the PSF-fit source catalog from the reference image as `seeds' for sources to be detected in the single epoch images. It then cross-matches every source detected in the single epoch stack (from the PSF-fit source catalogs) to the source catalog for the reference image to perform associations between sources and build up a complete light curve using  every visit for the respective field. 

The match file product stores the photometric measurements along with metadata for every exposure that contributed to the match file in multiple tables inside the output HDF5 file. The tables include an \textit{exposures}, \textit{sources} and \textit{sourcedata} table. The \textit{exposures} table stores metadata and observing conditions for every field visit that contributed to the match file, while the \textit{sources} table contains photometric measurements and quality flags of all sources detected in the reference image for that field sub-quadrant. Additionally, the \textit{sources} table stores statistics for the light curve of each source (if detected in the single epochs), including the average scatter, minimum, maximum and number of detections. The \textit{sourcedata} table contains individual photometric measurements of every source in the reference image that is detected in any single epoch image, storing the photometric measurements and quality flags. Each row in the \textit{sourcedata} table is associated with a unique object in the \textit{sources} table by a unique index in the \textit{sources} table to allow for light curve generation for each source. 

\section{On-sky performance}
\label{sec:performace}
The development period for the GDPS ended in June 2019, and the system was deployed for full survey operations on 2019 July 02, which marks the beginning of survey operations at the end of the commissioning period. We discuss the on-sky performance of the survey using all data acquired in the month of August 2019.

\subsection{Real-time pipeline success rate and latency}

\begin{figure*}
\centering
    \includegraphics[width=0.49\textwidth]{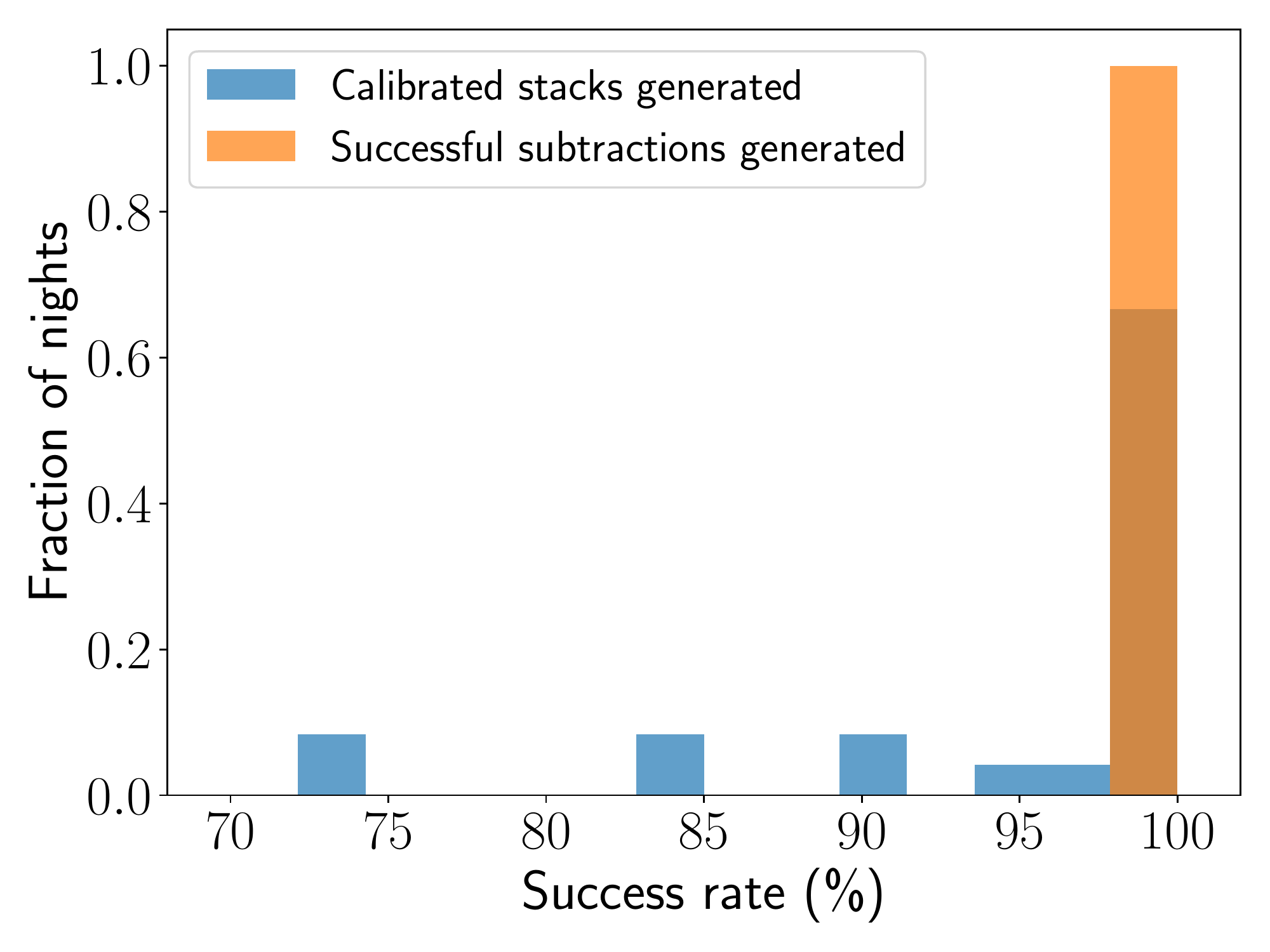}
    \includegraphics[width=0.49\textwidth]{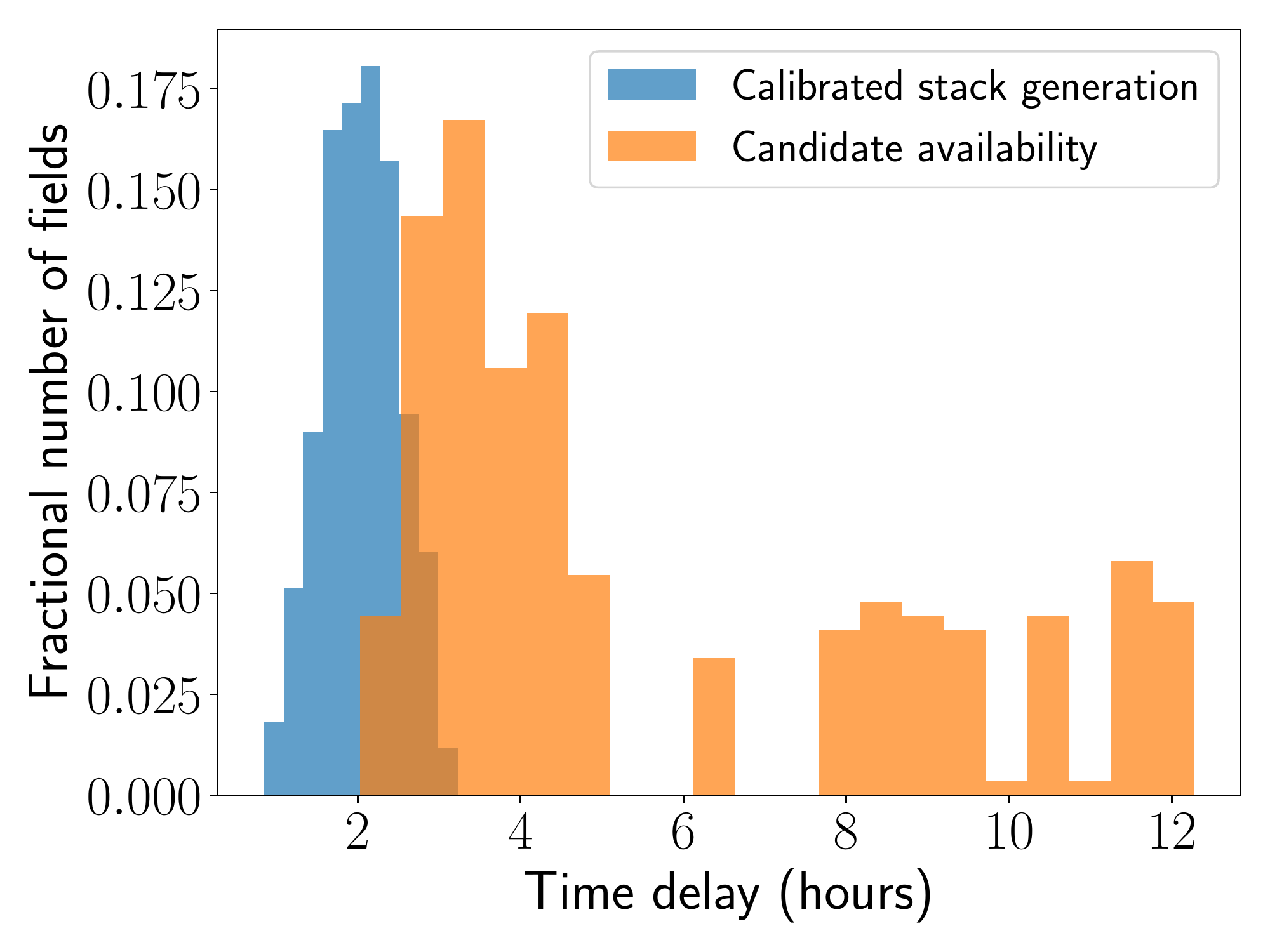}
    \caption{(Left) Distribution of the percentage success rate of data flow from raw images to photometrically calibrated stacks and transient candidates, as derived from data acquired in the month of August 2019. Primary causes for failure of the pipeline are observations taken under non-photometric conditions through clouds or at times of high humidity. (Right) Distribution of elapsed time between the end of field observation and the availability of photometrically calibrated stacks and transient candidates on the database server at Caltech. The median time for the availability of calibrated stacks is $\approx 2$ hours while the same for transient candidates is $\approx 3.8$ hours.}
    \label{fig:successtiming}
\end{figure*}

Figure \ref{fig:successtiming} shows a distribution of the success rate of the real-time pipeline processing in producing i) photometrically calibrated stacked data products on disk, and ii) in producing a usable subtraction for transient extraction. The success rate distribution was measured from all nights in August 2019. The median success rate of the production of photometrically calibrated stacks is $99.2$\%, while it is $ 99.4$\% for subtractions and transient extraction. The primary issue affecting the success rate of photometric stacks was observations taken under non-photometric conditions (e.g., through clouds, or through high humidity) leading to a significant fraction of astrometry and photometry failures (up to $\approx 30$\% of images on one of the nights in this period). The situation with regard to periods of high humidity is expected to improve with the recent installation of a window heater to avoid the accumulation of condensation.

Figure \ref{fig:successtiming} also shows a distribution of the elapsed time between the end of a field observation at the telescope and i) a photometrically calibrated stacked image being available on disk and ii) availablity of transient candidates in the DB for human vetting. The median time between the end of an observation and the generation of a photometrically calibrated image is $\approx 2$ hours, while all images are generally processed within $\approx 4$ hours from the end of an observation. Transient candidates are available within a median time of $\approx 3.8$ hours, although the distribution of elapsed time has a long tail extending out to $\approx 12$ hours. This primarily occurs due to fields in the Galactic plane where the high source density leads to the detection of a large number of candidates that strain the subsequent steps of PSF-fitting, photometry and external cross-matches. However, all candidates are available well before the start of the next night (the pipeline completes no later than noon of the following day on a typical night), and thus the processing is well suited for human vetting of transient candidates within a day of the observation and subsequent follow-up assignment.

\subsection{Astrometric accuracy}
\label{sec:astrometry_perf}

\begin{figure*}[ht]
    \centering
    \includegraphics[width=0.49\textwidth]{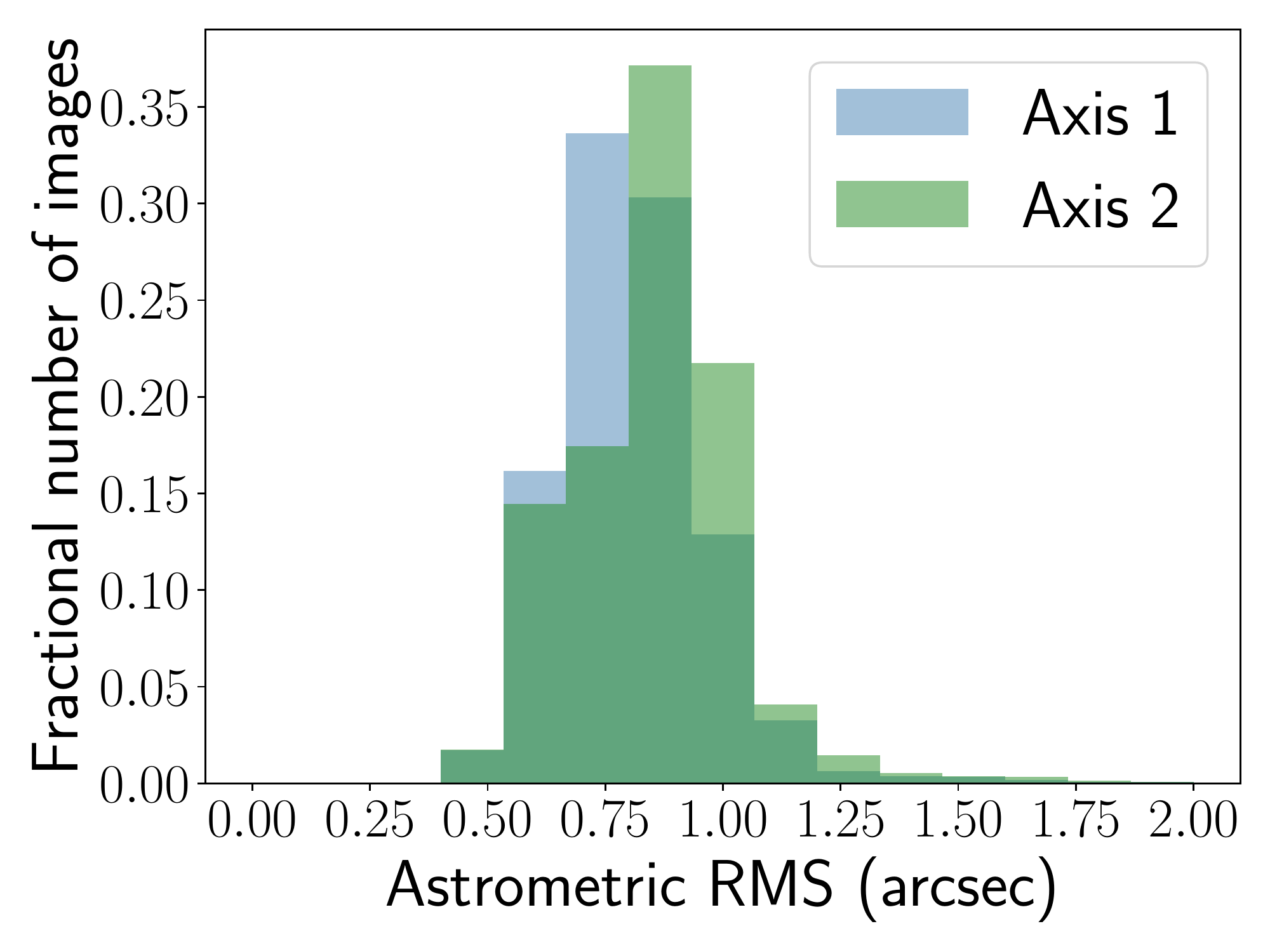}
    \includegraphics[width=0.49\textwidth]{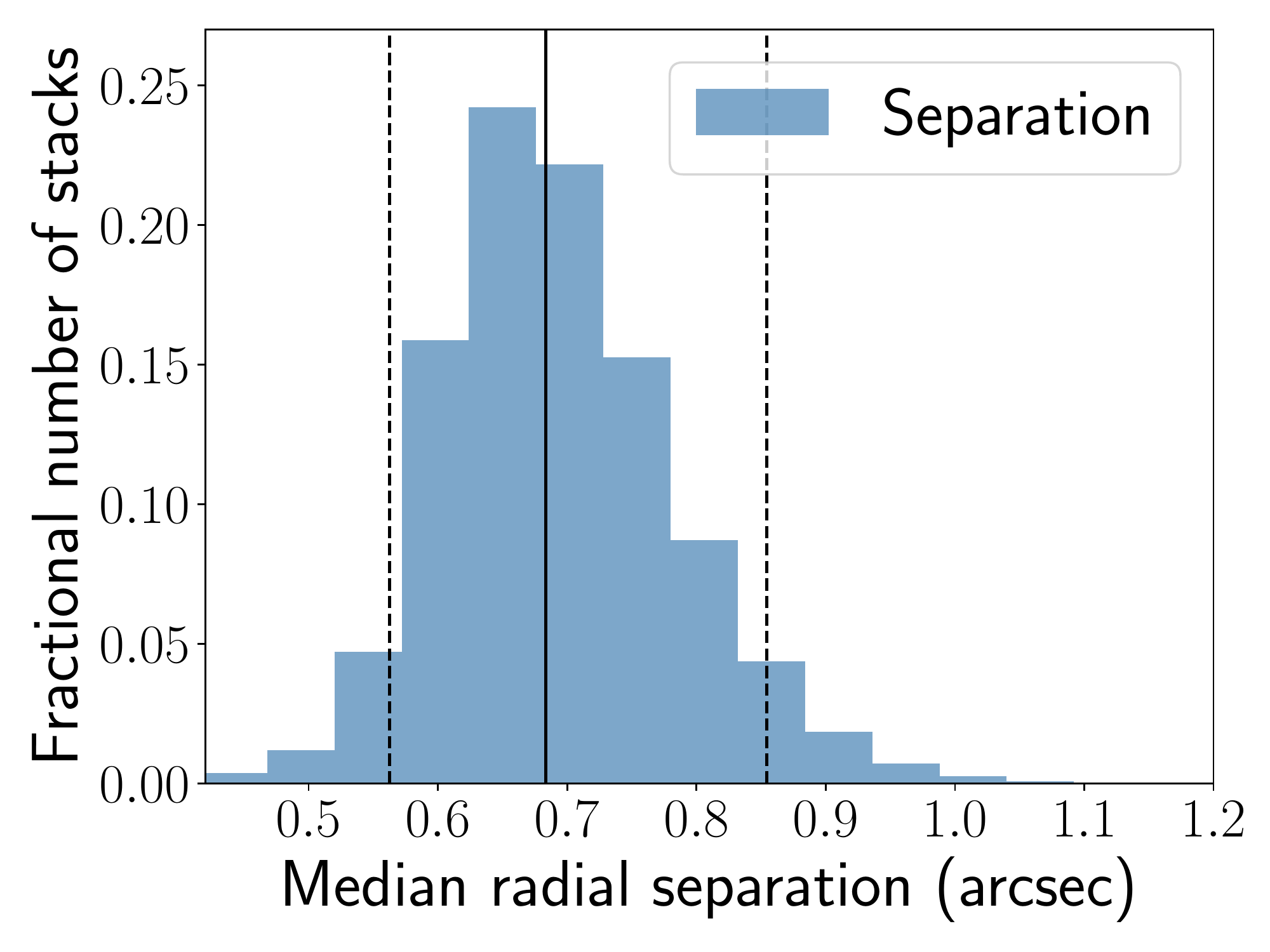}
    \caption{(Left) Distribution of the astrometric RMS derived from the scamp catalog fit for sources with S/N $>$ 10, using all images taken over all nights in the month of August 2019. (Right) Distribution of the radial RMS offset between cross-matched sources in drizzled stacks and the photometric catalog. The solid line is the median radial separation while the dashed lines show the 5th and 95th percentiles of the radial separation distribution.}
    \label{fig:astdistribution}
\end{figure*}

The astrometric calibration method described in Section \ref{sec:astrometry} leads to astrometric solutions which have typical accuracies of $\approx 0.8$\,\arcsec\, ($\approx 0.1$ pixels) over the entire sky. Figure \ref{fig:astdistribution} (left panel) shows the distribution of the astrometric RMS (per axis) achieved over a range of airmasses and observing conditions during all nights in the month of August 2019, as measured from the astrometric solutions of the native images (after de-trending) using \texttt{Scamp} for sources with S/N $> 10$. We show an equivalent plot for the stacked images produced using \texttt{Drizzle} on the right panel of Figure \ref{fig:astdistribution}, plotting the distribution of the median radial separation of sources detected in the Drizzled images cross-matched to the Gaia DR2 reference catalog. The median radial separation is $\approx 0.7$\,\arcsec.

\begin{figure*}
    \centering
    \includegraphics[width=0.49\textwidth]{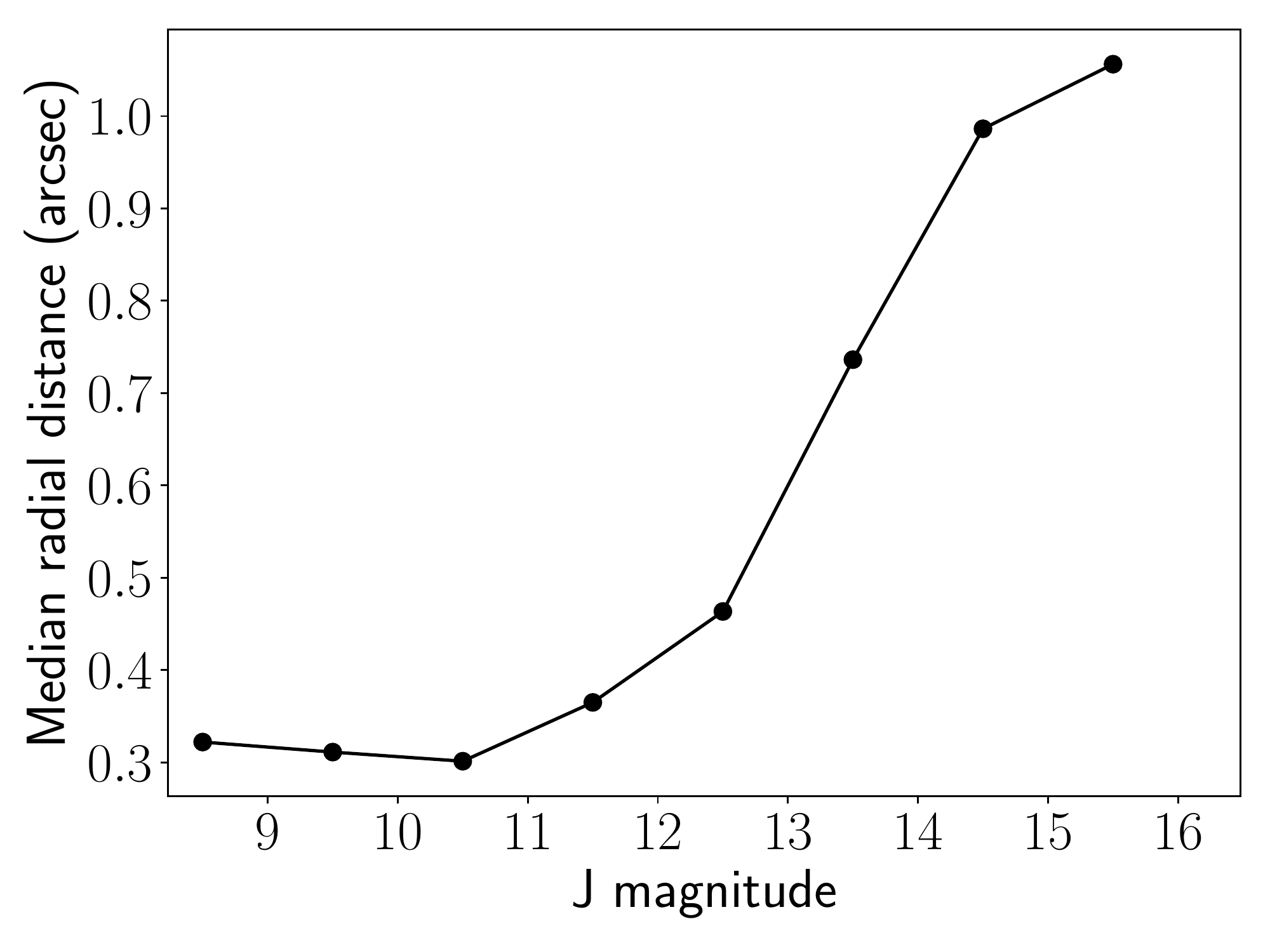}
    \includegraphics[width=0.49\textwidth]{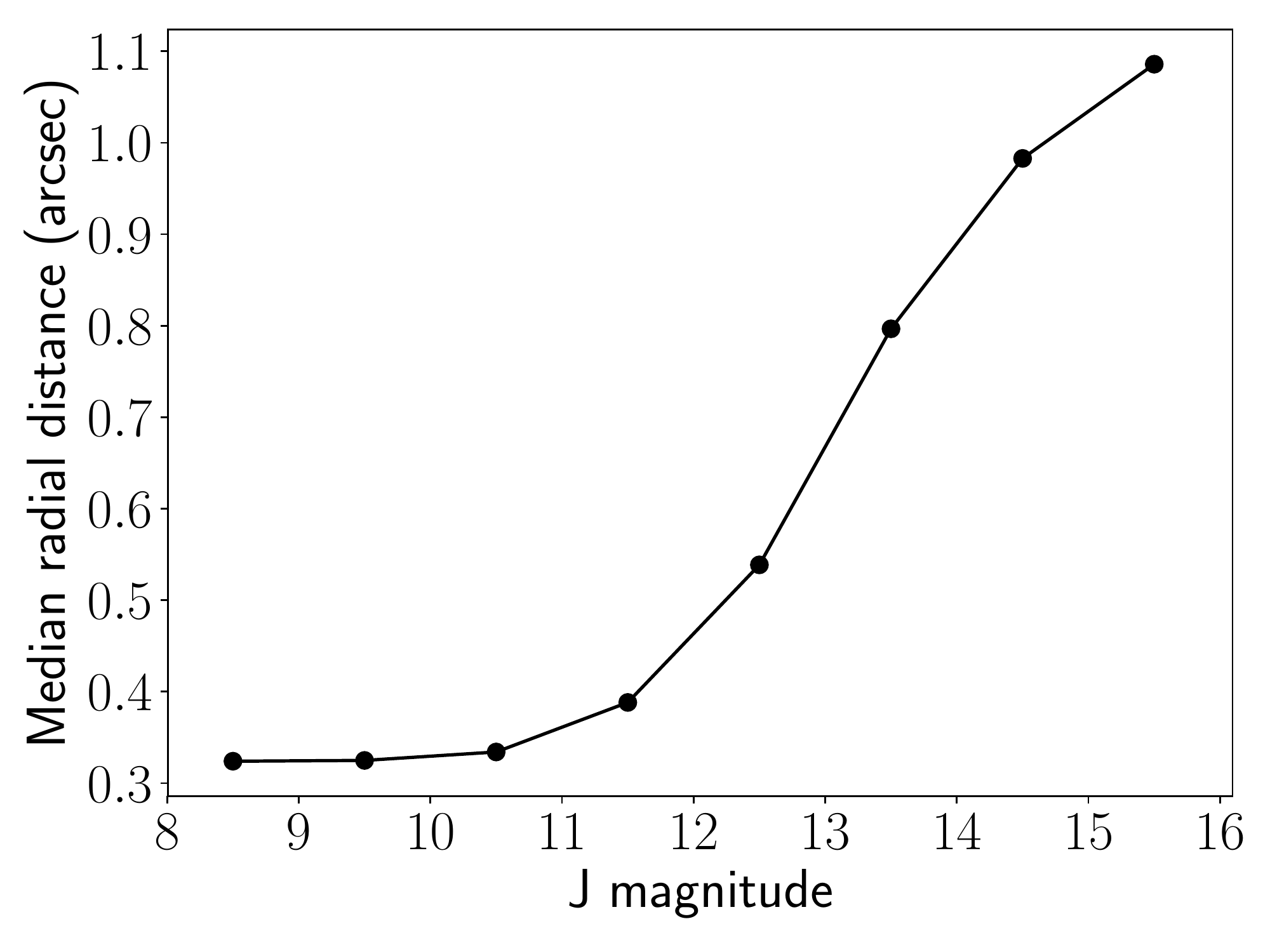}
    \caption{Median radial astrometric distance with respect to Gaia DR2 as a function of the J Vega magnitude of sources down to SNR = 5. The left plot was created from the astrometric solutions for 20 field visits of a high Galactic latitude field while the right plot corresponds to the same for 20 field visits of a low Galactic latitude field. The astrometric accuracy for sources brighter than $\approx 12$\,Vega mag is better than $\approx 0.4$\,\arcsec\, ($0.05$ native pixels).}
    \label{fig:astmagrms}
\end{figure*}

Note that Figure \ref{fig:astdistribution} shows the astrometric accuracy of \textit{all} sources detected above SNR $> 10$, where the astrometric measurements are prone to Poisson errors of centroiding for sources near SNR $\approx 10$. The true achievable precision is higher for the case of brighter sources and is depicted in Figure \ref{fig:astmagrms}, showing the median radial separation of sources from the reported Gaia DR2 position as a function of the source magnitude, both for a high Galactic latitude and a low Galactic latitude field. For sources brighter than $\approx 13$\,Vega mag, the astrometric precision achievable is better than $\approx 0.4$\,\arcsec\, ($0.05$ native pixels) and is representative of the achievable astrometric precision. The astrometric precision is likely limited by the measurement of accurate source positions in the presence of asymmetric and variable PSFs in the final stacked images.

\subsection{Photometric accuracy}
\label{sec:photometry_perf}
\begin{figure*}[!ht]
    \centering
    \includegraphics[width=0.49\textwidth]{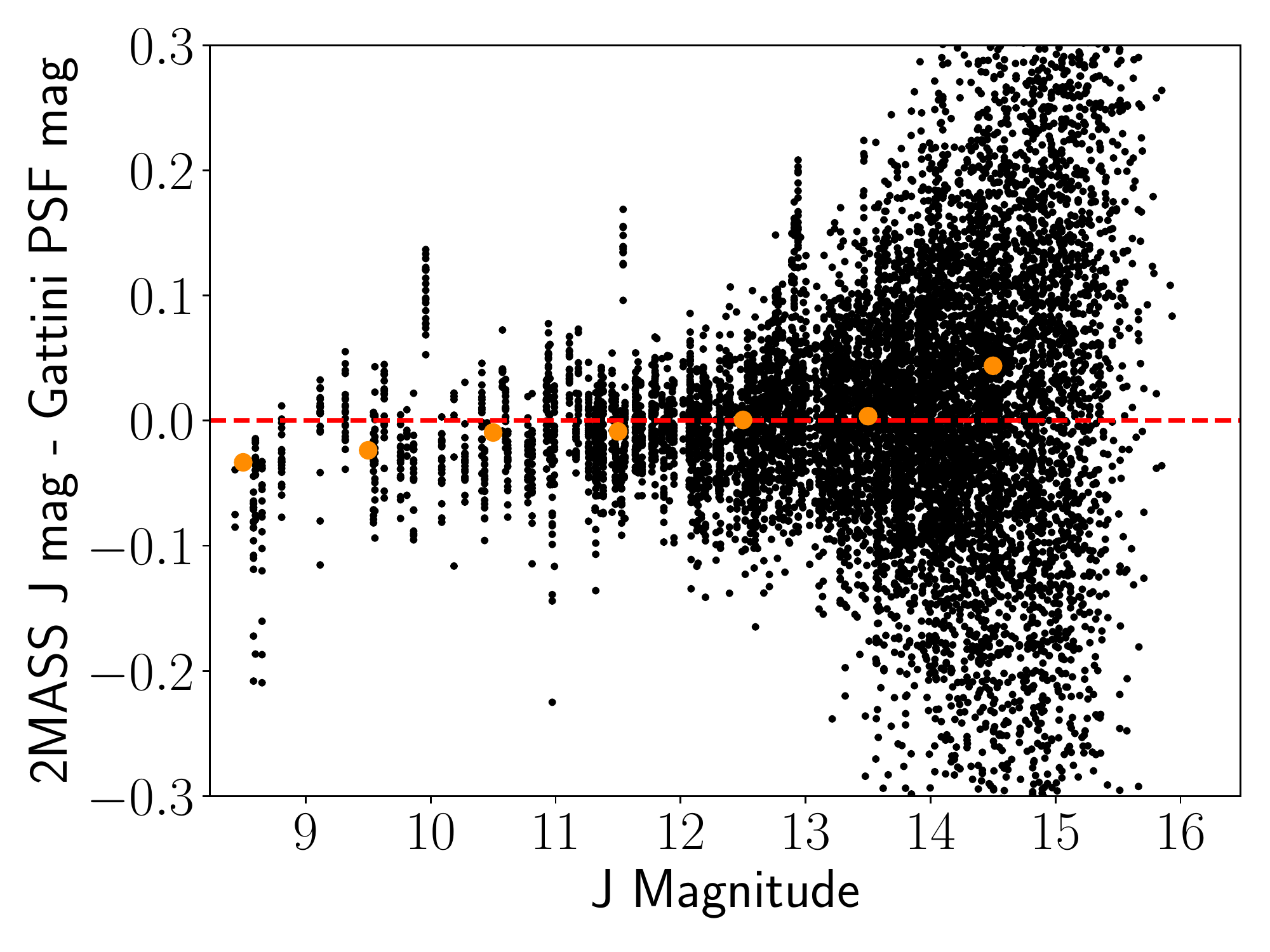}
    \includegraphics[width=0.49\textwidth]{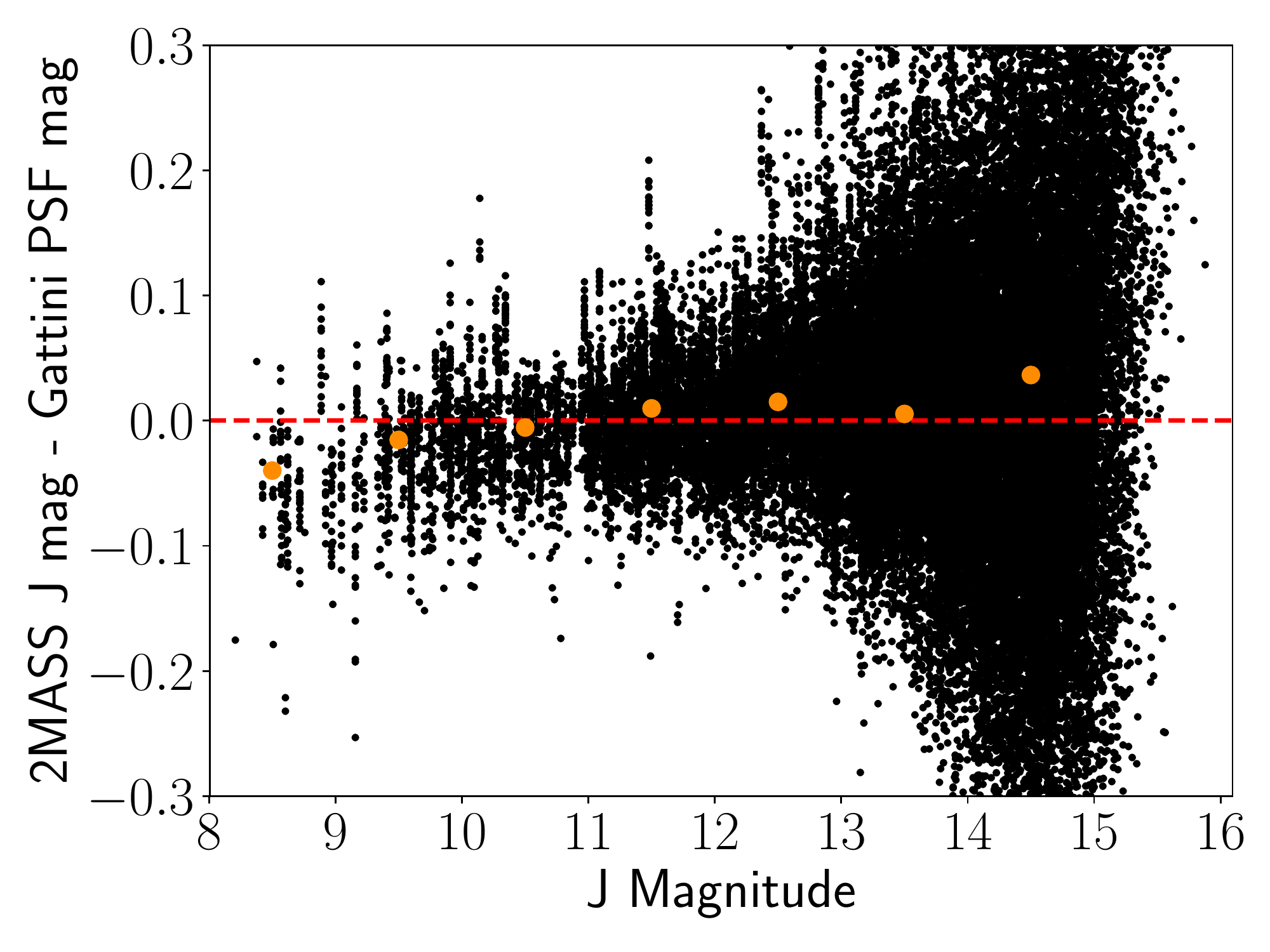}
    \includegraphics[width=0.49\textwidth]{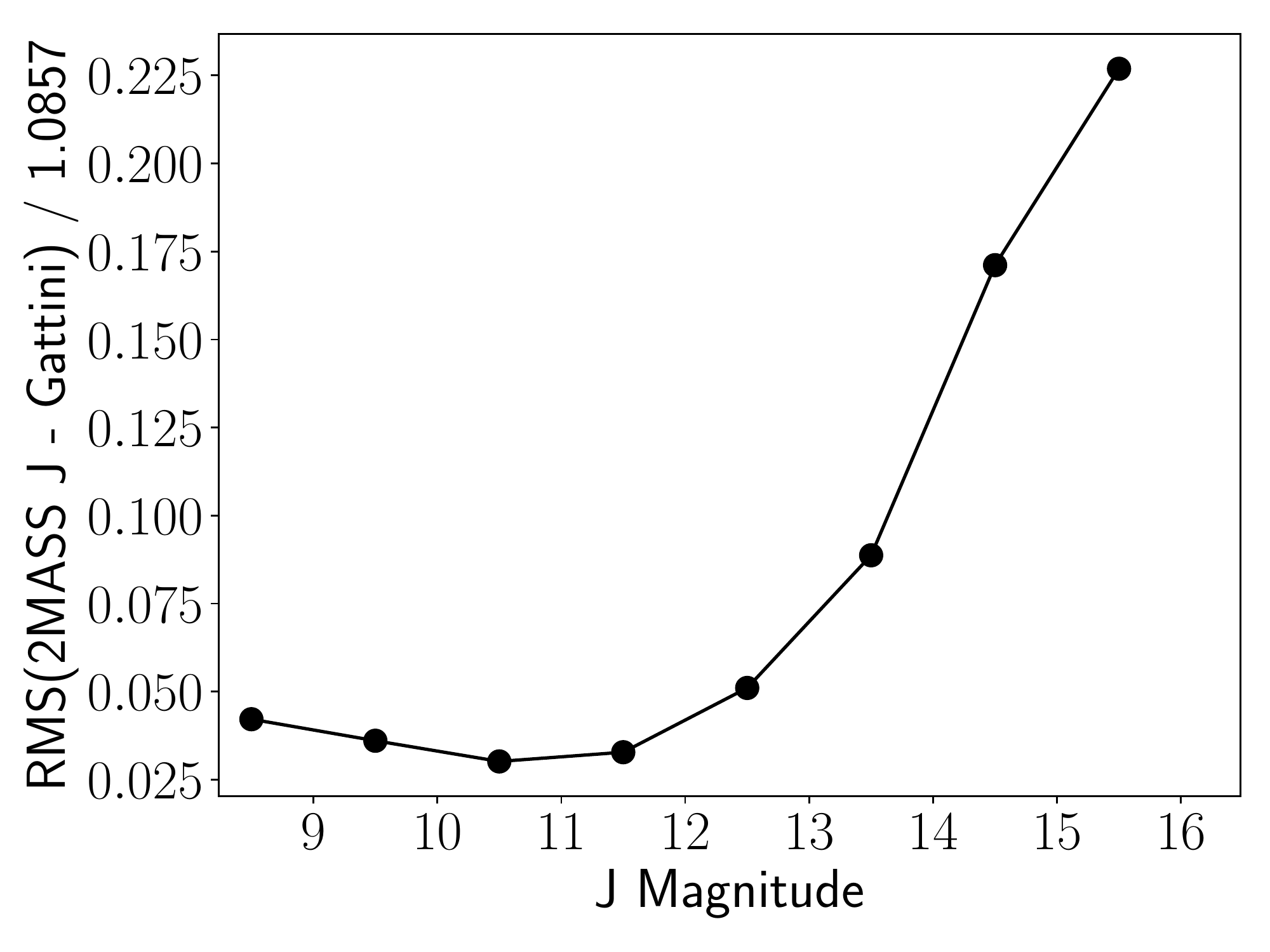}
    \includegraphics[width=0.49\textwidth]{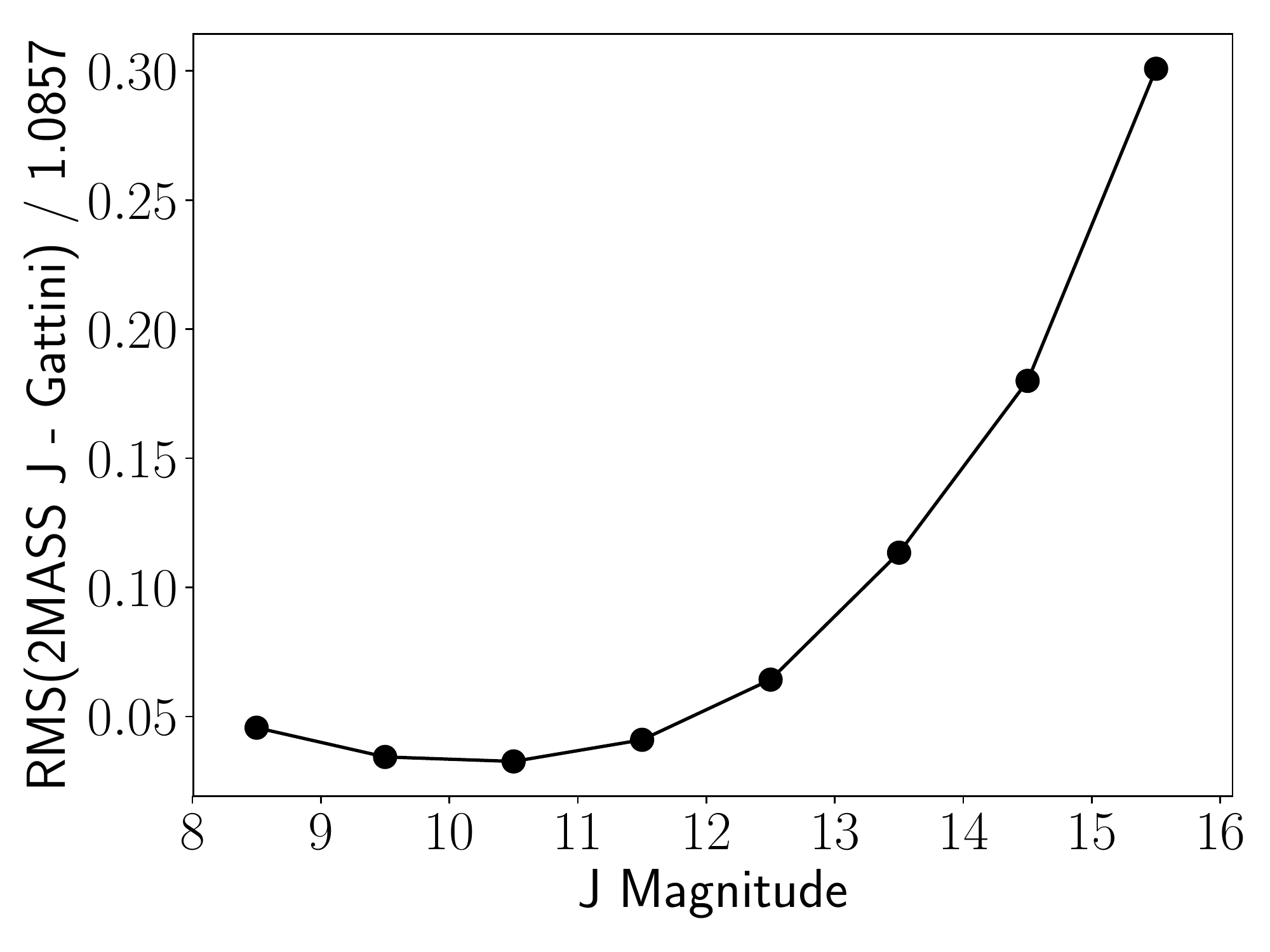}
    \caption{(Top) Difference between 2MASS and Gattini J band magnitudes, as a function of J band magnitude, from the photometric solutions derived in the GDPS. The left panel shows a high Galactic latitude field and the right panel shows a low Galactic latitude field, compiled from 20 visits of each field. The orange dots denote the median deviation in bins of 1 magnitude, showing evidence of a small systematic shift at the brightest and faintest ends. (Bottom) Relative flux RMS between 2MASS and Gattini as a function of source magnitude. The relative flux RMS is estimated from the magnitude RMS and then normalizing by 1.0857. The left and right panels are for the same single epoch fields shown in the upper panels.}
    \label{fig:zpsol}
\end{figure*}

Section \ref{sec:photometry} describes the photometric calibration procedure for stacked images against the 2MASS catalog. Figure \ref{fig:zpsol} shows the distribution of the difference between 2MASS and Gattini calibrated magnitudes (including the zero-point and a color term) measured from the epochal PSF-fit source catalogs for a stacked field sub-quadrant. The residuals are plotted as a function of the source magnitude combined over 20 visits of a high and low Galactic latitude field respectively. The left panels show the distributions for a high Galactic latitude field (low source density) and the right panels are for a low Galactic latitude field (high source density). The photometric scatter RMS increases for fainter sources and up to $\approx 20$\% for sources near the 5$\sigma$ limiting magnitude. The orange squares denote the median residual in bins of 1 magnitude, which show evidence of a small systematic deviation at the brightest and faintest ends of the distribution, likely due to uncorrected non-linearity in the detector pixel response.

The bottom panels show the dependence of the median flux RMS residuals against 2MASS as a function of the source magnitude, binned into groups of 1 magnitude. The best achievable photometric accuracy is $\approx 3$\% (30 mmag) at the bright end (brighter than $\approx 11$ mag), while it is better than $\approx 5$\% for sources brighter than $\approx 13$mag. The scatter is believed to be largely dominated by errors in background estimation, flat-fielding and PSF fitting over the field sub-quadrant (note that the PSF variation over a single field sub-quadrant can be significant at the edges of the focal plane). 

\subsection{Sensitivity and image quality}
\begin{figure*}[!ht]
    \centering
    \includegraphics[width = 0.49\textwidth]{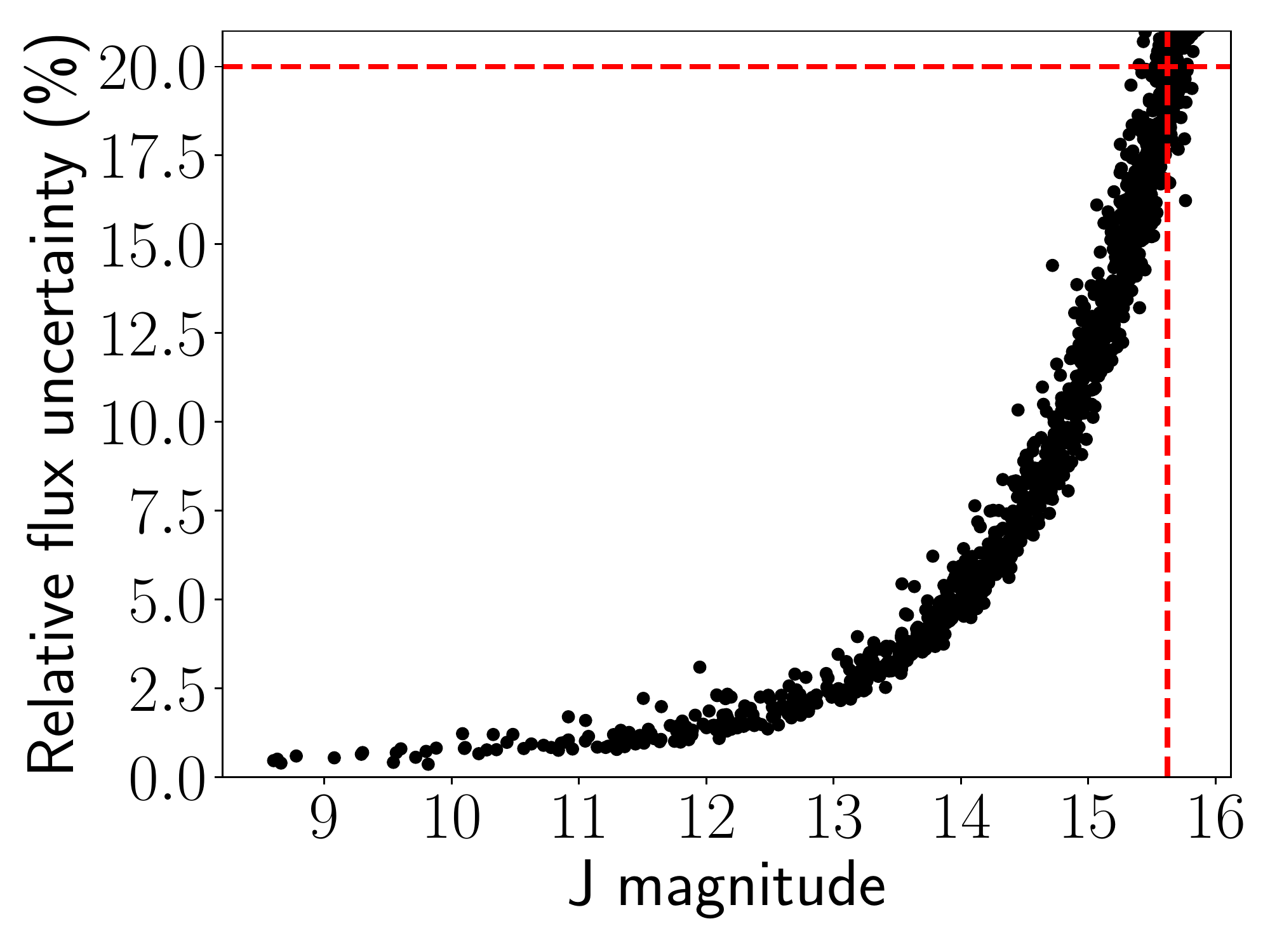}
    \includegraphics[width = 0.49\textwidth]{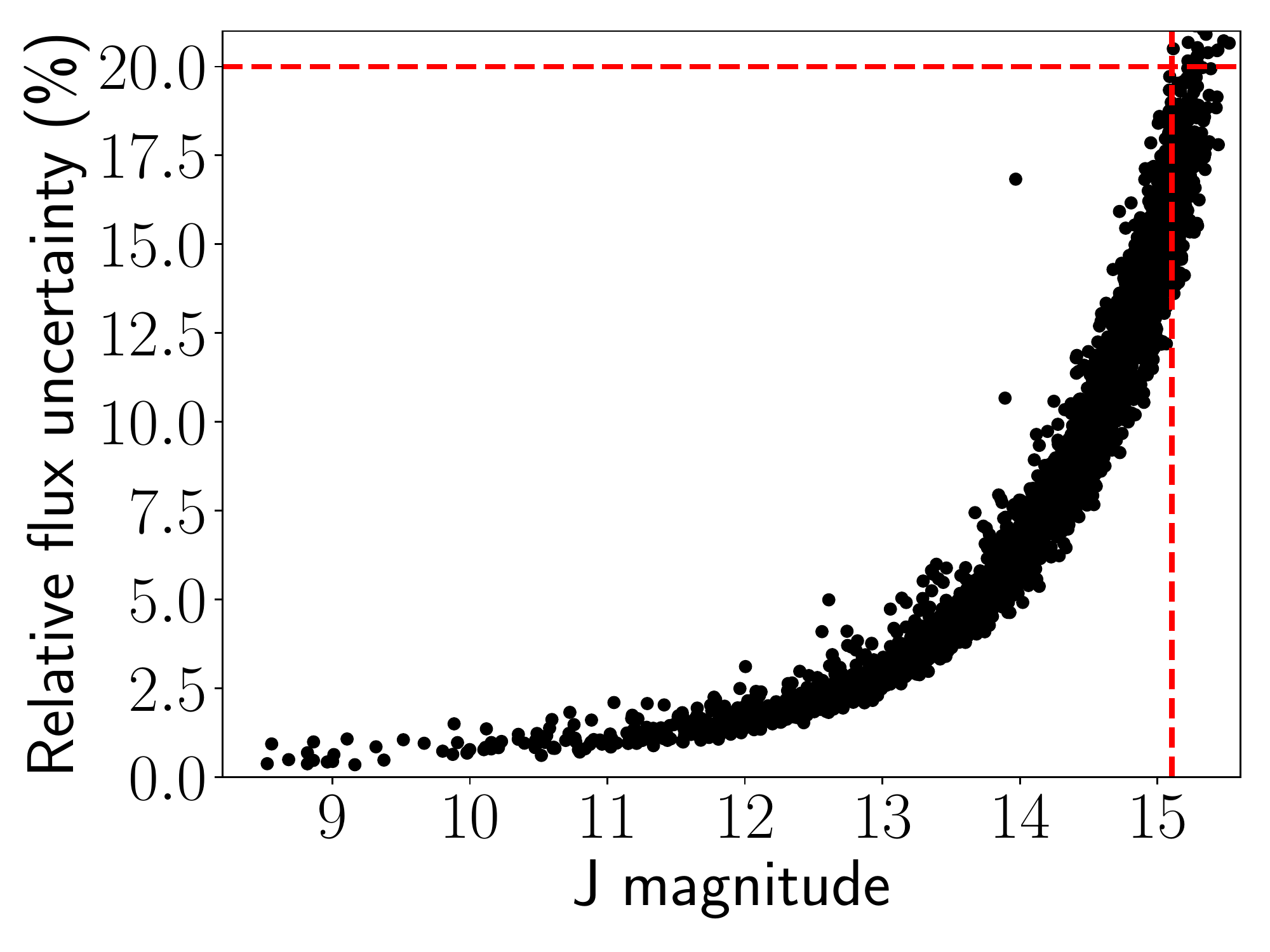}
    \caption{Relative flux uncertainty as a function of source magnitude from single epoch PSF-fit source catalogs in the drizzled stacked images. The red dashed vertical lines show the limit corresponding to SNR = 5 estimated from the background RMS, and are consistent with that estimated from 20\% flux uncertainty of the detected sources, indicated by the red dashed horizontal lines. The left plot is for a high Galactic latitude field and the right is for a low Galactic latitude field.}
    \label{fig:fluxunc_ep}
\end{figure*}{}

The $N\sigma$ limiting magnitude of each stacked image can be estimated using either an estimate of the median background RMS in the image and the size of the PSF, or measuring the observed magnitudes for sources in a narrow range of SNR around $N\sigma$. The GDPS measures the limiting magnitude of each image using both methods, while including the corrections for correlated noise in the image pixels (Section \ref{sec:correlatednoise}). Figure \ref{fig:fluxunc_ep} shows the relative flux uncertainty for sources detected in a single epoch field visit of a high Galactic latitude field (left) and a low Galactic latitude field (right). The 5$\sigma$ limiting magnitude corresponds to a relative flux uncertainty of 20\%. The vertical red dashed lines correspond to the limiting magnitude estimated using the background RMS and PSF size information, while the horizontal red dashed line marks the location of 20\% flux uncertainty. The intersection of the two lines overlap with the contour of sources in the flux uncertainty plane suggesting consistent limiting magnitude estimates from the two methods.

\begin{figure}[!ht]
    \centering
    \includegraphics[width = 0.49\textwidth]{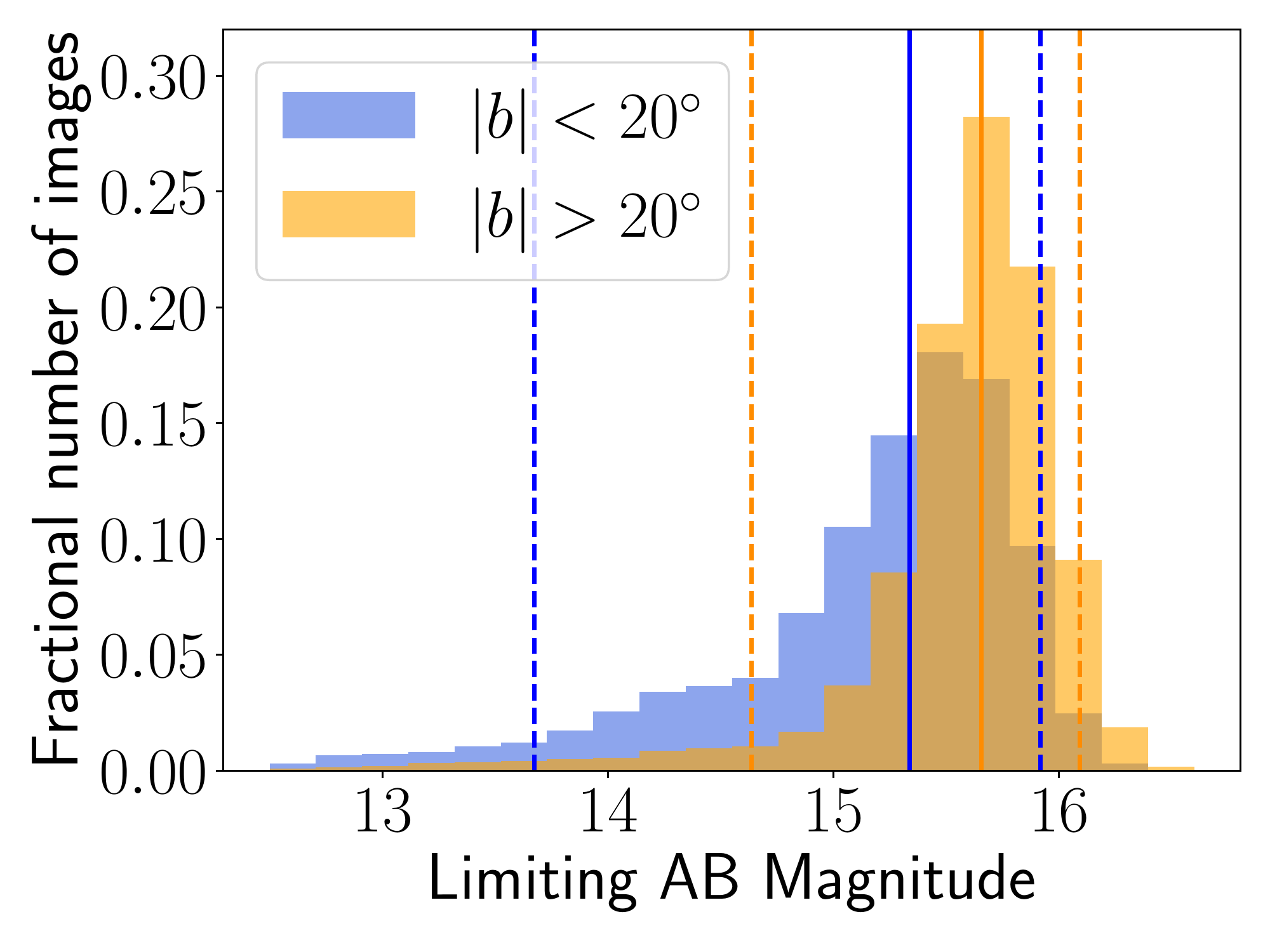}
    \caption{Distribution of 5$\sigma$ limiting magnitudes for low (blue) and high (orange) Galactic latitude fields for all observations taken in August 2019. The limiting magnitudes were estimated using the magnitudes of sources detected in narrow range of SNR aroung 5. The orange solid vertical line denotes the median limiting magnitude for the high Galactic latitude fields while the two dashed lines denote the 5th and 95th percentile limiting magnitudes. The blue vertical lines correspond to the same statistics for the low Galactic latitude fields.}
    \label{fig:limmagdist}
\end{figure}{}

As seen in Figure \ref{fig:fluxunc_ep}, the depth of the image for the low Galactic latitude field is shallower than that for the high Galactic latitude field as a result of confusion noise in regions of very high source density, given the large pixel scale of the detector. Confusion noise generally degrades the limiting magnitudes (estimated using the the magnitude of sources near SNR = 5) of observations in the Galactic plane fields. In Figure \ref{fig:limmagdist}, we show a distribution of the 5$\sigma$ limiting magnitudes of images taken over nights in August 2019 near the Galactic plane ($\left|b\right| < 20^\circ$) and outside the plane ($\left|b\right| > 20^\circ$). The median 5$\sigma$ limiting magnitude of images outside the plane is $\approx 15.7$\,AB mag while the same for low Galactic latitude fields is $\approx 15.3$ AB mag. The low Galactic latitude fields show a long tail in the distribution extending to shallow limiting magnitudes, corresponding to the most crowded fields in the Galactic plane.

\begin{figure*}
\centering
\includegraphics[width = 0.8\textwidth]{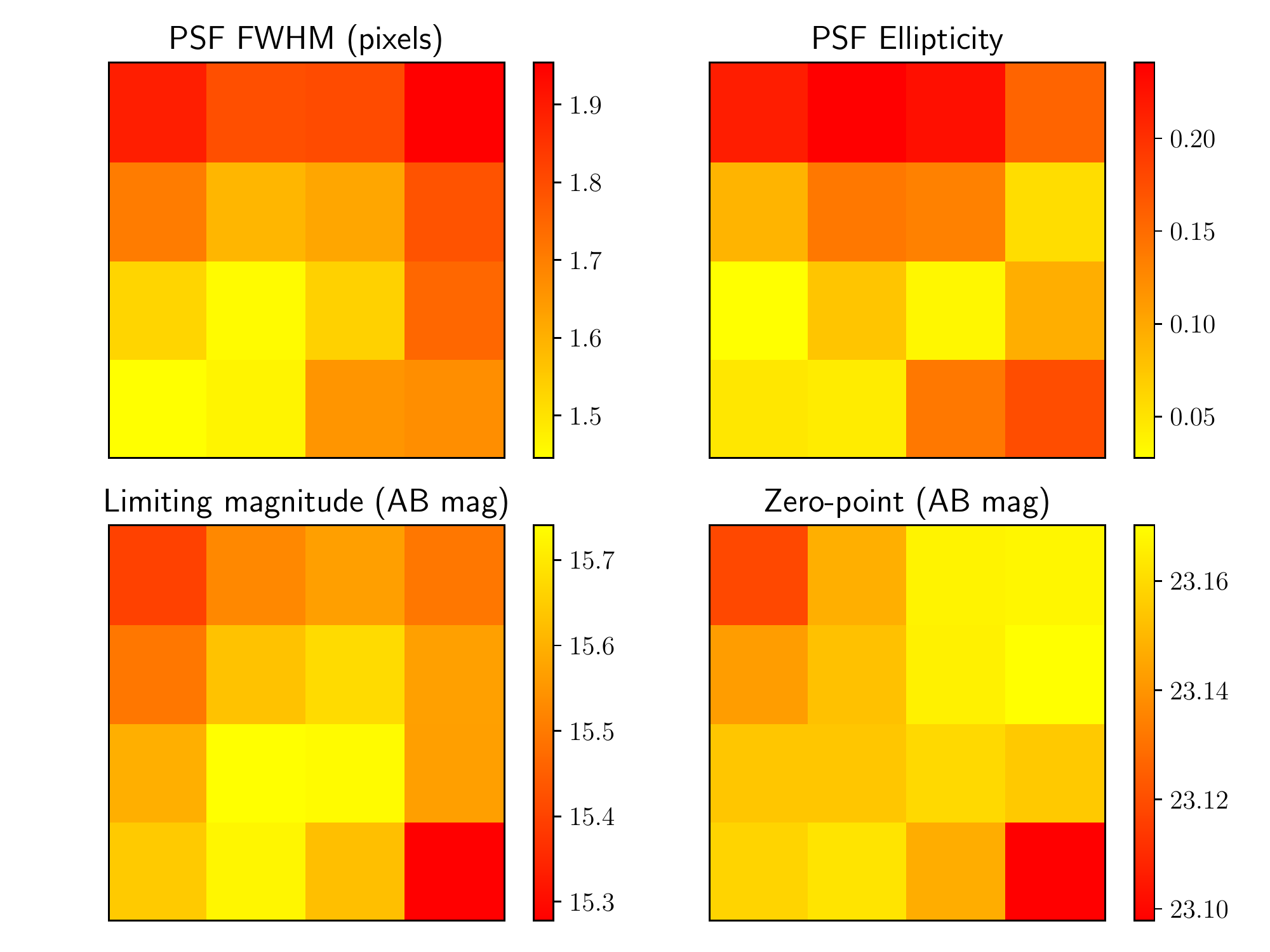}
\caption{PSF and zero-point variation aroung the Gattini focal plane, derived from all data taken in August 2019. (Top left) PSF FWHM variation around the focal plane in units of native detector pixels. (Top right) PSF ellipticity variation around the focal plane. (Bottom left) Median limiting magnitude as a function of position in the focal plane. (Bottom right) AB zero-point as a function of position in the decetor plane. }
\label{fig:foczp}
\end{figure*}

The limiting magnitude of the stacked field sub-quadrants also depends on its position in the plane of the detector due to large variations in the PSF size and shape across the detector. Larger PSFs lead to shallower limiting magnitudes given the high sky background. The top panels in Figure \ref{fig:foczp} show the variation of the PSF FWHM and ellipticity measured as a function of position on the detector, averaged over all nights of observations in August 2019. Sub-pixel image quality has not been achieved with the designed optical system. In the absence of a focusing mechanism (due to be installed in 2019), variations in the PSF FWHM and ellipticity have been observed as a function of the ambient temperature. The typical PSF FWHM size is $\approx 12 - 14$\,\arcsec\, ($\approx 1.5 - 1.7$ detector pixels), although it is worse (up to $\approx 17$\,\arcsec\, $\approx 2$ detector pixels) at the edges of the focal plane. As a result, the use of \texttt{Drizzle} with an output pixel scale half of the native pixel scale has been adequate to reconstruct Nyquist sampled images. 

The lower panels in Figure \ref{fig:foczp} show the median limiting magnitude (including both low and high Galactic latitude observations) and zero-point as a function of position in the detector plane. The zero-point of the detector is largely uniform across the plane except for the bottom right corner, where the zero-point (sensitivity) is lower by $\approx 0.05$\,mag due to the absence of a AR coating. The distribution of limiting magnitude as a function of position in the detector plane is influenced by a combination of the local zero-point and PSF size, such that larger zero-points and smaller PSF FWHMs lead to deeper limiting magnitudes. Since the sky background from the PSF foortprint is the dominant noise contribution limiting the image depths, the depths of the images are expected to improve with the installation of a focus mechanism in the second half of 2019.

\subsection{Reference images}

\begin{figure*}
\centering
\includegraphics[width = 0.9\textwidth]{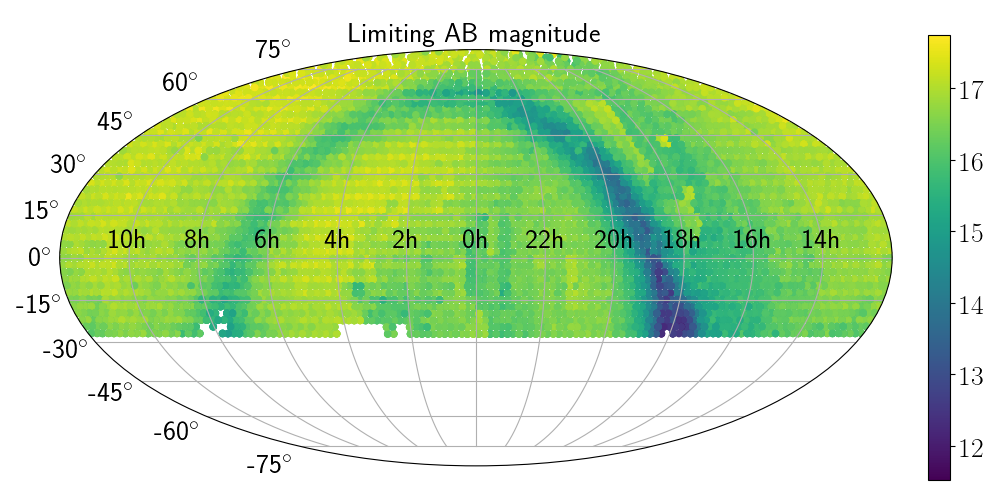}
\caption{Distribution of the depths of reference images as a function of sky position, as of June 2019 (just before the start of survey operations in July 2019). The deepest reference images achieve depths of $\approx 18$ AB mag, while the depths are limited by confusion noise in the Galactic plane.}
\label{fig:refdepth}
\end{figure*}

Section \ref{sec:referencegen} discusses the generation of reference images by stacking all exposures of a given field acquired under photometric conditions. Although reference image availability was limited during the commissioning phase of the survey due to extended periods of bad weather and poor observing conditions, reference maps were re-built at the end of June 2019 with data acquired between November 2018 and June 2019. The resulting reference maps have 99.3\% coverage of the visible sky from Palomar and the sky distribution of the reference limiting magnitudes are shown in Figure \ref{fig:refdepth}. The number of images stacked in each field varies between 40 individual dithers ($5\times$ the nominal survey field visits) and 300 individual dithers, with fields near the north pole having the largest number of images stacked due to their near continuous visibility from Palomar Observatory. The resulting limiting magnitudes of the reference stacks vary between $\approx 14.5$\,AB mag in crowded regions in the Galactic plane (where the depths are limited by confusion) to $\approx 18$\,AB mag outside the plane. The reference images are deeper than 16.5\,AB mag for $60$\% of the sky.

\subsection{Difference images and transient recovery}\label{sec:recovery}

\begin{figure}
    \centering
    \includegraphics[width=0.44\textwidth]{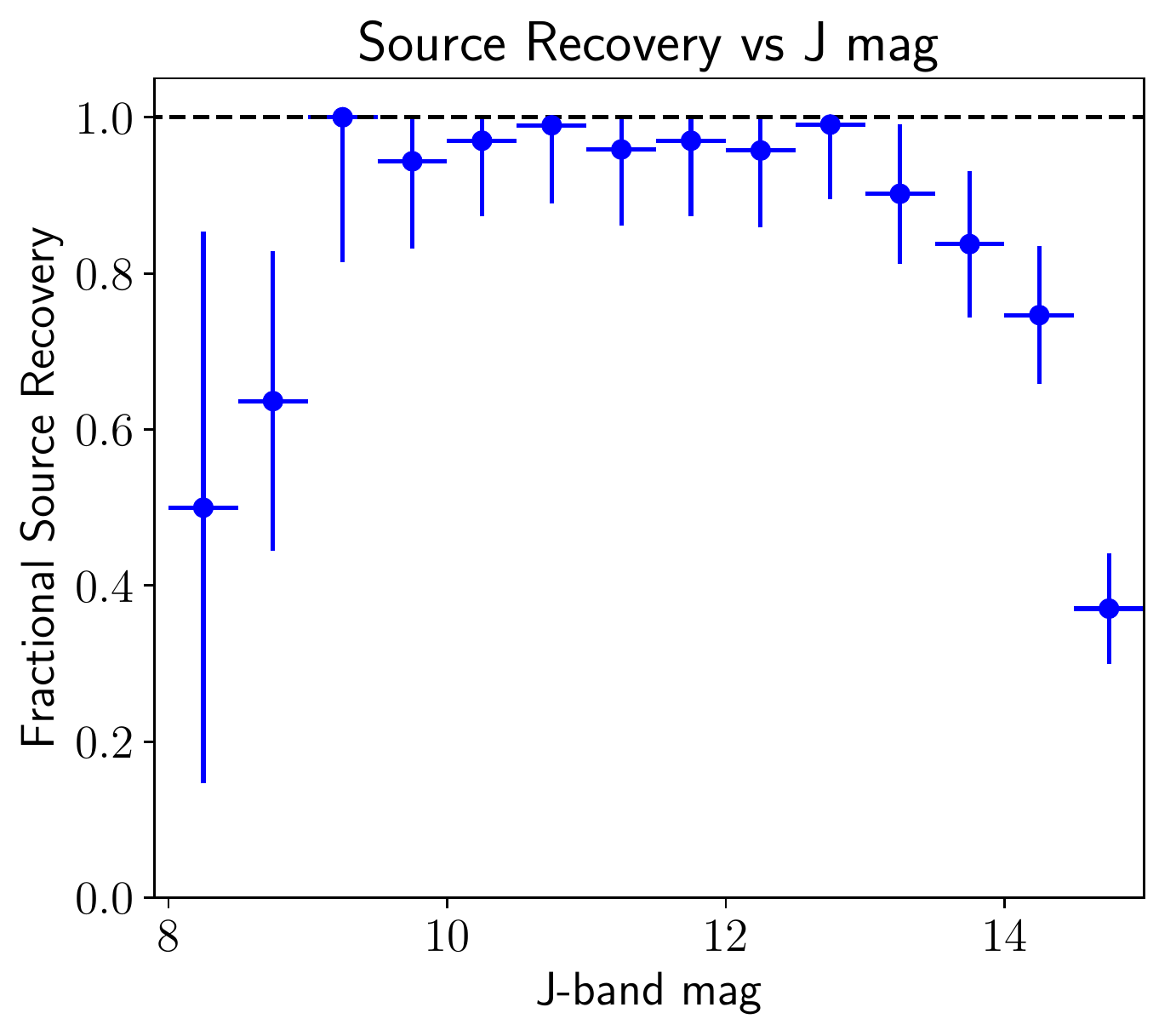}
    \caption{Fractional recovery of injected sources as a function of their J-band magnitude. The y error bars assume counting statistics for the number of sources in each bin. The black dashed line indicates the maximum possible fractional recovery value.}
    \label{fig:FakeRecovery1}
\end{figure}

In order to evaluate the efficacy of the difference imaging pipeline we performed various tests with injected fake sources. First, a set of test data was created using a copy of the nightly survey data. A total of 992 fake sources were inserted in these images using a PSF model for the image generated using PSFex and scaled to a specific magnitude. The magnitude of the fake sources were randomly drawn from a uniform brightness distribution between 0 and 5 magnitudes above the limiting magnitude of the science image. The positions for the fake sources were randomly drawn from a uniform distribution of x \& y coordinates in the image, with checks to ensure that sources were not placed in low-weight portions of the maps (i.e., regions near the edge which were not sampled by at least half the dithers of the dither sequence). The subtraction pipeline was run on these test images to check the fractional source recovery, which are shown in Figure \ref{fig:FakeRecovery1}.

Injected sources which were missed by the pipeline can be broadly characterized into six groups: sources that were flagged as saturated (2), sources that were coincident with the masked area of a nearby bright star (11), sources masked as possible artifacts from the missing AR coating on a corner of the detector (23), sources which failed to achieve a peak Scorr value of 5 (38), sources contained in images that failed quality criteria for candidate extraction (2), and sources that were initially detected but cut as part of candidate filtering criteria (17). The values listed after each category indicate the number of missed sources in that category. In total, 899 sources out of 992 possible sources were recovered, yielding an overall recovery fraction of 90.6\%. 

\begin{figure}
    \centering
    \includegraphics[width=0.45\textwidth]{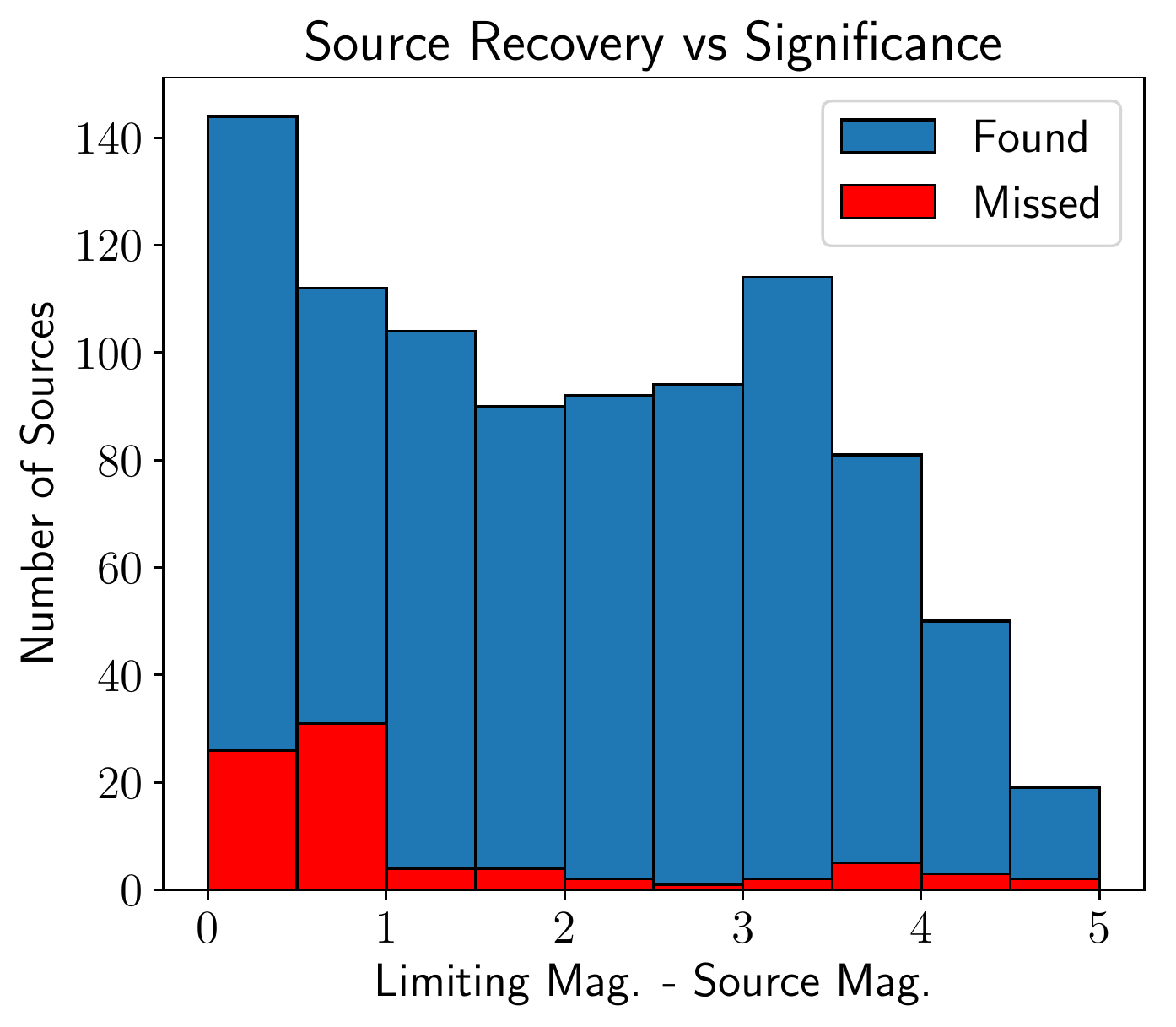}
    \caption{Histogram of recovered and missed sources binned by the difference of the source magnitude and limiting magnitude of the image. These data are the same from Figure \ref{fig:FakeRecovery1}. }
    \label{fig:FakeRecovery2}
\end{figure}

\begin{figure}
    \centering
    \includegraphics[width=0.44\textwidth]{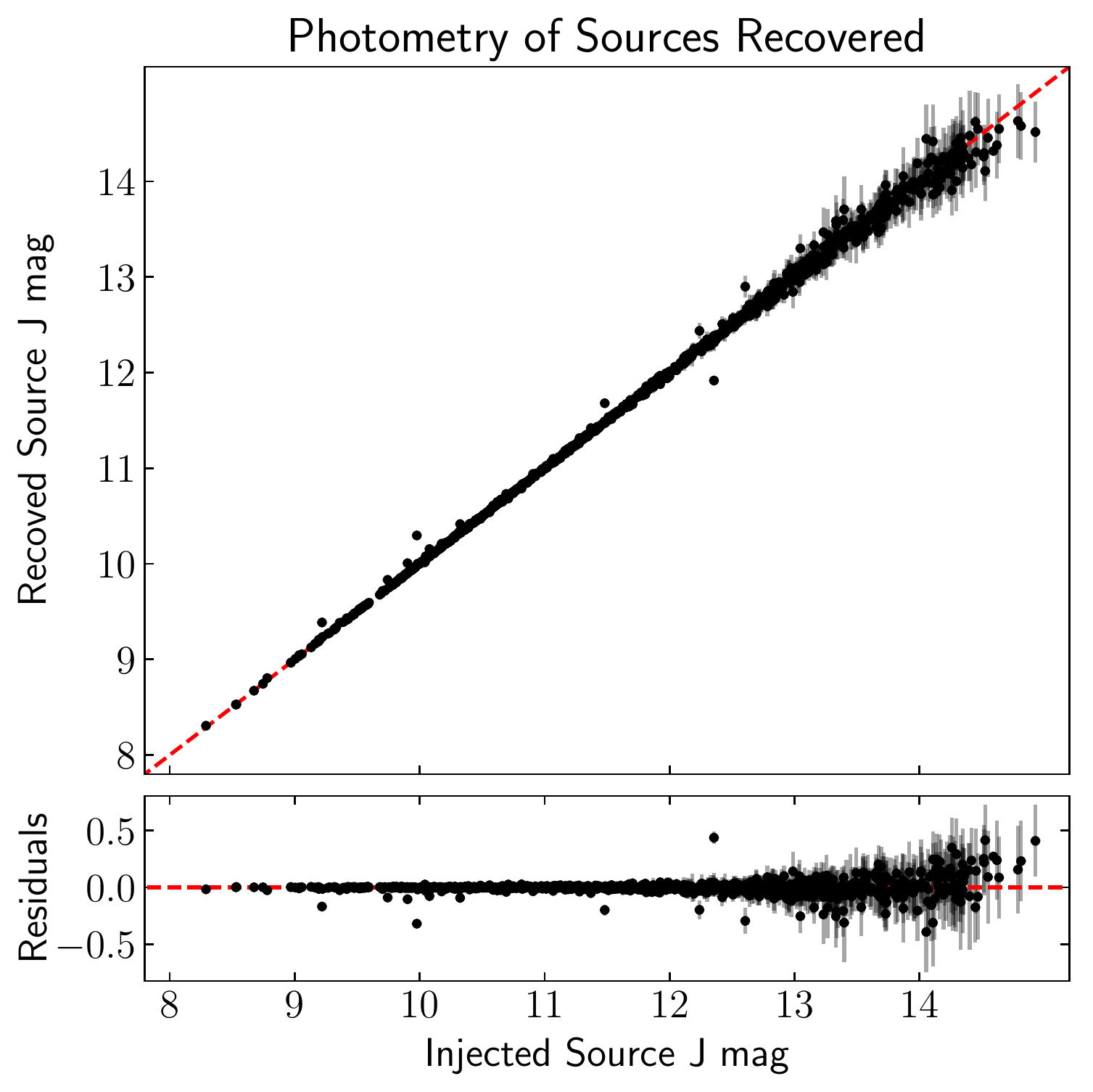}
    \caption{Injected vs. recovered magnitude of sources by the subtraction pipeline. The dashed red line in the top plot shows x=y, which the data generally follows. The bottom panel shows the source residuals with the red dashed line indicating the zero level.}
    \label{fig:RecoveredMag}
\end{figure}

\begin{figure}
    \centering
    \includegraphics[width=0.47\textwidth]{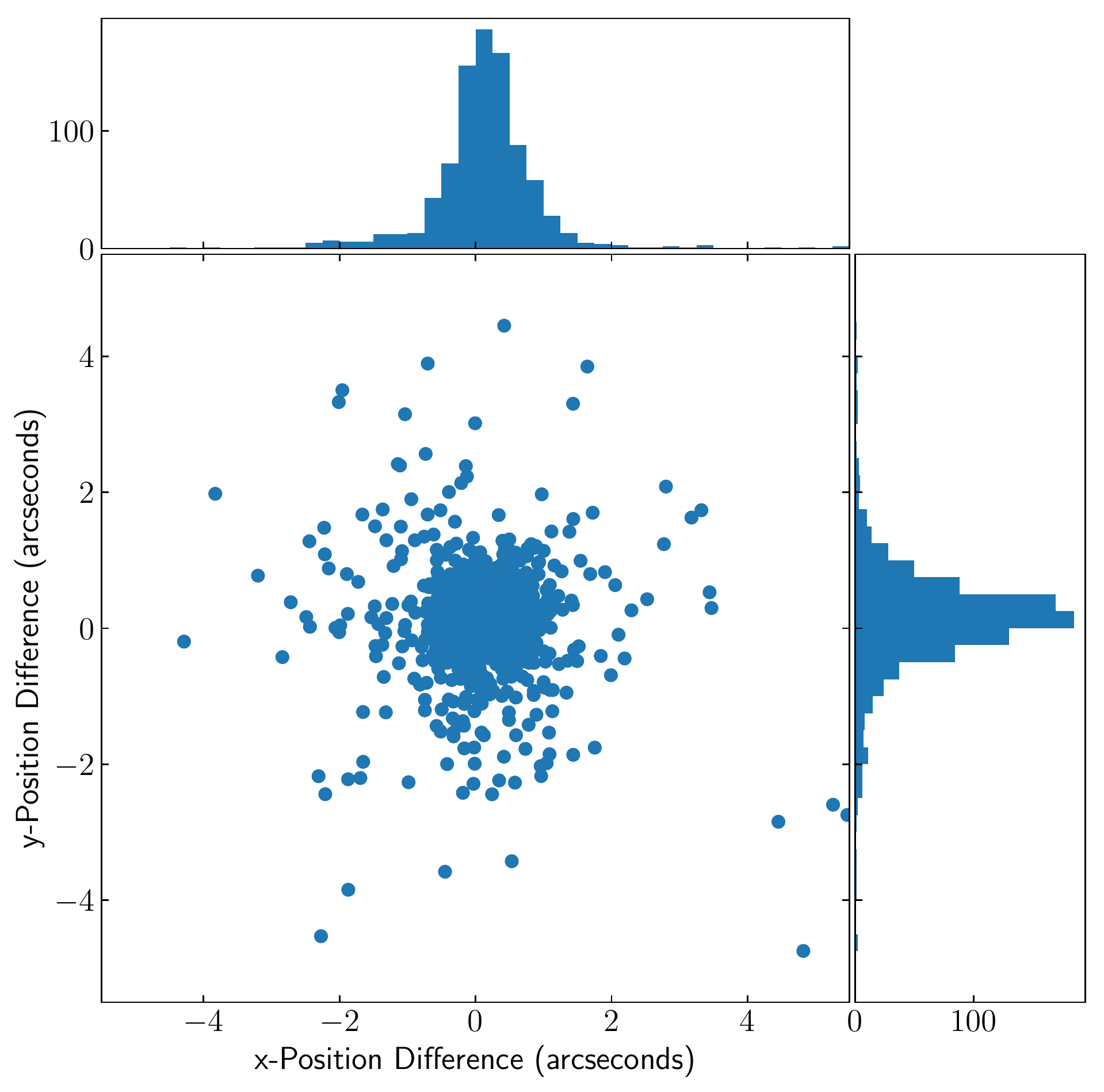}
    \caption{The astrometric precision of recovered candidates. The central scatter plot shows the difference of the injected source position and the recovered source position. The histograms show the values for the scatter plot with respect to the x and y directions. }
    \label{fig:RecoveredPosition}
\end{figure}

The recovered magnitude for the injected sources were compared with their injected magnitude to ensure consistency. Additionally, the coordinates of the injected sources were compared with their corresponding recovered candidate coordinates. The results of these tests can be found in Figures \ref{fig:RecoveredMag} and \ref{fig:RecoveredPosition}. The photometric precision for sources brighter than 13th Vega mag is $\approx 3$\%, while the astrometric RMS is 0.9\,\arcsec\, in each axis for all recovered sources (down to S/N = 5).

In addition to the first set of test data, we performed a second evaluation of the subtraction pipeline to assess its efficacy for finding sources in close proximity to nearby galaxies. An additional set of test data was generated using the same procedure described above, except that the positions for the fake sources were randomly selected to be within a 30\,\arcsec\, radius from nearby galaxies that are part of the Census of the Local Universe (CLU; \citealt{Cook2019}) catalog. The subtraction pipeline was run on this test data to check the rate of injected sources recovered as a function of the surface brightness at the location of the injected source. The test yielded what was at first an unexpected result, because the sky background level at J-band dominates over the galaxy light. This makes the `standard' test of surface brightness vs. recovery somewhat more difficult to interpret. Overall, the performance of this test was comparable to the earlier randomly placed fake sources. The pipeline found 366 out of 382 possible sources. This slightly improved rate of 95.8 \% sources found is likely due to the fact that CLU galaxies are found in regions where the J-band source density is not particularly high, unlike the Galactic plane fields which were included in the first test. Regions with high source densities make transient detection much more difficult because the data is typically not as deep due to the effects of confusion and there is higher chance of being masked by coincidence. Taking these into account, the improved performance of the second test makes sense and bodes well for finding transients in nearby galaxies with Palomar Gattini-IR.

\section{First results}
\label{sec:results}

Commissioning operations of Palomar Gattini IR started in November 2018 and continued until the end of June 2019. The quality of the data taken during the first few months was affected by extended periods of bad weather and high humidity, during which the data processing and transient discovery system were extensively tested and modified to produce better quality data products. As the data reduction procedures were finalized, data from the start of the survey were re-processed to produce the complete baseline of observations available from the acquired data. Survey operations of Palomar Gattini-IR began on 2019 Juy 02, to survey the entire celestial sphere visible from Palomar Observatory. We present initial science results from transients and variables detected in the commissioning phase. Candidates detected each night are accessible from the GROWTH Marshal \citep{Kasliwal2019} where they are vetted by an on-duty astronomer on the following day for assignment of follow-up optical / NIR imaging and spectroscopy with the Palomar 200-inch telescope using the optical Double Beam Spectrograph (DBSP) and the NIR Triple Spectrograph (TripleSpec). Follow-up is prioritized for sources coincident with nearby galaxies in the CLU catalog (candidate supernovae in nearby galaxies), host-less transients (candidate novae or dwarf novae) and large amplitude variables (candidate flaring stellar sources such as young stellar objects).

\subsection{Transient science}

\begin{figure*}
\centering
\includegraphics[width=0.87\textwidth]{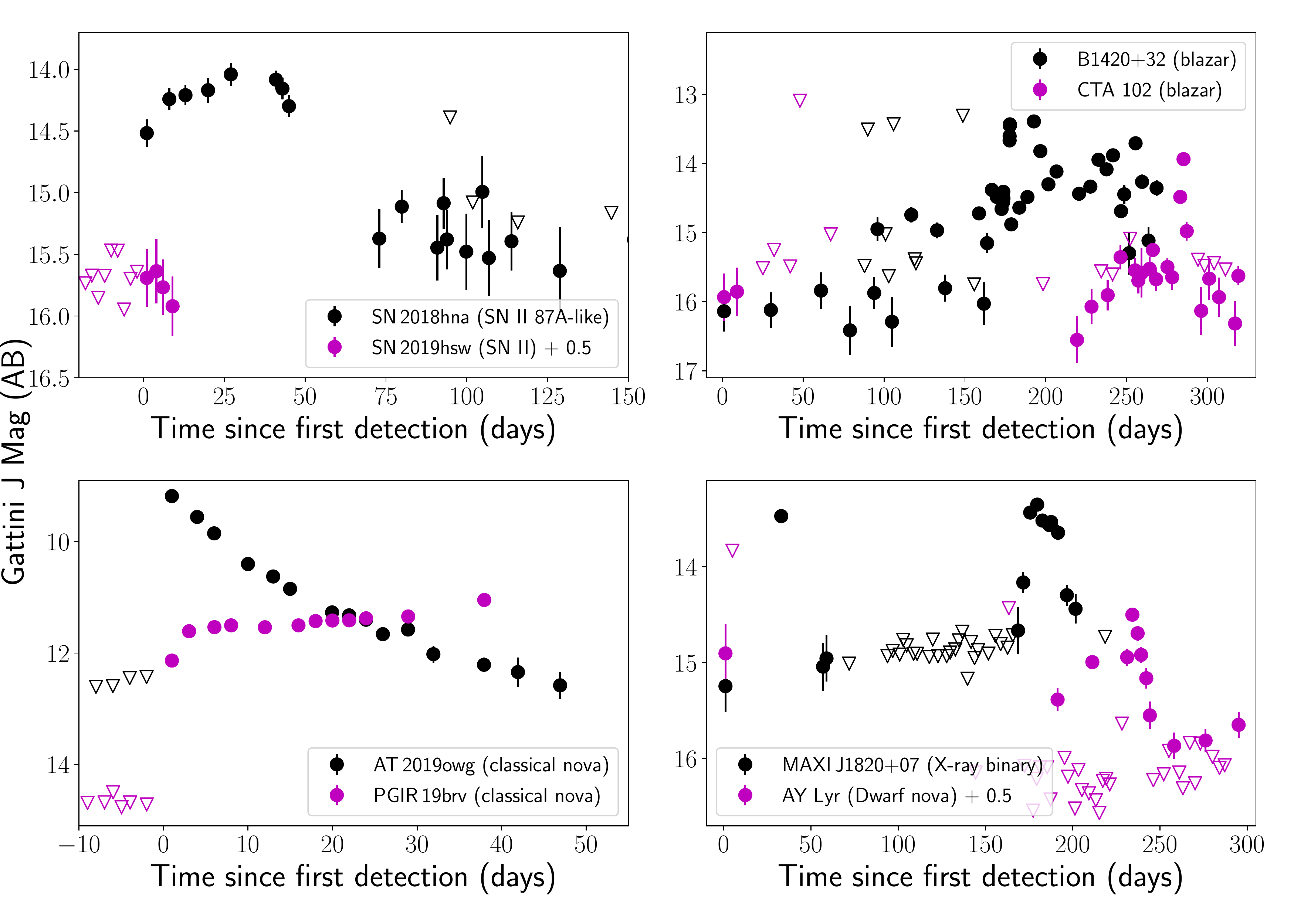}
\includegraphics[width=0.84\textwidth]{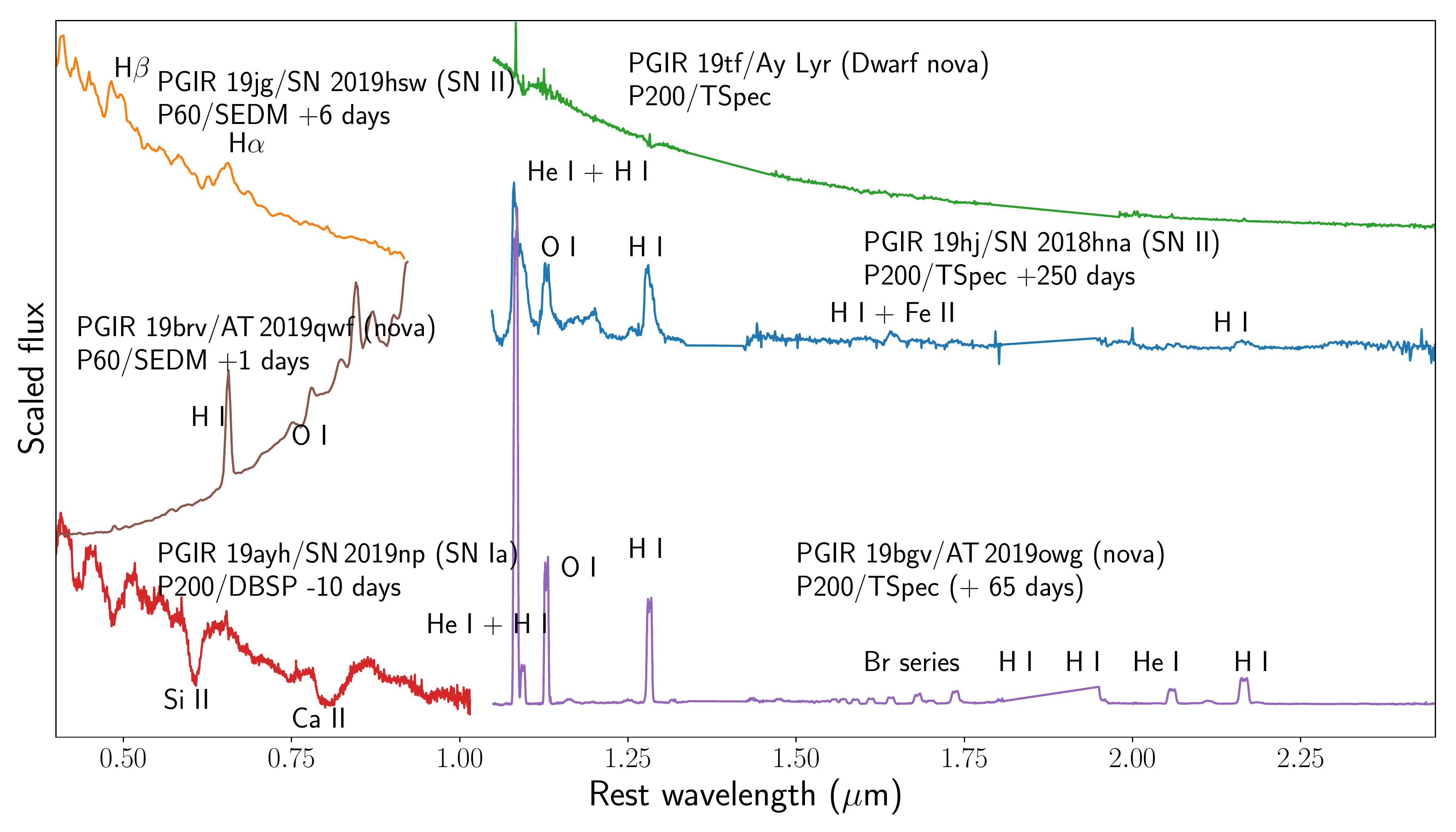}
\caption{(Top panel) Collage of light curves of extragalatic and Galactic transients deteced with Gattini. Circles denote detections while inverted triangles are 5$\sigma$ upper limits. The extragalactic transients include the Type II SN\,1987A-like SN\,2018hna (PGIR\,19hj) and Type II SN\,2019hsw (PGIR\,19jg), as well as the large amplitude flaring blazars B21420+32 (PGIR\,19c) and CTA 102 (PGIR\,19ayd). The Galactic transients include the reddened classical novae PGIR\,19brv (AT\,2019qwf) and PGIR\,19bgv (AT\,2019owg), a flaring X-ray binary MAXI\,J1820+07 (PGIR\,19auj) and a recurrent dwarf nova Ay Lyr (PGIR\,19tf). (Lower panel) Optical and NIR spectra obtained as follow-up for some of these sources, marking the prominent spectral lines in each source. In addition, we show an optical spectrum of the Type Ia SN\,2019np (PGIR\,19ayh).}
\label{fig:trans_lcspec}
\end{figure*}

Since the start of the commissioning period, several bright supernovae and extragalactic transients were recovered in the Gattini transient stream. These include the SNe II 2018hna, 2019hsw and SN Ia 2019np. In addition, large amplitude NIR flaring was detected from several high redshift blazars, a subset of which were followed up with optical / NIR spectroscopy on the Palomar 200-inch telescope. In Figure \ref{fig:trans_lcspec}, we show a collage of the J band light curves and spectra of extragalactic transients detected in the commissioning phase. SN\,2018hna was reported by K. Itagaki to the Transient Name Sever (TNS) early in the commissioning phase of Gattini-IR and brightened to be detectable in the NIR soon after explosion. Gattini observed the field as a part of regular survey operations with photometry capturing the entire rise to the radioactive peak and subsequent decline of this SN\,1987A-like Type II SN (Figure \ref{fig:trans_lcspec}, top left panel). SN\,2019hsw (ASASSN\,19pn / PGIR\,19jg) is a Type II SN at 25 Mpc detected in the Gattini-IR commissioning data, and shows a slow J band decline at early time (Figure \ref{fig:trans_lcspec}, top left panel). The SN went close to the sun soon after discovery and was not covered as a part of the nightly Gattini-IR observations. SN\,2019np (PGIR\,19ayh) was a nearby Type Ia SN reported to TNS \citep{Itagaki2019} and subsequently detected in Gattini data. Only two detections of the SN were recovered due to an extended period of poor weather surrounding the detection of the SN. PGIR\,19c is a large ampltiude flaring blazar (B1420+326) detected as a transient over several weeks of operations (Figure \ref{fig:trans_lcspec}, top right panel). The fast variability and large amplitude flaring detected in the Gattini-IR data were announced via the Astronomer's Telegram \citep{De2019a}. Gattini-IR detected a new NIR flare of the blazar CTA 102 (Figure \ref{fig:trans_lcspec}, top right panel), which was saved with the internal name PGIR\,19ayd and announced publicly \citep{De2019b}.

Figure \ref{fig:trans_lcspec} also shows light curves of several Galactic transients detected in the data -- highly reddened Galactic novae AT\,2019qwf and AT\,2019owg (Figure \ref{fig:trans_lcspec}, bottom left panel), the outbursting X-ray binary MAXI J1820+07 / ASASSN-18ey (\citealt{Tucker2018}; the NIR brightening was reported in \citealt{Hankins2019c}) and the dwarf nova PGIR\,19tf (Figure \ref{fig:trans_lcspec}, bottom right panel). AT\,2019qwf (PGIR\,19brv) was first discovered and reported as a bright ($\approx 11$\,mag) NIR transient at Galactic latitude of 0.2 degrees by Gattini-IR, and classified as a Galactic nova with optical spectroscopy \citep{De2019d}. Another reddened classical nova AT\,2019owg (initially reported as Gaia\,19dum to TNS) was detected as a bright NIR transient (at $\approx 8$th mag), close to the saturation magnitude of the instrument \citep{De2019c}. Between July and September 2019, Gattini-IR detected a total of four Galactic classical novae -- V3890 Sgr, Gaia\,19dum/AT\,2019owg, PGIR\,19brv/AT\,2019qwf and V2860\,Ori. Both MAXI\,J1820+07 and MAXI\,J1807+32 were detected as NIR transients coincident with a brightening detected in the optical wavebands. After re-processing older data taken during initial commissioning, a previous outburst of MAXI\,J1820+07 in 2019 was also recovered. Several recurrent outbursts from the dwarf nova Ay Lyr (PGIR\,19tf) were detected in the commissioning, and are shown in Figure \ref{fig:trans_lcspec}. Gattini-IR detected a reddened binary microlensing event in the Galactic plane (PGIR\,19btb / Gaia\,19dqj / AT\,2019odt) which was announced in \citealt{De2019e}.

Figure \ref{fig:trans_lcspec} also shows a collage of optical (from P60 + SED Machine / P200 + DBSP) and NIR spectra (from P200 + TripleSpec) obtained for these transients detected during the commissioning phase. PGIR\,19jg and PGIR\,19hj exhibit typical features of Type II SNe, including broad P-Cygni lines of H, He, O I and Fe II. A peak light spectrum of PGIR\,19ayh shows typical features of Type Ia SNe near peak -- Si II, S II, Ca II and Fe II. PGIR\,19brv (AT\,2019qwf) was followed up with rapid low resolution spectroscopy on the SED Machine spectrograph on the Palomar 60-inch telescope \citep{Blagorodnova2018}. The spectrum showed a reddened continuum and strong emission lines of H and O, classifying this source as a reddened Galactic classical nova \citep{De2019d}. We also obtained a NIR spectrum of the reddened Galactic nova AT\,2019owg, which shows broad emission lines of He and H along with strong emission lines of O I. The NIR spectrum of Ay Lyr (PGIR\,19tf) shows several narrow absorption lines of H, typical of dwarf nova outbursts.

Given the large field of view of Gattini, it has also been performing targeted follow-up of the localization regions of several alerts announced by LIGO/Virgo in O3. Gattini has demonstrated the capability to tile large fractions of the error regions of the localization regions, ranging from 32\% of the poorly localized single detector detection of LIGO/Virgo S190425z \citep{Coughlin2019b}, and of the localization region of the candidate NS-BH mergers S190426c (92\%; \citealt{Hankins2019a, Hankins2019b}) and S190814bv (89.5\%; \citealt{Hankins2019d}). In the case of S190426c, Gattini tiled the localization region a total of $\approx 20$ times over the course of one week after the merger, while each field in the localization region of S190814bv was observed for $\approx 2.5$ hours during one week after the merger. Given the longer timescale ($\sim 1$ week) of the infrared emission in kilonova counterparts \citep{Kasen2017}, stacking multiple epochs of data will allow the first constraints on infrared emission from compact binary mergers \textit{independent} of optical searches.

\subsection{Variable science}

\begin{figure*}[!ht]
\includegraphics[width=\textwidth]{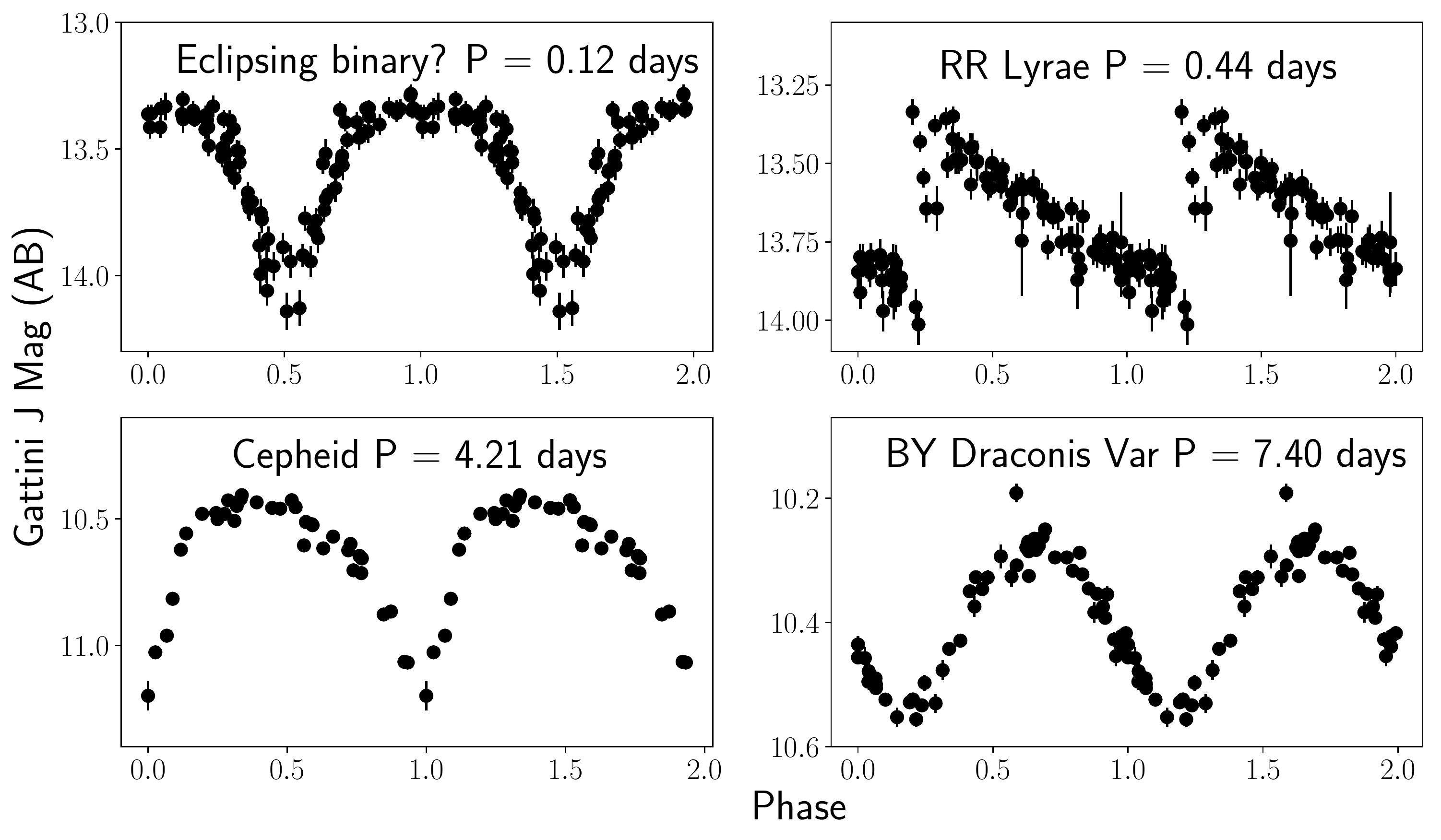}
\caption{Periodic variables recovered in the Gattini data. The periods were recovered by running a blind period finding algorithm on match file products generated by GDPS, and folding at the best period. This figure represents a sample of periodic sources recovered in the period search. All sources apart from the top left panel had previous classifications in SIMBAD, while the top left panel represents a candidate eclipsing binary with a short period of 0.12 days.}
\label{fig:pervariables}
\end{figure*}

As a part of nominal survey operations, Gattini-IR will obtain J-band light curves of sources brighter than J $\approx 16$ AB mag. The photometric measurements from these observations are readily available in the epochal PSF-fit source catalogs, which are cross-matched across epochs to produce match files for every detected source. In order to demonstrate the quality of light curves and the variable science potential of Gattini, we ran a blind period search algorithm on all sources detected more than 30 times. In Figure \ref{fig:pervariables}, we show a sample of periodic variables recovered from the blind period search, some of which already had known variable counterparts in SIMBAD. These include a candidate short period ($\approx 3$ hour) eclipsing binary, an RR-Lyrae type variable showing a distinct saw-tooth shaped light curve, a Cepheid variable and a BY Draconis type variable, demonstrating the photometric quality of the data and the capability for blind period searches. 

\begin{figure*}[!ht]
\includegraphics[width=\textwidth]{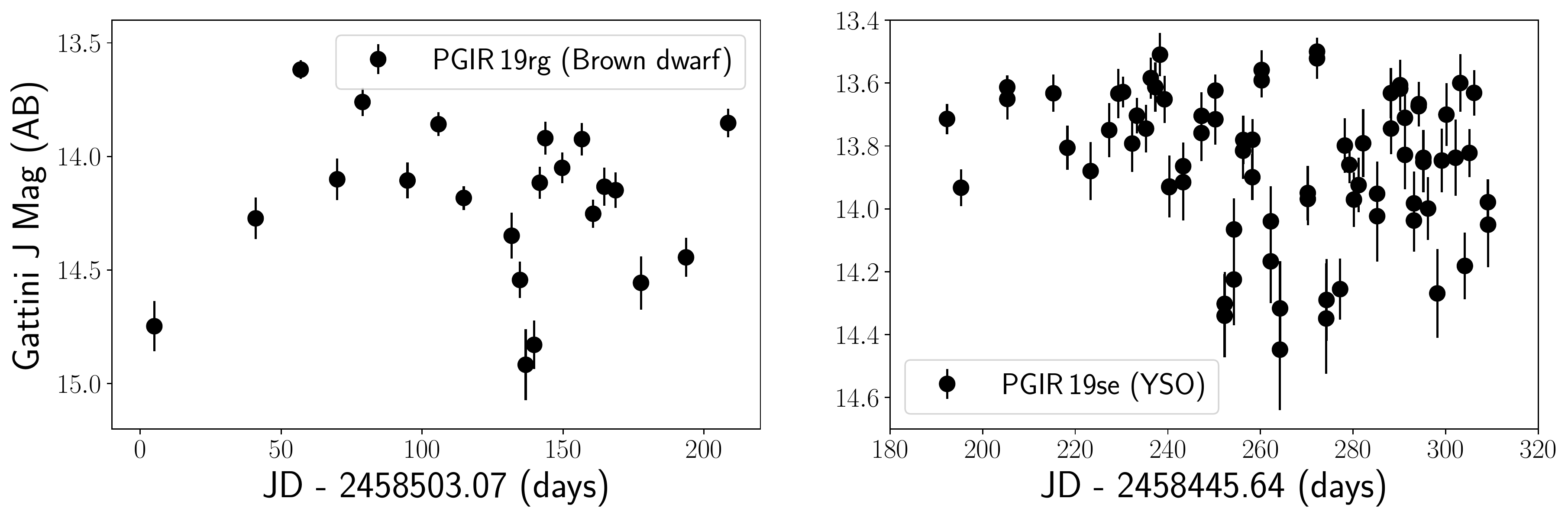}
\caption{Large amplitude Galactic variables identified as transients in the subtraction pipeline. The left panel shows a M3.3 type brown dwarf reported in \citealt{Dawson2014}. The right panel shows large amplitude variability from a candidate YSO PGIR\,19se, for which the classification was confirmed with optical spectroscopy on the Palomar 200-inch telescope.}
\label{fig:variables}
\end{figure*}

Figure \ref{fig:variables} also shows light curves of Galactic variables that were detected in the subtraction pipeline -- a candidate brown dwarf and a known Young Stellar Object (YSO). While the brown dwarf is undetected in the optical due to its red color, the variability from the YSO is undetected in the optical due to extinction. Operating in J band, Gattini will probe variability in the coolest and dustiest stars in the galaxy (such as brown dwarfs and asymptotic giant branch stars) that are bright in the infrared but faint in the optical. It will also be particularly sensitive to stellar variability in the most dust extinguished lines of sight in the Galactic plane where optical time domain surveys become insensitive. 

\section{Summary}
\label{sec:summary}
Palomar Gattini-IR is a new wide-field (25 square degree field of view) NIR time domain survey using a 30 cm telescope at Palomar Observatory. Gattini-IR operates in J band with a H2RG detector and a pixel scale of $\approx 8.7$\,\arcsec. Under the nominal survey, Gattini-IR scans the entire northern sky to a median 5$\sigma$ depth of 16 AB mag (outside the Galactic plane) with a median cadence of two nights. We presented the performance of the robotic scheduling system that runs the survey. The observing system scans $\approx 7500$ degrees of the sky every night with a median observing efficiency of $\approx 60$\%. We outline the design and operations of a real-time data processing system that produces science quality stacked and calibrated images from dithered raw images taken on sky, as well as transient candidates identified from subtractions. The calibrated science images are delivered within a median time of $\approx 2$ hours from the end of an observation while transient candidates are delivered within a median time of $\approx 4$ hours. 

The median astrometric accuracy of the stacked images (calibrated to Gaia DR2) is $\approx 0.7$\,\arcsec for sources with SNR $> 10$, while it is close to $\approx 0.3$\,\arcsec for sources brighter than 13 mag. The achieved photometric precision (calibrated against 2MASS) is $\approx 3$\% for sources brighter than 12 mag. Reference images were generated for the entire visible sky at the start of survey operations, and cover 99.3\% of the visible sky with 60\% of the reference image coverage having depths $> 16.5$ AB mag. As with the epochal science images, the reference image depths are limited by confusion noise near the Galactic plane. The efficiency of transient detection was estimated using fake sources injected into the data stream and found to be $\approx 90$\% for sources down to the 5$\sigma$ limiting magnitude. The photometric recovery precision (RMS) for injected sources is 3\% for transients brighter than 13\,Vega mag, while the astrometric recovery precision (RMS) is $\approx 0.9$\,\arcsec.

Survey operations for Palomar Gattini-IR began in July 2019, marking the end of the commissioning period. While the quality of the data during the commissioning period was affected by long periods of high humidity and bad weather causing condensation on a window in the OTA, this issue has been fixed with the recent installation of a window heating mechanism. Additionally, the planned installation of an automated focus mechanism is expected to yield better image quality and image depths moving into the second half of the first year survey. Additional planned improvements include using "sample up the ramp" to read out the detector, which will increase the dynamic range of the instrument for bright sources by 2.5 magnitudes.

With the largest field of view of any NIR imaging instrument, Gattini-IR is a pathfinder of time domain astronomy in the NIR. In addition to the stream of known optically bright transients and variables, Gattini will be sensitive to the reddest and dustiest explosions in the nearby universe and the stellar variability from the most dust extinguished regions of the galaxy that are inaccessible to current optical time domain surveys. As a demonstration of the science capabilities, we present sample results from transients and large amplitude variables detected since the start of the commissioning period. Gattini-IR is already discovering dust extinguished novae in the Galactic plane, and is expected to be sensitive to transients behind large columns of extinction within the galaxy where optical time domain surveys lose sensitivity. Over the course of the nominal two year survey, Gattini-IR will explore the phase space of transients and variables in the dynamic infrared sky for the first time with an untargeted, all-sky sampling at two day cadence. As the first working demonstration for wide-field NIR time domain astronomy, Gattini-IR will lead the way for future IR time domain experiments like WINTER at Palomar Observatory \citep{Simcoe2019}, and DREAMS \citep{Soon2018} at Siding Spring Observatory. 

\section*{Acknowledgements}
We thank A. Fruchter, F. Masci, S. R. Kulkarni, C. Steidel and M. J. Graham for valuable discussions on this work. MMK and EO acknowledge the US-Israel Bi-national Science Foundation Grant 2016227. MMK and JLS acknowledge the Heising-Simons foundation for support via a Scialog fellowship of the Research Corporation. MMK and AMM acknowledge the Mt Cuba foundation. J. Soon is supported by an Australian Government Research Training Program (RTP) Scholarship. SED Machine is based upon work supported by the National Science Foundation under Grant No. 1106171.

KD and MJH thank the hospitality of the astrophysics group at the Weizmann Institute of Science, Rehovot, Israel, where part of this work was carried out. This work was supported by the GROWTH (Global Relay of Observatories Watching Transients Happen) project funded by the National Science Foundation under PIRE Grant No 1545949. GROWTH is a collaborative project among the California Institute of Technology (USA), University of Maryland College Park (USA), University of Wisconsin Milwaukee (USA), Texas Tech University (USA), San Diego State University (USA), University of Washington (USA), Los Alamos National Laboratory (USA), Tokyo Institute of Technology (Japan), National Central University (Taiwan), Indian Institute of Astrophysics (India), Indian Institute of Technology Bombay (India), Weizmann Institute of Science (Israel), The Oskar Klein Centre at Stockholm University (Sweden), Humboldt University (Germany), Liverpool John Moores University (UK) and University of Sydney (Australia).  

The High Performance Wireless Research \& Education Network (HPWREN; \url{https://hpwren.ucsd.edu}) is a project atthe University of California, San Diego and the National Science Foundation (grant numbers 0087344 (in 2000), 0426879 (in 2004), and 0944131 (in 2009)). This publication makes use of data products from the Two Micron All Sky Survey, which is a joint project of the University of Massachusetts and the Infrared Processing and Analysis Center/California Institute of Technology, funded by NASA and the National Science Foundation. This work has made use of data from the European Space Agency (ESA) mission Gaia (\url{https://www.cosmos.esa.int/gaia}),processed by the Gaia Data Processing and Analysis Consortium (DPAC, \url{https://www.cosmos.esa.int/web/gaia/ dpac/consortium}). Funding for the DPAC has been provided by national institutions, in particular the institutions participating in the Gaia Multilateral Agreement. This work has also made use of the Pan-STARRS1 (PS1) Surveys (\url{http://www.ifa.hawaii.edu/pswww/}) and the PS1 public science archive (\url{https://panstarrs.stsci.edu}). 

\software{\texttt{astropy} \citep{Astropy2013}, \texttt{matplotlib} \citep{Hunter2007}, \texttt{scipy} \citep{Virtanen2019}, \texttt{pandas} \citep{McKinney2010}, \texttt{SExtractor} \citep{Bertin1996}, \texttt{scamp} \citep{Bertin2006}, \texttt{PSFEx} \citep{Bertin2011}, \texttt{pysedm} \citep{Rigault2019}, \texttt{pyraf-dbsp} \citep{Bellm2016b}, \texttt{spextool} \citep{Cushing2004}, \texttt{xtellcor} \citep{Vacca2003}}

\appendix
\section{General purpose optical / NIR image reduction pipeline}
\label{sec:appendix_drp}
We provide a brief description of the general purpose optical / NIR image reduction pipeline developed initially for optical (WASP) and NIR (WIRC) imaging instruments on the Palomar 200-inch. The code is highly modular to allow support for a large range of optical and NIR instruments, and was subsequently updated to support data from MOSFIRE on the Keck-I telescope and the FOURSTAR camera on the Magellan telescope. The code is written completely in \texttt{python}, using several functions from \texttt{astropy} and the astromatic suite of software for source extraction, astrometry and stacking. The code will be publicly released on \texttt{github} for general use. The parameters of the reduction are controlled using a configuration file that can be modified for custom reductions. We outline the various implemented steps implemented for the reduction of optical / NIR data .
\begin{enumerate}
    \item The code uses the headers of the raw files to create a log for all images recorded in the night. This log is used to generate a list of science and calibration exposures to be processed.
    \item A master dark image is created using darks found in the list of calibration frames. The master dark is subtracted from all the images prior to processing.
    \item A master flat image is created using either dome flats (if available) or sky frames (using the bright sky background in the NIR). Since individual exposures can contain extended galaxies\footnote{The nominal observing strategy in such cases is to interleave the target exposures with sky exposures outside the galaxy field.} (if the observing program involves a nearby galaxy) which do not reflect the flatness of the detector, the pipeline runs \texttt{SExtractor} to generate a list of detected sources, and checks for large extended sources in the field occupying more than 20\% of the image. If an extended galaxy is detected, the specific image is not used in the flat-field generation. Images with no large extended sources are marked to be included in the flat-field generation. 
    \item The pipeline generates a first-pass flat-field by performing a median combination of the sky frames,  which is used to flat-field all the target exposures. In case separate dome flat or twilight flats are available, those images are used to generate a median combined flat frame. 
    \item For each science exposure, the pipeline uses a median combination of the nearest sky (without extended sources) exposures to create a sky image which is subtracted from the science frame. 
    \item Following the generation of flat-fielded and sky-subtracted science frames, the code proceeds to generating a preliminary astrometric solution using the initial WCS from the image header and the \texttt{autoastrometry}\footnote{\url{http://www.astro.caltech.edu/~dperley/programs/autoastrometry.py}} code, with 2MASS (or SDSS, if available) as the reference catalog. The initial astrometric solution only includes a distortion free CD matrix.
    \item The \texttt{scamp} code is then used to derive a refined astrometric solution using Gaia DR2 as the reference catalog including a 3rd order distortion solution in the field. Given the small field of view (few arc-minutes) of most imaging instruments, the two-step astrometric solution (fitting distortions in the second pass astrometry after obtaining an initial solution) was found to have a higher success rate for astrometric calibration.
    \item The astrometric solutions are used to generate a first-pass stacked image using the \texttt{Swarp} package. \texttt{Sextractor} is run on the first pass image to generate a list of detected sources. 
    \item The source map from the first pass stacked image is used to mask sources in the individual images by mapping positions in the stacked image to the individual science images. The master flat-field image (in case the flat-field was generated using sky images) and sky frames are then re-generated for the individual target exposures.
    \item The science images are flat fielded and sky subtracted again using the new flat and sky images and then stacked using the existing astrometric solution using \texttt{Swarp}. This stacked image serves as the final stacked image product.
    \item A photometric solution is derived on the final stacked image using the relevant catalog for the observed filter. \texttt{PSFEx} is used to generate a PSF model and corresponding PSF zero-points followed by aperture corrections for several apertures. The photometric solution is written to the header of the final stacked image product.
\end{enumerate}{}
In the case of optical instruments, dome flats are directly used for producing flat field calibrations and a median spatial filter is used for sky subtraction.

\end{document}